\documentclass[12pt,preprint]{aastex}

\shorttitle{ALMA observations of NGC 1808}
\shortauthors{Salak et al.}

\begin{document}

\title{Gas dynamics and outflow in the barred starburst galaxy NGC 1808 revealed with ALMA}

\author{Dragan Salak}

\affil{Department of Physics, School of Science and Technology, Kwansei Gakuin University, Gakuen 2-1 Sanda, Hyogo 669-1337, Japan}
\email{d.salak@kwansei.ac.jp}

\author{Naomasa Nakai and Takuya Hatakeyama}

\affil{Division of Physics, Faculty of Pure and Applied Sciences, University of Tsukuba, Tennodai 1-1-1, Tsukuba, Ibaraki 305-8571, Japan}

\and

\author{Yusuke Miyamoto}
\affil{Nobeyama Radio Observatory, National Astronomical Observatory of Japan, 462-2 Nobeyama, Minamimaki, Minamisaku, Nagano 384-1305, Japan}

\begin{abstract}
NGC 1808 is a nearby barred starburst galaxy with an outflow from the nuclear region. To study the inflow and outflow processes related to star formation and dynamical evolution of the galaxy, we have carried out \(^{12}\)CO (\(J=1-0\)) mapping observations of the central \(r\sim4\) kpc of NGC 1808 using the Atacama Large Millimeter/submillimeter Array (ALMA). Four distinct components of molecular gas are revealed at high spatial resolution of 2\arcsec (\(\sim100\) pc): (1) a compact (\(r<200\) pc) circumnuclear disk (CND), (2) \(r\sim500\) pc ring, (3) gas-rich galactic bar, and (4) spiral arms. Basic geometric and kinematic parameters are derived for the central 1-kpc region using tilted-ring modeling. The derived rotation curve reveals multiple mass components that include (1) a stellar bulge, (2) nuclear bar and molecular CND, and (3) unresolved massive (\(\sim10^7~M_\sun\)) core. Two systemic velocities, 998 km s\(^{-1}\) for the CND and 964 km s\(^{-1}\) for the 500-pc ring, are revealed, indicating a kinematic offset. The pattern speed of the primary bar, derived by using a cloud-orbit model, is \(56\pm11\) km s\(^{-1}\) kpc\(^{-1}\). Non-circular motions are detected associated with a nuclear spiral pattern and outflow in the central 1-kpc region. The ratio of the mass outflow rate to the star formation rate is \(\dot{M}_\mathrm{out}/SFR\sim0.2\) in the case of optically thin CO (1-0) emission in the outflow, suggesting low efficiency of star formation quenching.
\end{abstract}


\section{Introduction}

Inflows and outflows of baryons play important roles in galaxy evolution. In addition to circular rotation, observations of the interstellar medium (ISM) using molecular spectroscopy have revealed significant non-circular motions of gas clouds in galaxies. These motions are often observed in galaxies that exhibit non-axisymmetric gravitational potential due to spiral and bar structures, and in those that undergo tidal interaction with their neighbours. We consider the following four-step evolutionary process: (1) cold gas is transported toward the galactic center by gravitational torques, i.e., by losing angular momentum, or by accretion from intergalactic medium (IGM). One of the consequences of gas infall is the crowding of giant molecular clouds (GMCs) in nuclear rings at the outermost inner Lindblad resonance (radius a few 100 pc), resulting in (2) efficient feeding of star formation activity (e.g., \citealt{Ath92b,BC96,PR00}). Bursts of star formation (starburst episodes), in turn, may trigger (3) outflows of ISM material (feedback) that remove molecular gas via supernova explosions and stellar winds, thereby suppressing further star formation in some regions while igniting it in others by outflow-induced shocks (e.g., \citealt{CC85,Sch85,MQT05,MMT11,VCB05,NS09,Bol13a}). Nuclear rings are also important gas reservoirs that may fuel the central accretion disks via nuclear (secondary) bars or spiral arms down to a scale of \(\sim10\) pc (e.g., \citealt{GB05,Com14}), though the mechanism of angular momentum removal from kiloparsec to sub-parsec scale is not fully understood. Finally, the expelled ISM can either fall back onto the galaxy as a fountain, or escape and pollute the IGM with heavy elements, eventually (4) accreting onto another galaxy.

In order to study gas dynamics related to the inflow/outflow cycle in starburst galaxies and the evolution of GMCs in barred galactic disks, comprehensive observations are necessary. While GMC-scale studies require angular resolution \(\sim100\) pc or better, diffuse outflows can be detected only in measurements with high surface-brightness sensitivity. This is now readily achievable with the superb imaging capabilities of the Atacama Large Millimeter/submillimeter Array (ALMA) (e.g., \citealt{Com13,GB14,Sak14}). In this work, we have used ALMA to image the molecular gas traced with the rotational line \(^{12}\)CO (\(J=1-0\)) in the nearby (10.8 Mpc; \citealt{Tul88}) starburst galaxy NGC 1808. As briefly described below, the galaxy is a promising case-study target that shows signatures of the feeding/feedback evolution in the nearby Universe.

\begin{figure}
\epsscale{1}
\plotone{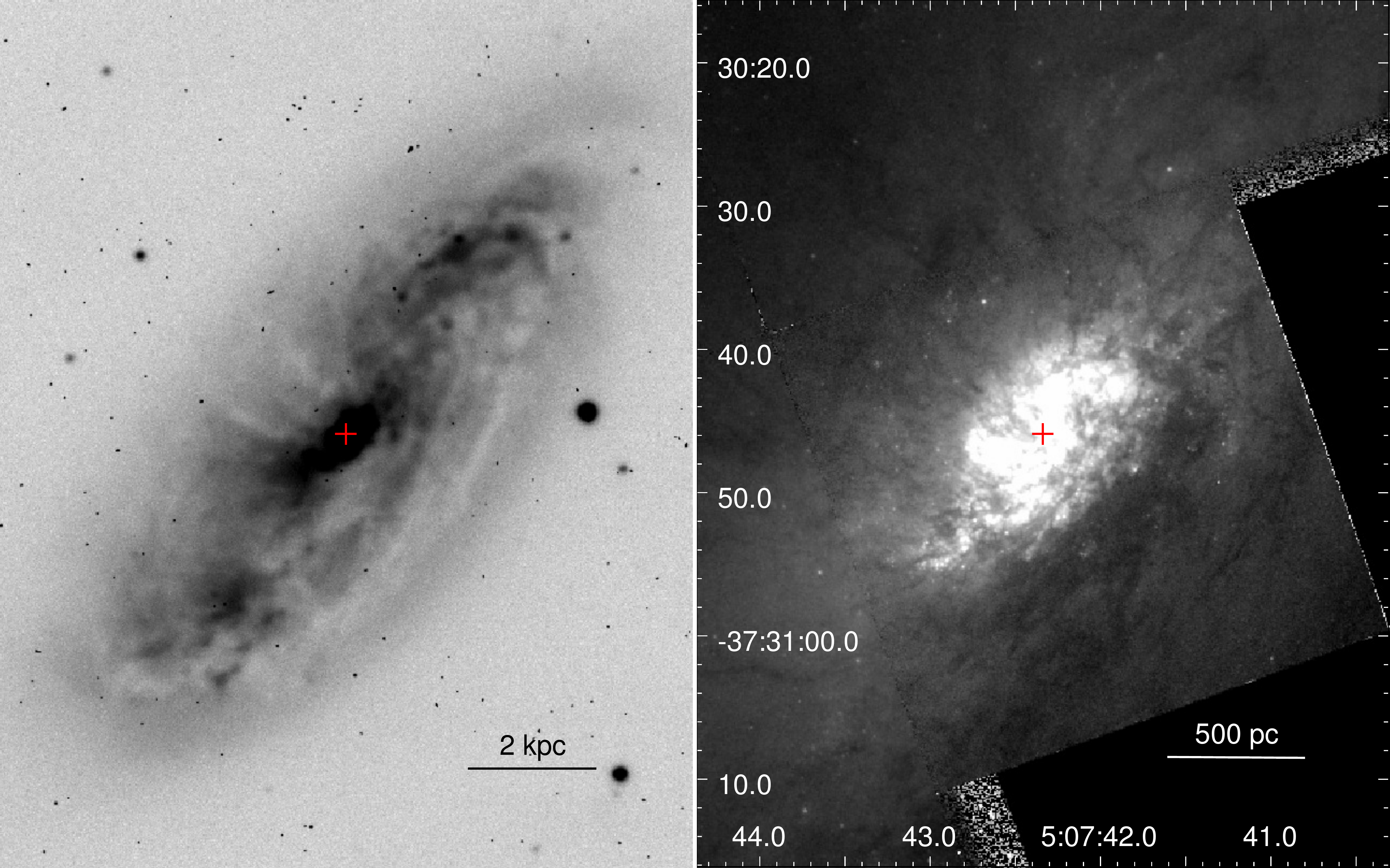}
\caption{\emph{Left.} Large-scale \(R\)-band negative image of NGC 1808 \citep{Meu02}. The cross marks the galactic center determined from our continuum observations (section 3.1). Dust lanes perpendicular to the galactic disk suggest outflows. \emph{Right.} HST image (675W) of the central region (acquired from Hubble Legacy Archive).}
\label{fig1}
\end{figure}

NGC 1808 is a (R)SAB(s)a barred spiral galaxy \citep{deV91} known for peculiar ``hot spots'' in the star-forming central 500-pc region \citep{Mor58,SP65}, and prominent polar dust lanes (figure \ref{fig1}) revealed in optical studies \citep{BB68,Phi93} that indicate an outflow of dust and neutral gas. The central region is dotted with radio and infrared sources revealing supernova remnants and young star clusters \citep{Sai90,For92,Kra94,Kot96,TG96,GA08}, while the galactic center harbors a candidate for a low-luminosity active galactic nucleus (AGN) \citep{VV85,Awa96,Jim05}. Atomic gas traced with the H\textsc{i} 21-cm line is concentrated in the galactic bar, disk, and a warped outer ring, indicating a tidal interaction in the past \citep{Kor93,Kor96}. The galaxy has been observed in CO before, although only with single-dish telescopes at low angular resolution of \(\gtrsim22\arcsec\) \citep{Dah90,Aal94,Sal14}. In this work, we present the first high-resolution CO (1-0) and 2.8-mm band continuum data.

The paper is organized as follows. In section 2, we describe the observations and data reduction. The results are presented in section 3, including the properties of the continuum and CO (1-0) line emission on the galaxy scale, followed by discussion (sections 4, 5, and 6) where we focus on the central 1-kpc molecular zone (CO gas distribution, kinematics, and dynamics) and briefly address the dynamics of the large-scale bar. A dynamical model is introduced in section 4, extended by an analysis of the molecular gas outflow in section 5. Finally, star formation is briefly discussed in section 6. Analyses of the galaxy-scale GMC properties and star formation in the bar are the subject of a forthcoming paper that includes combined 12-m array, 7-m array (ACA), and total power data (Salak et al. in prep.). The work is summarized in the last section.

The basic parameters of NGC 1808 are listed in table \ref{tab1}. The velocity in this paper is defined with respect to the local standard of rest (LSR) in radio definition, and all images are displayed in equatorial coordinates \((\alpha,\delta)_\mathrm{J2000.0}\).

\begin{table}
\begin{center}
\caption{Basic parameters of NGC 1808.}\label{tab1}
\begin{tabular}{ll}
\tableline\tableline
Right ascension (J2000.0)\tablenotemark{a} & \(05^\mathrm{h}07^\mathrm{m}42\fs343\) \\
Declination (J2000.0)\tablenotemark{a} & \(-37\degr30\arcmin45\farcs95\) \\
Distance\tablenotemark{b} & 10.8 Mpc \\
Adopted linear scale & \(1\arcsec=52\) pc \\
Systemic velocity (LSR)\tablenotemark{c} & \(963.9\pm2.5\) km s\(^{-1}\) (global) \\
& \(998.4\pm2.5\) km s\(^{-1}\) (\(r<235\) pc) \\
Position angle\tablenotemark{d} & \(324\degr\) \\
Inclination\tablenotemark{e} & \(57\degr\) \\
Morphological type\tablenotemark{f} & (R)SAB(s)a \\
Activity\tablenotemark{g} & H\textsc{ii}, Seyfert 2 (?) \\
\tableline
\end{tabular}
\tablenotetext{a}{Derived in this work from the peak of the 2.8-mm continuum emission (see table \ref{tab3}).}
\tablenotetext{b}{\cite{Tul88}.}
\tablenotetext{c}{Derived in this paper from the CO (1-0) line profile.}
\tablenotetext{d}{Derived in this work for the central 1 kpc from CO kinematics and a \(K_s\)-band image; receding side of the galaxy measured from N toward E. The derived value is close to the global one found in optical studies (\(323\arcdeg\); \citealt{LV89}).}
\tablenotetext{e}{\cite{Rei82}.}
\tablenotetext{f}{\cite{deV91}.}
\tablenotetext{g}{NED classification.}
\end{center}
\end{table}

\clearpage

\section{Observations and data reduction}

\begin{figure}
\epsscale{0.75}
\plotone{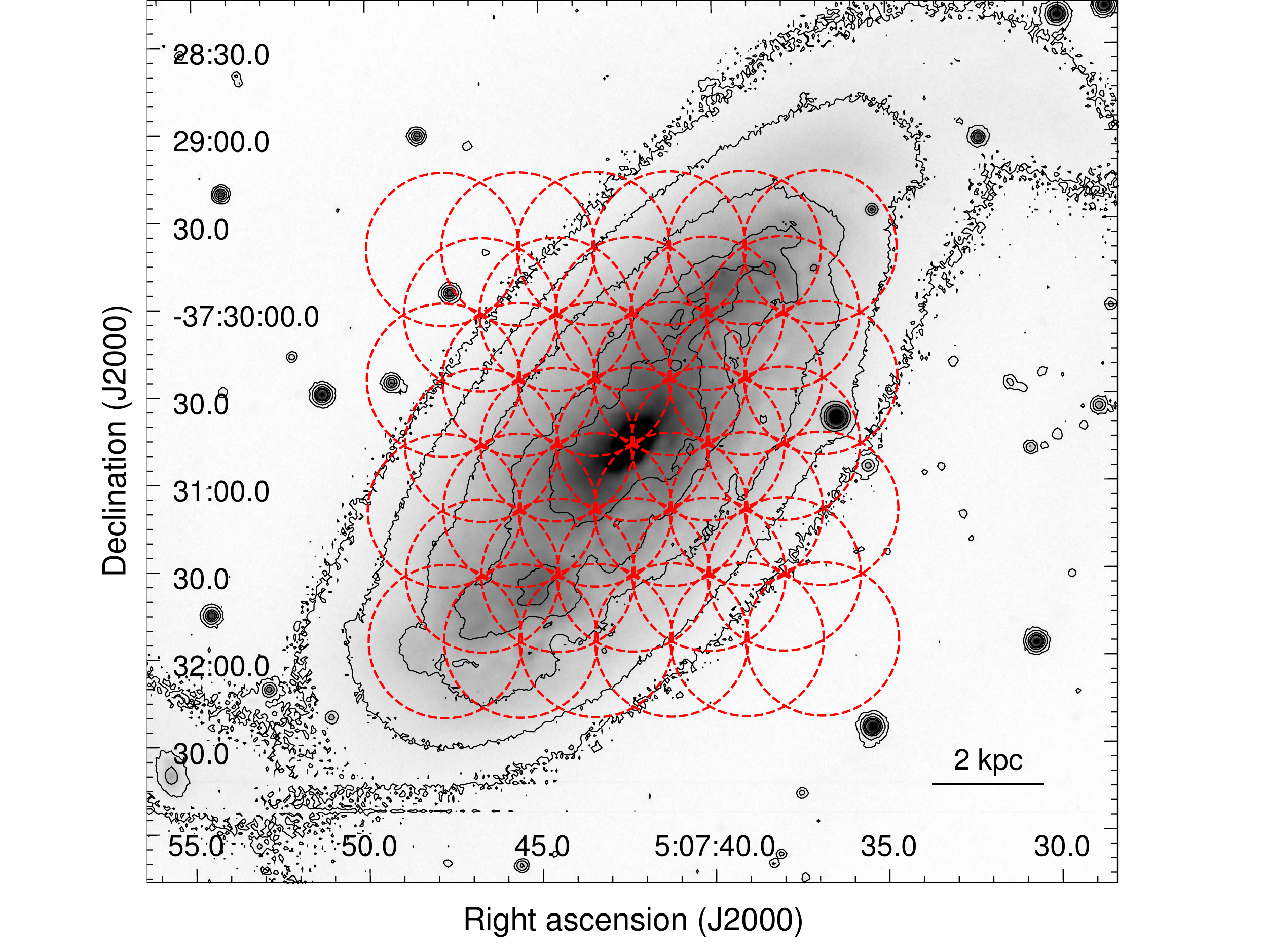}
\caption{Observed fields (39 pointings) superimposed on an \(R\)-band negative image of NGC 1808 (same as in figure \ref{fig1}); the circle size is the primary beam with a FWHM of \(52\arcsec\) (approx. \(1.13\lambda/D\) where \(D=12\) m). Contours are plotted on a square root scale from 0.1\% to 2\% of the peak intensity.}
\label{fig2}
\end{figure}

\begin{figure}
\epsscale{0.6}
\plotone{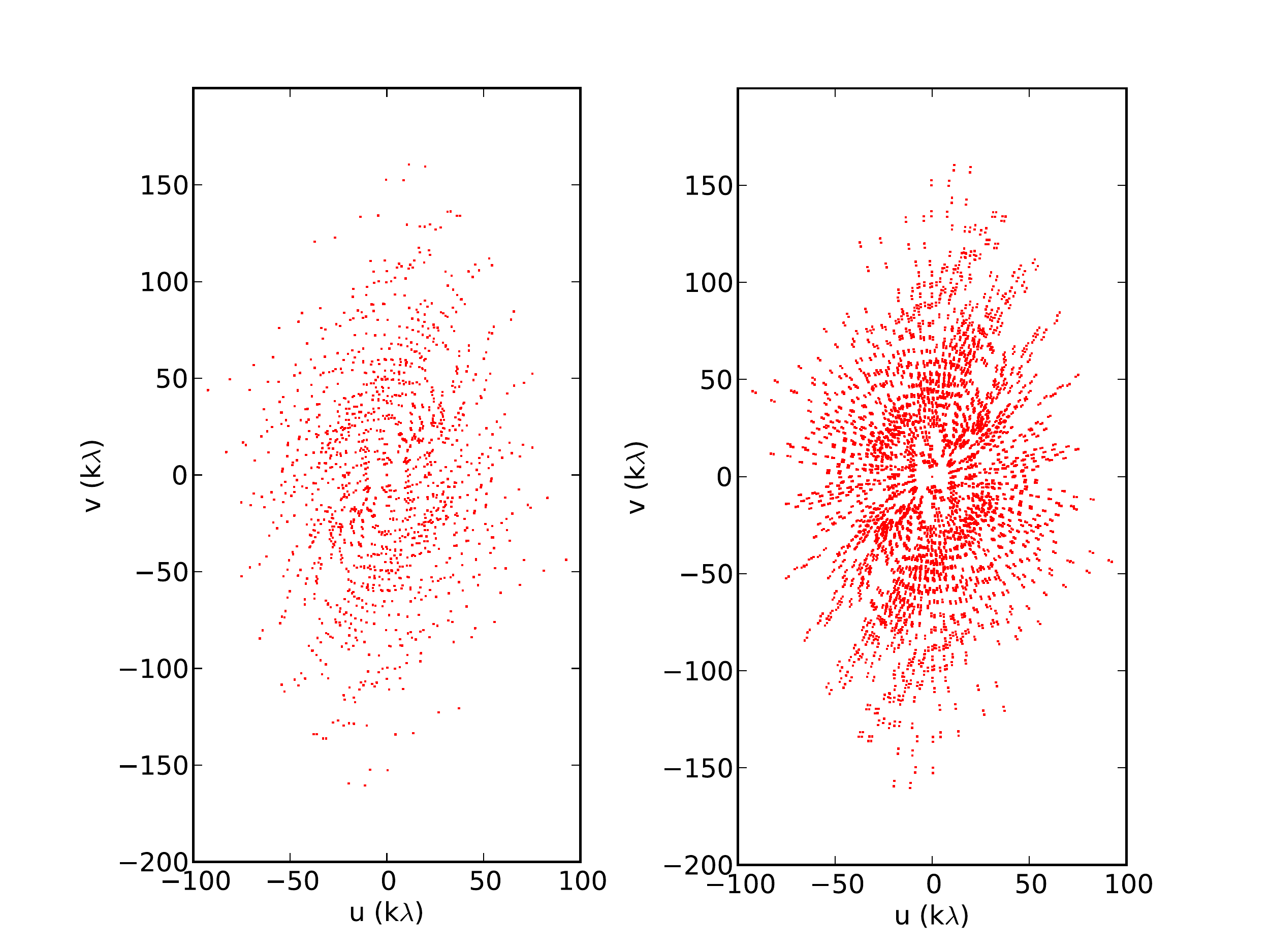}
\caption{\((u,v)\) coverage of the central pointing of the CO (1-0) observation (left), and continuum observation (right).}
\label{fig3}
\end{figure}

Observations were conducted on 2014 March 8, with twenty seven antennas of the ALMA 12-m array as part of our cycle 1 project. The imaging was performed toward 39 fields combined into a \(150\arcsec\times150\arcsec\) mosaic (about \(7.5\times7.5\) kpc\(^2\) at the adopted distance of 10.8 Mpc) shown in figure \ref{fig2}. The shortest projected baseline of the 12-m array was \(L_\mathrm{min}\approx15\) m, hence the visibilities that correspond to \((u,v)\lesssim6\) k\(\lambda\), where \(\lambda\) is the observed wavelength, were not sampled (see figure \ref{fig3}). The maximum recoverable scale per pointing is \(\theta_\mathrm{MRS}\simeq0.6\lambda/L_\mathrm{min}\approx21\arcsec\); any structure larger than this is entirely filtered out. The longest baseline of the 12-m array was \(L_\mathrm{max}=423\) m, allowing sampling of high spatial frequencies (angular resolution given by \(\theta=k\lambda/L_\mathrm{max}\), where \(k\) is a constant that depends on the weighting function). The cycle 1 project includes supplement observations using the Atacama Compact Array (ACA) and total power (TP) arrays needed to recover extended emission and total flux, and the combined data will be presented in a future paper. The sources used as calibrators were: Callisto (flux density), J0609-1542 (bandpass), and J0522-3627 (phase). The total on-source time was about 41 minutes, but the integration time per pointing was about 1 minute. Four spectral windows at the rest frequencies of (1) 101.271, (2) 103.200, (3) 113.271, and (4) 115.271202 GHz (rest frequency of the rotational transition CO 1-0) were observed simultaneously, each with a bandwidth of 2 GHz except window 4 which had a bandwidth of 1.870 GHz (4892.3 km s\(^{-1}\)) and high spectral resolution of 976.56 kHz (2.548 km s\(^{-1}\)) to resolve the CO (1-0) line.

Data reduction was carried out using the Common Astronomy Software Applications (CASA) package \citep{McM07}. The data were calibrated, continuum-subtracted, deconvolved using the CLEAN algorithm with Briggs weighting. The CLEAN procedure was carried out interactively until the residual map resembled pure noise. The iteration was then proceeded automatically until a threshold of about 1 \(\sigma\) was reached. This threshold was adopted by inspecting emission-free channels in the dirty map. As a result, despite a relatively coarse \((u,v)\) coverage, side lobes were successfully suppressed to negligible levels comparable to 1 \(\sigma\) r.m.s. Two images were produced to analyze different galactic scales: (1) high-resolution image with \(\mathrm{robust}=0.5\), velocity resolution of 2.548 km s\(^{-1}\), and spatial resolution [the full-width at half maximum (FWHM) of the synthesized beam] of \(2\farcs26\times1\farcs23\) (\(117~\mathrm{pc}\times64~\mathrm{pc}\)) for the galactic central region where the signal-to-noise ratio is highest, and (2) high-sensitivity image with \(\mathrm{robust}=2\) and velocity resolution 10.192 km s\(^{-1}\) to increase the sensitivity of extended structure over the entire mosaic. The angular resolution in the final data cube (\(\mathrm{robust}=2\)) is \(2\farcs55\times1\farcs41\) (\(133~\mathrm{pc}\times73~\mathrm{pc}\)). The r.m.s. sensitivity of the CO image in emission-free channels is 5.5 mJy beam\(^{-1}\) per velocity bin of 10.2 km s\(^{-1}\), or about 0.14 K (1 \(\sigma\)) in brightness temperature units. The final continuum image produced by multi-frequency synthesis using four spectral windows after subtracting the contribution from CO (1-0) has a sensitivity of 0.20 mJy beam\(^{-1}\). The continuum image was created with \(\mathrm{robust}=0.5\) weighting to achieve a compromise between angular resolution and sensitivity expecting that emission is confined to the galactic nuclear region. The images presented in figures are not corrected for primary beam attenuation, except those of the central region. The images were corrected, however, in order to derive fluxes and masses.

Observational parameters are summarized in table \ref{tab2}.

\begin{table}
\begin{center}
\caption{Observational summary.}\label{tab2}
\begin{tabular}{ll}
\tableline\tableline
ALMA 12-m array & \\
\tableline
Number of antennas & 27 \\
Observation date & 2014 March 8 \\
Flux calibrator & Callisto \\
Bandpass calibrator & J0609-1542 \\
Phase calibrator & J10522-3627 \\
Map size & \(150\arcsec\times150\arcsec\) (\(7.8\times7.8\) kpc\(^2\)) \\
Mosaic & 39 fields (Nyquist sampling) \\
Primary beam FWHM (\(\lambda=2.7\) mm) & \(52\arcsec\) \\
Synthesized beam & \(\theta_\mathrm{maj}\times\theta_\mathrm{min}\) \\
~~~High sensitivity, FWHM\tablenotemark{a} & \(2\farcs55\times1\farcs41\) \\
~~~High resolution, FWHM\tablenotemark{b} & \(2\farcs26\times1\farcs23\) \\
Total time on source & 41 min \\
Sensitivity & 1 \(\sigma\) \\
~~~High sensitivity, CO (1-0)\tablenotemark{a} & 5.5 mJy beam\(^{-1}\) \\
~~~High resolution, CO (1-0)\tablenotemark{b} & 10.0 mJy beam\(^{-1}\) \\
~~~Continuum\tablenotemark{c} & 0.20 mJy beam\(^{-1}\) \\
\tableline
\end{tabular}
\tablenotetext{a}{Briggs weighting with \(\mathrm{robust}=2\) and velocity resolution \(\Delta v=10.2\) km s\(^{-1}\).}
\tablenotetext{b}{Briggs weighting with \(\mathrm{robust}=0.5\) and velocity resolution \(\Delta v=2.5\) km s\(^{-1}\).}
\tablenotetext{c}{Bandwidth per spectral window \(B=2\) GHz.}
\end{center}
\end{table}

\clearpage

\section{Results}

\subsection{Continuum emission}

\begin{figure}
\epsscale{0.9}
\plotone{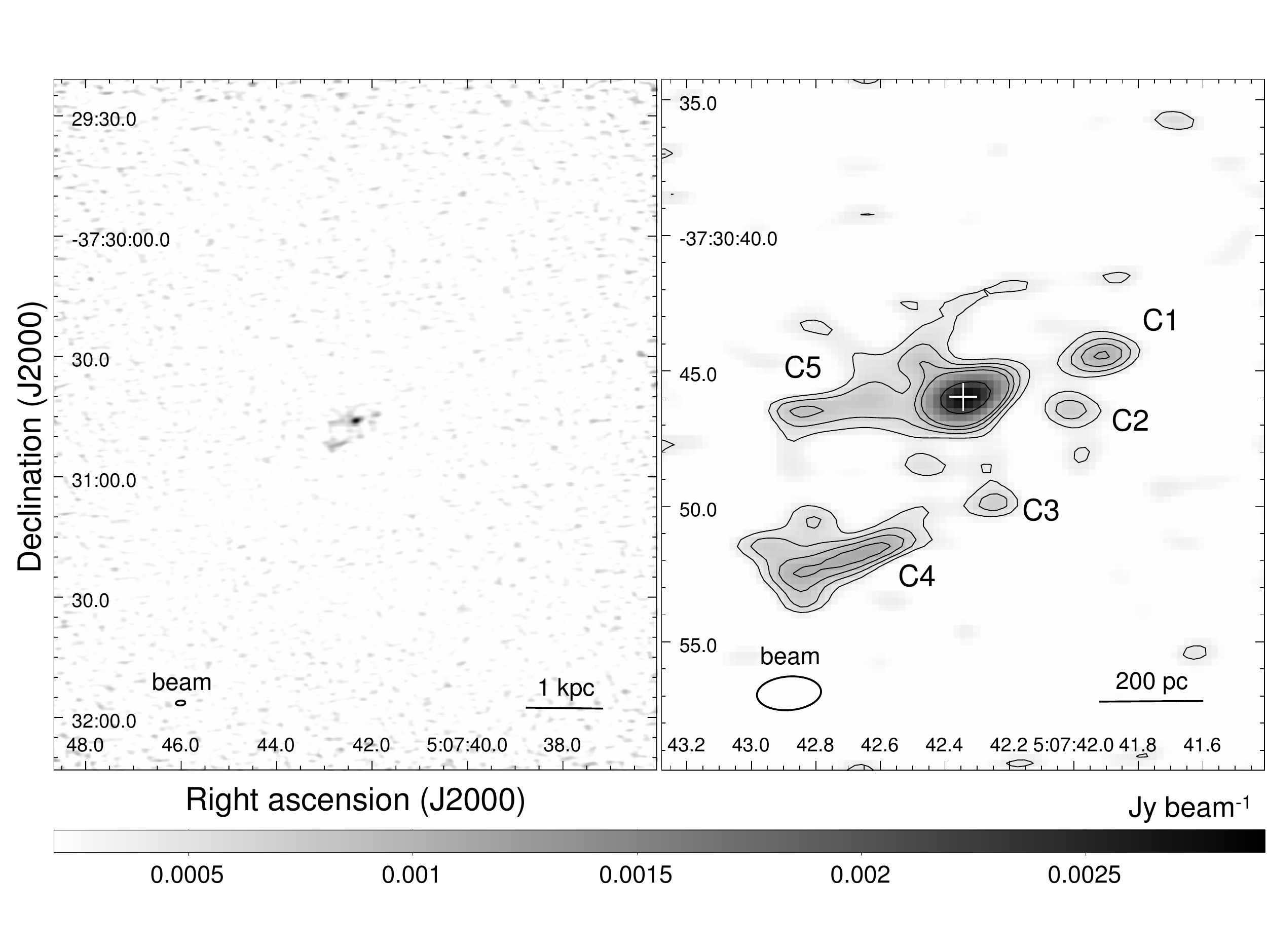}
\caption{Continuum emission synthesized from emission-free channels of four spectral windows: total image (left) and the central 1 kpc (right). The color is shown on a linear scale starting from 0.3 mJy beam\(^{-1}\) (1.5 \(\sigma\)) with contours at 0.4, 0.6, 0.8, 1, 2 mJy beam\(^{-1}\). The cross marks the galactic center determined from Gaussian fitting (see table \ref{tab3}). The synthesized beam is shown at the bottom left corner.}
\label{fig4}
\end{figure}

Radio continuum in NGC 1808 was previously observed at low frequencies (20, 6, 3.6, and 2 cm) that revealed several ``hot spots'' in the galactic central region (e.g., \citealt{Con87,Sai90,Col94,Kot96}). Our simultaneous observations of four spectral windows (sub-bands with frequencies between 100 and 115 GHz) yielded wide-field, sensitive, high-resolution continuum images that add new information about the complex starburst region (figure \ref{fig4}). In total, six sources were detected within \(r<400\) pc from the galactic center with a significance of \(>3~\sigma\) and three sources with \(>5~\sigma\) including the nucleus (\(13~\sigma\)) and two circumnuclear hot spots. The circumnuclear sources are denoted by C1-C5 in figure \ref{fig4}. The locations and flux densities of all sources are derived by 2-dimensional Gaussian fitting and the results are presented in tables \ref{tab3} and \ref{tab4}. Some of the sources are unresolved as their shapes resemble the synthesized beam, but the nucleus appears slightly extended toward east and north-east, as well as the elongated source C4 south-east of the nucleus. The emission peak is located at \((\alpha,\delta)_\mathrm{J2000.0}=(5^\mathrm{h}7^\mathrm{m}42\fs343,-37\arcdeg30\arcmin45\farcs95)\).

\begin{table}
\begin{center}
\caption{Properties of the central continuum source derived from 2-dimensional Gaussian fitting.}\label{tab3}
\begin{tabular}{ll}
\tableline\tableline
Peak position \(\alpha_\mathrm{J2000}\) & \(05^\mathrm{h}07^\mathrm{m}42\fs3434\) \(\pm\) \(0\fs0097\) \\
Peak position \(\delta_\mathrm{J2000}\) & \(-37\degr30\arcmin45\farcs955\) \(\pm\) \(0\fs059\) \\
Major axis (FWHM)\tablenotemark{a} & \(1\farcs67\pm0\farcs49\) (\(87\pm25\) pc) \\
Minor axis (FWHM)\tablenotemark{a} & \(1\farcs22\pm0\farcs77\) (\(63\pm40\) pc) \\
Position angle\tablenotemark{a} & \(130\degr\pm52\degr\) \\
Total flux density & \(4.68\pm0.57\) mJy \\
Peak flux density & \(2.64\pm0.22\) mJy beam\(^{-1}\) \\
\tableline
\end{tabular}
\tablenotetext{a}{Deconvolved from a clean beam with size (FWHM) \(2\farcs37\times1\farcs24\) and position angle \(PA=-85\fdg74\) at reference frequency \(\nu_\mathrm{ref}=107.8849\) GHz. The fitting region is a circle of diameter \(4\arcsec\) centered on the flux density peak of the convolved image.}
\end{center}
\end{table}

\begin{table}
\begin{center}
\caption{Properties of circumnuclear continuum sources detected with \(>3\sigma\) derived from 2-dimensional Gaussian fitting.}\label{tab4}
\begin{tabular}{ccccc}
\tableline\tableline
Source & \(\alpha_\mathrm{J2000}\) & \(\delta_\mathrm{J2000}\) & Peak intensity [mJy beam\(^{-1}\)] & Detection significance [\(\sigma\)] \\
\tableline
C1 & \(05^\mathrm{h}07^\mathrm{m}41\fs92\) & \(-37\degr30\arcmin44\farcs41\) & \(1.024\pm0.050\) & 5.1 \\
C2 & \(05^\mathrm{h}07^\mathrm{m}42\fs01\) & \(-37\degr30\arcmin46\farcs43\) & \(0.680\pm0.041\) & 3.4 \\
C3 & \(05^\mathrm{h}07^\mathrm{m}42\fs25\) & \(-37\degr30\arcmin49\farcs77\) & \(0.716\pm0.060\) & 3.6 \\
C4 & \(05^\mathrm{h}07^\mathrm{m}42\fs71\) & \(-37\degr30\arcmin51\farcs93\) & \(1.114\pm0.054\) & 5.6 \\
C5 & \(05^\mathrm{h}07^\mathrm{m}42\fs82\) & \(-37\degr30\arcmin46\farcs59\) & \(0.856\pm0.078\) & 4.3 \\
\tableline
\end{tabular}
\end{center}
\end{table}

\begin{figure}
\epsscale{0.9}
\plotone{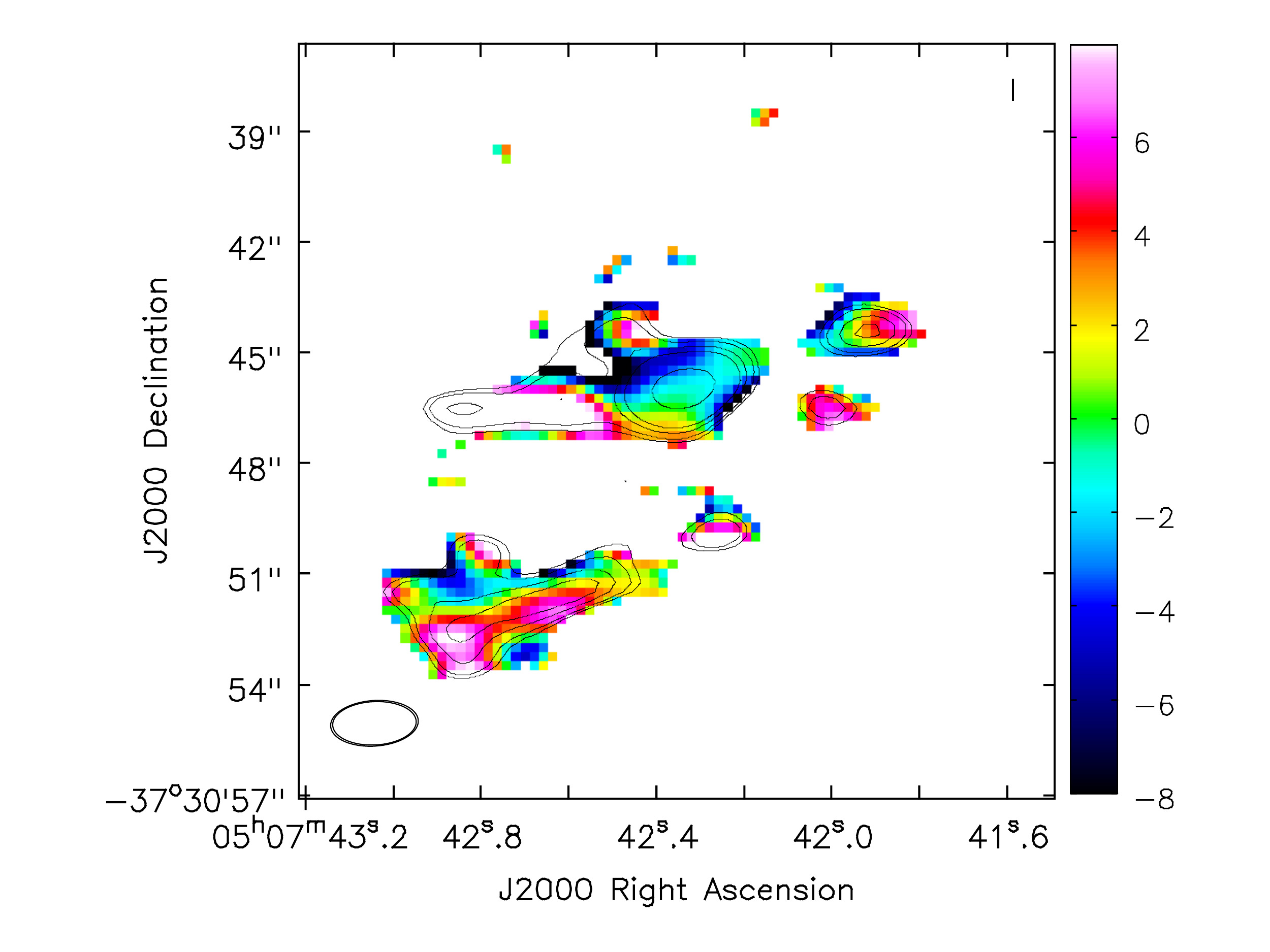}
\caption{Spectral index distribution calculated from synthesized images of the upper sub-bands at 113 and 115 GHz (\(\nu_\mathrm{u}\)) and the lower sub-bands at 101 and 103 GHz (\(\nu_\mathrm{l}\)). The contours are the same as in figure \ref{fig4}.}
\label{fig5}
\end{figure}

The flux density \(S_\nu\) of the continuum emission can be expressed as a function of frequency \(\nu\), \(S_\nu\propto\nu^\alpha\), where \(\alpha\) is the spectral index. Depending on the mechanism that generates the continuum emission at the observed frequency, the spectral index can vary from negative to positive values (e.g., \citealt{Con92}). At 2.8 mm, we expect a combination of free-free emission (nearly flat frequency dependence), blackbody emission from dust grains (positive \(\alpha\)), and non-thermal synchrotron emission from accelerated electrons in magnetic fields (negative \(\alpha\)). To investigate the nature of the continuum emission in the nuclear region, we derived a spectral index distribution from two images, each produced by a synthesis of two spectral windows: the upper-frequency image was produced from spectral windows 3 and 4 (113 and 115 GHz), denoted \(\nu_\mathrm{u}\), while the lower-frequency image was produced from spectral windows 1 and 2 (101 and 103 GHz), denoted \(\nu_\mathrm{l}\). The spectral index, as the slope of the spectrum between the two frequency sets, is defined by

\begin{equation}
\alpha(\nu)\equiv\frac{\log(S_\mathrm{u}/S_\mathrm{l})}{\log(\nu_\mathrm{u}/\nu_\mathrm{l})},
\end{equation}
and the resulting \(\alpha\) image is shown in figure \ref{fig5}. In the central \(r<100\) pc, the spectral index varies in the range \(-3\lesssim\alpha\lesssim1\) with \(\alpha_\mathrm{core}\simeq-1\) at the galactic center position. These values indicate mixed free-free radiation in hot ionized gas, where \(\alpha\) is typically close to 0 in optically thin H\textsc{ii} regions, and non-thermal synchrotron radiation from fast electrons accelerated in, e.g., supernova remnants. In addition to free-free emission, flattening of a non-thermal radiation spectrum can occur due to self-absorption of synchrotron radiation. The derived spectral index is similar to that of \(\alpha_\mathrm{core}=-0.9\) found by \citet{Dah90} by comparing 20 and 6 cm images. By contrast, the compact sources that surround the nucleus on average exhibit \(\alpha\gtrsim0\). Such values of \(\alpha\) can be produced in thermally-cooling regions with free-free and dust-grain emission. In fact, the compact sources except the core are not detected at the lower frequency, resulting in lower limits. The spectral index in these obscured regions may also be affected by optical depth and other variations in dust properties.

From the derived values of \(\alpha\), we suggest that the nucleus is dominated by a high-energy source, e.g., nuclear starburst accompanied with supernova explosions and/or a low-luminosity AGN [indicated from optical and X-ray studies too, e.g., \cite{VV85}, \cite{Awa96}, and \cite{Jim05}], confirming the results from previous radio-continuum observations (e.g., \citealt{Col94}). The circumnuclear sources may be compact H\textsc{ii} regions, possibly associated with young star clusters embedded in dusty clouds detected at infrared wavelengths \citep{GA08}.

\begin{figure}
\epsscale{0.75}
\plotone{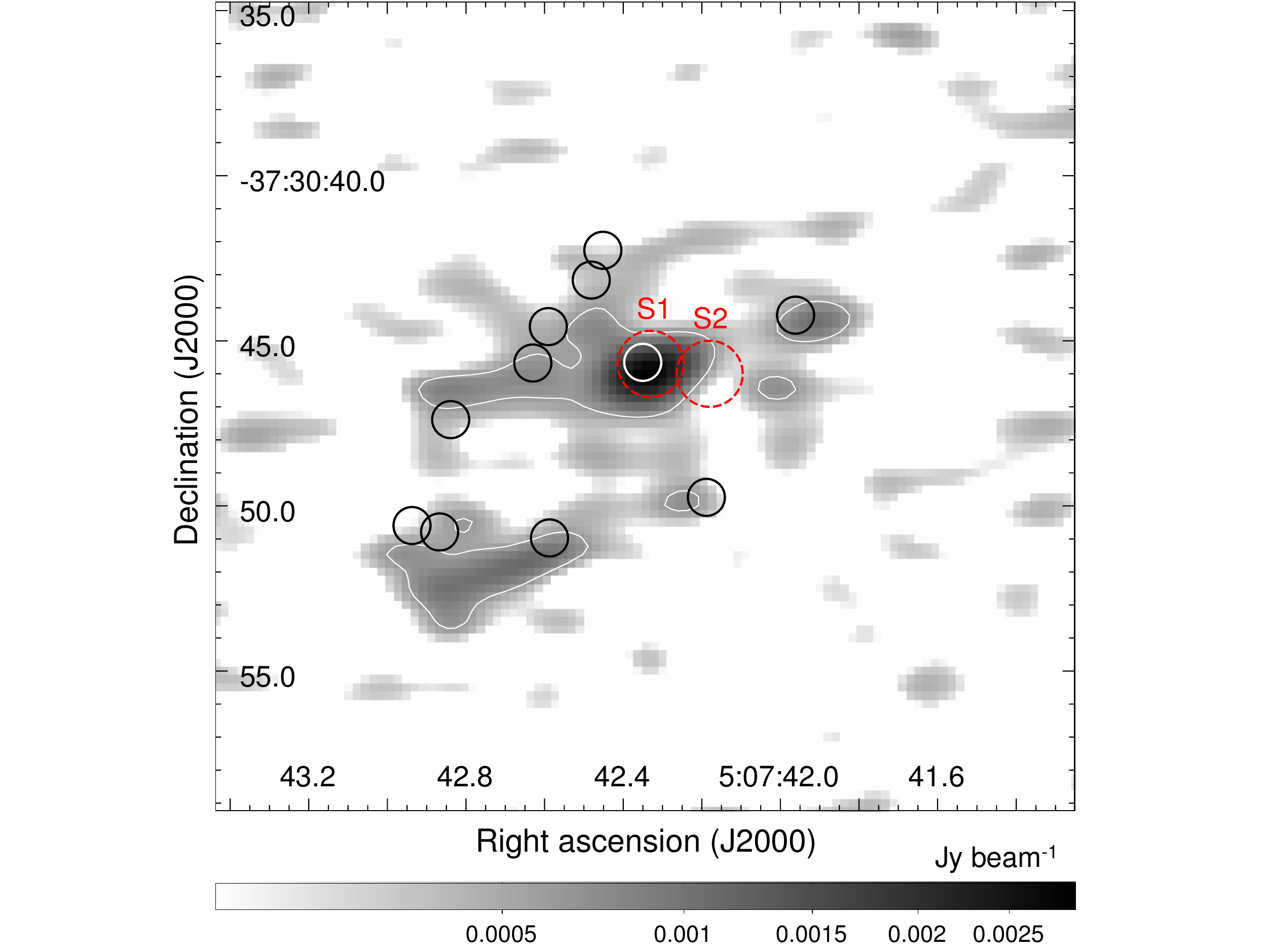}
\caption{Compact 3.6-cm sources (full circles) from \citet{Col94} and luminous X-ray sources (dashed circles S1 and S2) from \citet{Jim05} and \citet{HA07} superimposed on our 2.8-mm continuum image (grey scale). S1 is the location of the obscured low-luminosity AGN candidate. The contour is plotted at \(3~\sigma\).}
\label{fig6}
\end{figure}

In figure \ref{fig6} we show a comparison of our 2.8-mm continuum data with 3.6-cm continuum sources and the most luminous nuclear X-ray sources at 0.3-0.9 keV (S1) and at 2-10 keV (S2). The spectral indices of low-frequency sources indicate that their nature is supernova remnants \citep{Sai90,Col94}. However, the 2.8-mm sources revealed in ALMA images are only partially coincident with these objects, suggesting that the 2.8-mm and 3.6-cm sources are not identical in most cases.

As for the X-ray data, S1 is obscured by a hydrogen column density of \(N_\mathrm{H}\sim0.6\times10^{22}\) cm\(^{-2}\) and harbors a source with luminosity \(L_\mathrm{X}\sim1.5\times10^{39}\) erg s\(^{-1}\) in the 2-10 keV band \citep{Jim05,HA07}, somewhat lower than the low-luminosity AGN in nearby galaxies [\((0.4\mathrm{-}50)\times10^{40}\) erg s\(^{-1}\); \citealt{Ter02}]. Although both sources emit radiation in the hard X-ray band (S2 being more luminous with \(L_\mathrm{X}\sim9.2\times10^{39}\) erg s\(^{-1}\)), only source S1 has a counterpart at 2.8 mm and longer wavelengths.

The distribution of radio sources around the nucleus (figure \ref{fig6}) indicates the presence of a partial ionized gas ring (radius \(r_\mathrm{ion}\simeq7.5\arcsec\) or 390 pc), inclination \(i\simeq60\arcdeg\), and position angle \(PA\simeq310\arcdeg\), consistent with the global parameters of the galaxy (derived in section 4). As we show below, the ionized gas ring lies on the inner edge of a pseudo-ring of molecular gas detected in CO (1-0), and it is this inner region where most of the star-forming activity and its feedback (supernova explosions) is taking place. Since our data were taken over a wide mosaic field, imaging was not restricted by the field of view (except the lack of short-spacing baselines); no 2.8-mm continuum emission is detected beyond the ionized ring.

\clearpage

\subsection{Molecular gas distribution}

In this section we present the first high-resolution interferometric observations of CO in NGC 1808. Molecular gas traced by CO (1-0) emission is detected throughout the galactic disk (moment 0 defined as \(I_\mathrm{CO}\equiv M_0=\Delta v\sum \mathcal{S}_i\), where \(\Delta v=10.2\) km s\(^{-1}\) and \(\mathcal{S}_i\) is the CO (1-0) flux density per beam, shown in figure \ref{fig7}a), with four distinct regions: (1) compact circumnuclear disk (CND) or torus confined to \(r<200\) pc (peak emission \(I_\mathrm{CO}^\mathrm{max}=41\) Jy beam\(^{-1}\) km s\(^{-1}\) in the high-sensitivity image), (2) pseudo-ring\footnote{The ring is part of a two-arm spiral pattern hence ``pseudo''.} at 500 pc from the center, (3) giant molecular clouds (GMCs) and associations (GMAs) in the large-scale bar, and (4) GMCs in the disk where compact CO (1-0) emission is detected predominantly in spiral arms. An illustration of the main structures of molecular gas is shown in figure \ref{fig7}. The sensitivity of the CO (1-0) image (5.5 mJy beam\(^{-1}\)) is equivalent to \(\sim7\times10^4~M_\odot\), where the luminosity-to-mass conversion was done using a Galactic CO-to-H\(_2\) conversion factor of \(X_\mathrm{MW}=2\times10^{20}~\mathrm{cm}^{-1}(\mathrm{K~km~s}^{-1})^{-1}\) \citep{Bol13b}. The resolution and sensitivity are sufficient to detect typical Galactic GMCs (\(4\times10^5~M_\odot\); \citealt{YS91}) in the disk with \(5~\sigma\) significance. A comparison with a 2MASS (Two-Micron All Sky Survey; \citealt{Jar03}) image in \(K_s\)-band tracing old and middle-aged stellar populations (figure \ref{fig7}b) shows that molecular gas is concentrated in the bulge region and the leading side of the large-scale bar (figure \ref{fig7}c). In the central starburst zone, CO is detected in the region between the 500-pc ring and the CND in a nuclear disk with complex morphology and kinematics. We begin by describing the details of the large-scale morphology.

\subsubsection{Large-scale bar and spiral arms}

\begin{figure}
\epsscale{1}
\plotone{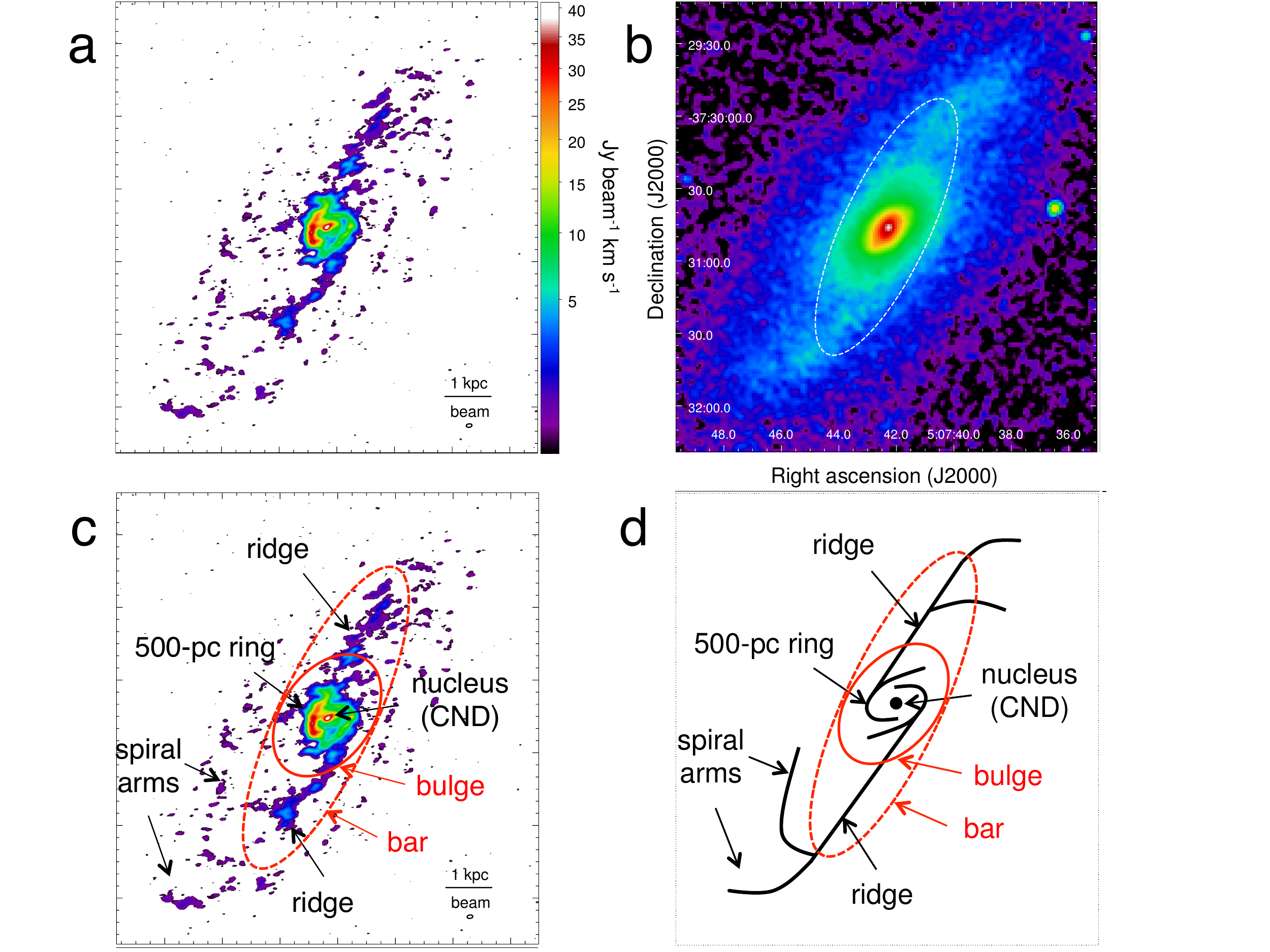}
\caption{(a) High-sensitivity image of CO (1-0) integrated intensity (moment 0) presented on a square root scale from 0.22 Jy beam\(^{-1}\) km s\(^{-1}\) (pixels below \(4~\sigma\), where \(1~\sigma=5.5\) mJy beam\(^{-1}\), were masked) to the peak value of 41 Jy beam\(^{-1}\) km s\(^{-1}\). The synthesized beam (\(2\farcs55\times1\farcs41\) at \(PA=-82\arcdeg\)) is shown at the lower right corner. (b) 2MASS image in \(K_s\)-band presented on a logarithmic scale \citep{Jar03}. The bar region is indicated with a dashed ellipse (semi-major axis \(a_\mathrm{b}=3\) kpc). (c)-(d) Illustration of the galactic structure: the bulge and bar are shown in red, and the major molecular-gas components (nucleus, 500-pc ring, ridge, and spiral arms) in black color.}
\label{fig7}
\end{figure}

Outside the central starburst region, molecular gas is concentrated in a bar (semi-major axis \(a_\mathrm{b}\simeq3\) kpc estimated from the \(K_s\)-band image; figure \ref{fig7}b) and global spiral arm pattern (illustrated in figure \ref{fig7}c). The position angle of the bar is \(PA_\mathrm{bar}\simeq335\arcdeg\), about \(10\arcdeg\) higher than the position angle of the galactic disk.

The most prominent feature of CO (1-0) emission in the bar is the concentration of molecular clouds along the ``offset ridges'' spatially correlated with the dust lanes on the leading side of the bar. These structures have been observed in numerous barred galaxies and have been reproduced by numerical and analytical calculations including the cloud orbit and hydrodynamical shock-wave models (e.g., \citealt{Con80,Bin91,Ath92a,Ath92b,LL94,Wad94,Sak99,She00,Hir14}; and others). In the cloud orbit model, the lanes are a consequence of oval orbits of gas clouds in the bar potential. Upon crossing the bar region, the clouds orbiting in the galactic disk gain inward motion as seen from the rotating frame. Orbit crowding leads to increased gas concentration manifested in stronger CO (1-0) emission and dust extinction.

At other wavelengths, the bar is also rich in H\textsc{i} gas and dotted with H\textsc{ii} regions revealed in H\(\alpha\) \citep{Kor96}. Enhanced intensity of the higher CO transition \(J=3\rightarrow2\) was also detected along the bar using the single-dish telescope ASTE \citep{Sal14}.

\begin{figure}
\epsscale{1}
\plotone{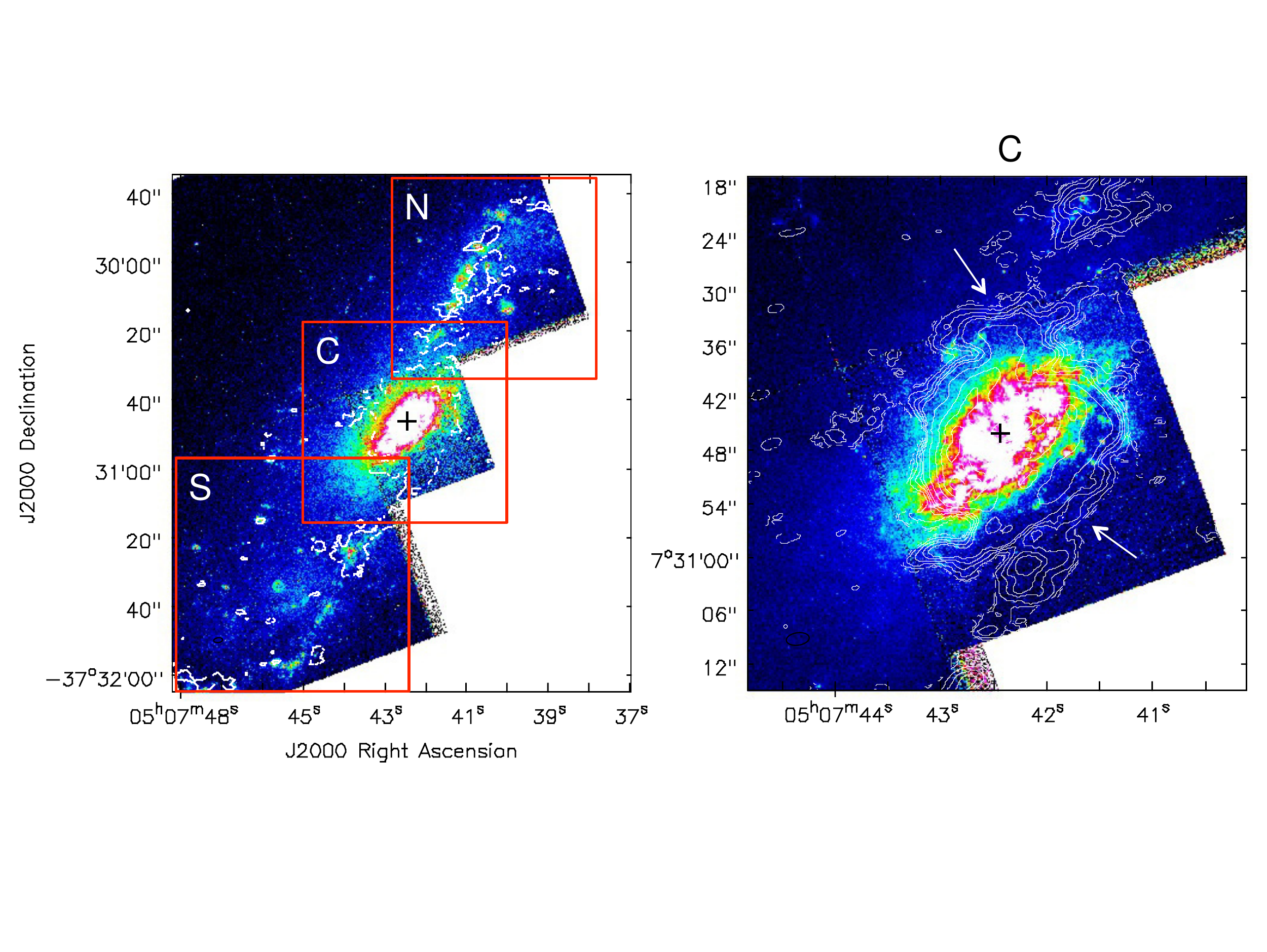}
\caption{\emph{Left.} H\(\alpha\) emission in NGC 1808 (Hubble Legacy Archive) shown in color. The contour is CO (1-0) intensity at 1\% of the peak 41 Jy beam\(^{-1}\) km s\(^{-1}\). The rectangles mark the regions shown in greater detail in panel on the right and in figure \ref{fig9}. \emph{Right.} Enlargement of the central region C. The contours are plotted at 0.005, 0.01, 0.025, 0.05, 0.075, 0.1, 0.2, 0.4, 0.6, 0.8 times the peak. The white arrows mark the regions where the ridges of the bar connect to the 500-pc ring.}
\label{fig8}
\end{figure}

\begin{figure}
\epsscale{1}
\plotone{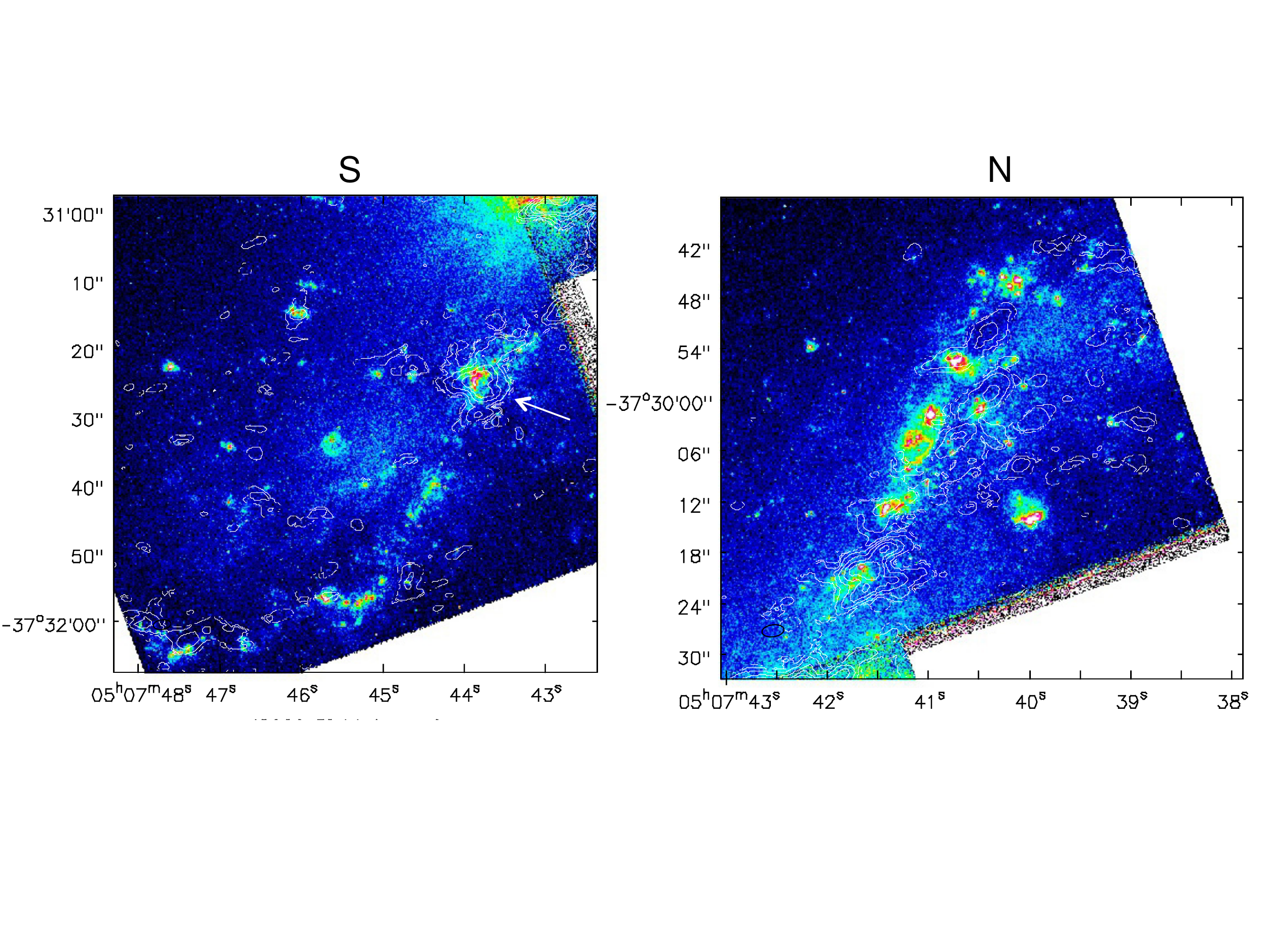}
\caption{Regions S (south) and N (north) from figure \ref{fig8}, with the same contours. The white arrow marks the location of a GMA with an associated giant H\textsc{ii} region (see text). Similar massive molecular complexes with star-forming regions are detected in the north too.}
\label{fig9}
\end{figure}

In figures \ref{fig8} and \ref{fig9}, we show a large-scale comparison of CO (1-0) observed by ALMA and a high-resolution HST image of H\(\alpha\). Outside the luminous starburst nucleus, we find that the bar harbors a number of large concentrations of molecular gas spatially correlated with H\(\alpha\). Consider, for example, a clump of CO gas marked with arrow in figure \ref{fig9} (S panel). The integrated intensity enclosed within \(r<10\arcsec\) from the CO emission peak of this cloud is \(I_\mathrm{CO}=46\) Jy km s\(^{-1}\). Using a CO-to-H\(_2\) conversion formula (with a Galactic conversion factor), introduced in the next section, this yields a mass of \(M_\mathrm{mol}\approx6\times10^7~M_\sun\). The object is a giant molecular association (GMA), and the comparison with H\(\alpha\) shows that it is a site of star formation inside the bar. Similar masses can be derived for the GMAs in the north too, where H\textsc{ii} regions are located in nearly a straight line aligned with the major axis of the bar. This behavior has been reproduced in recent numerical simulations where molecular clouds in bars form star-forming GMAs through merging and collisions \citep{Fuj14}.

Outside the bar, individual GMCs can be identified in the disk, many of which are concentrated in structures that resemble a spiral pattern (figure \ref{fig7}c). In particular, several arms are connected to the bar and emerge from its ends, with notable asymmetry in the density of GMCs and H\textsc{ii} regions: there are more molecular clouds and star-forming regions on the trailing side of the bar, especially around the NW semi-major axis. We leave a more quantitative analysis of the relation between CO and star-forming tracers to a future work that includes the short-spacing correction.

\clearpage

\subsubsection{Gas surface density in the central starburst region}

\begin{figure}
\epsscale{1}
\plotone{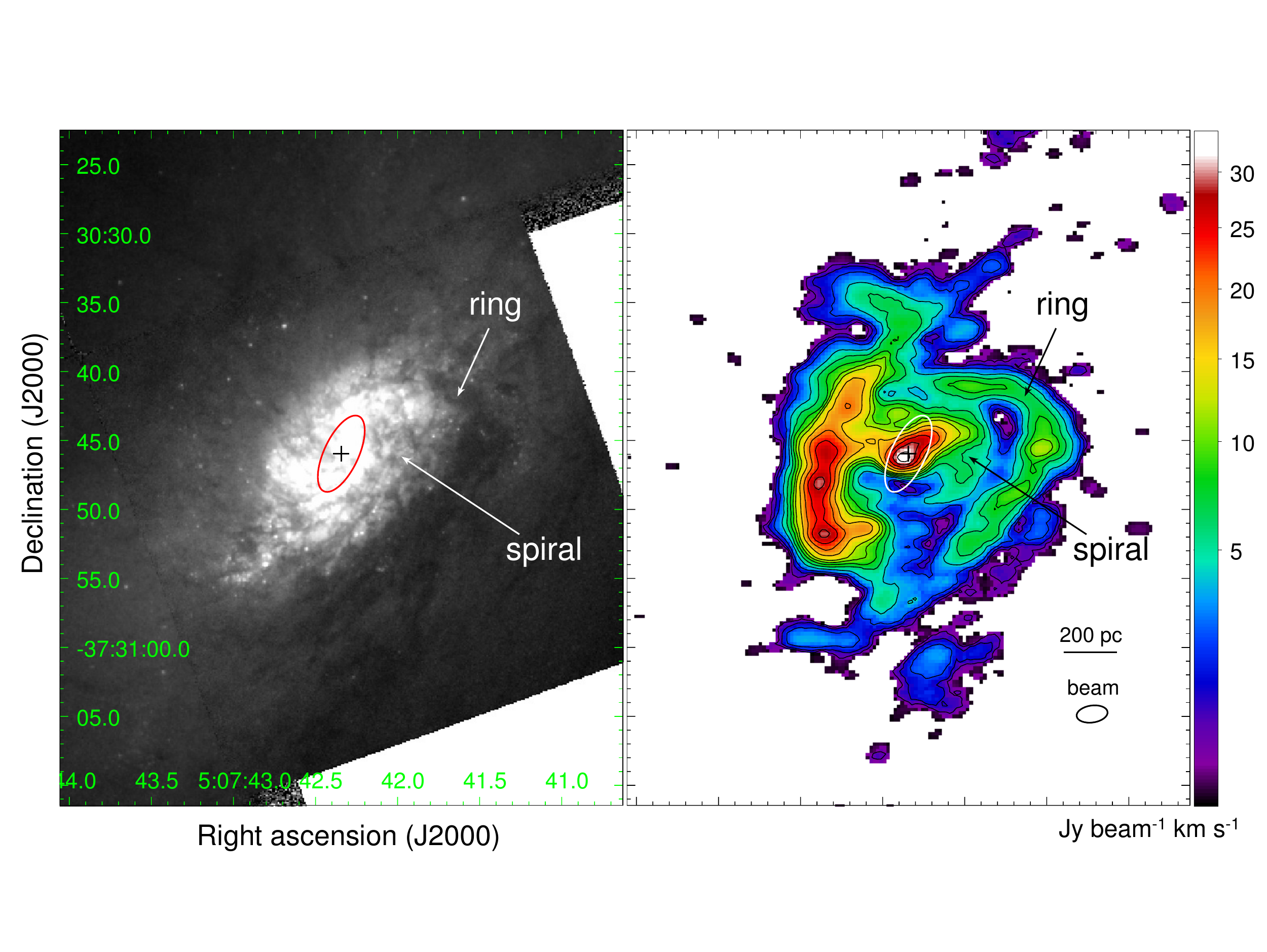}
\caption{\emph{Left.} HST \(R\)-band image (Hubble Legacy Archive). The black cross marks the position of the 2.8-mm continuum peak. The 500-pc ring and nuclear spiral are indicated with arrows; the ellipse in the center marks the region of the nuclear bar. \emph{Right.} Distribution of the CO (1-0) integrated intensity displayed on a square root scale from 0.127 Jy beam\(^{-1}\) km s\(^{-1}\) (pixels below \(5~\sigma\) in this high-resolution image, where \(1~\sigma=10\) mJy beam\(^{-1}\), were masked) to the peak value of 34 Jy beam\(^{-1}\) km s\(^{-1}\) with contours at 0.5, 1, 2, 4, 6, 8, 10, 12, 16, 20, 24, 28, 32 Jy beam\(^{-1}\) km s\(^{-1}\). The synthesized beam (\(2\farcs26\times1\farcs23\)) is shown at the bottom right corner.}
\label{fig10}
\end{figure}

\begin{figure}
\epsscale{0.75}
\plotone{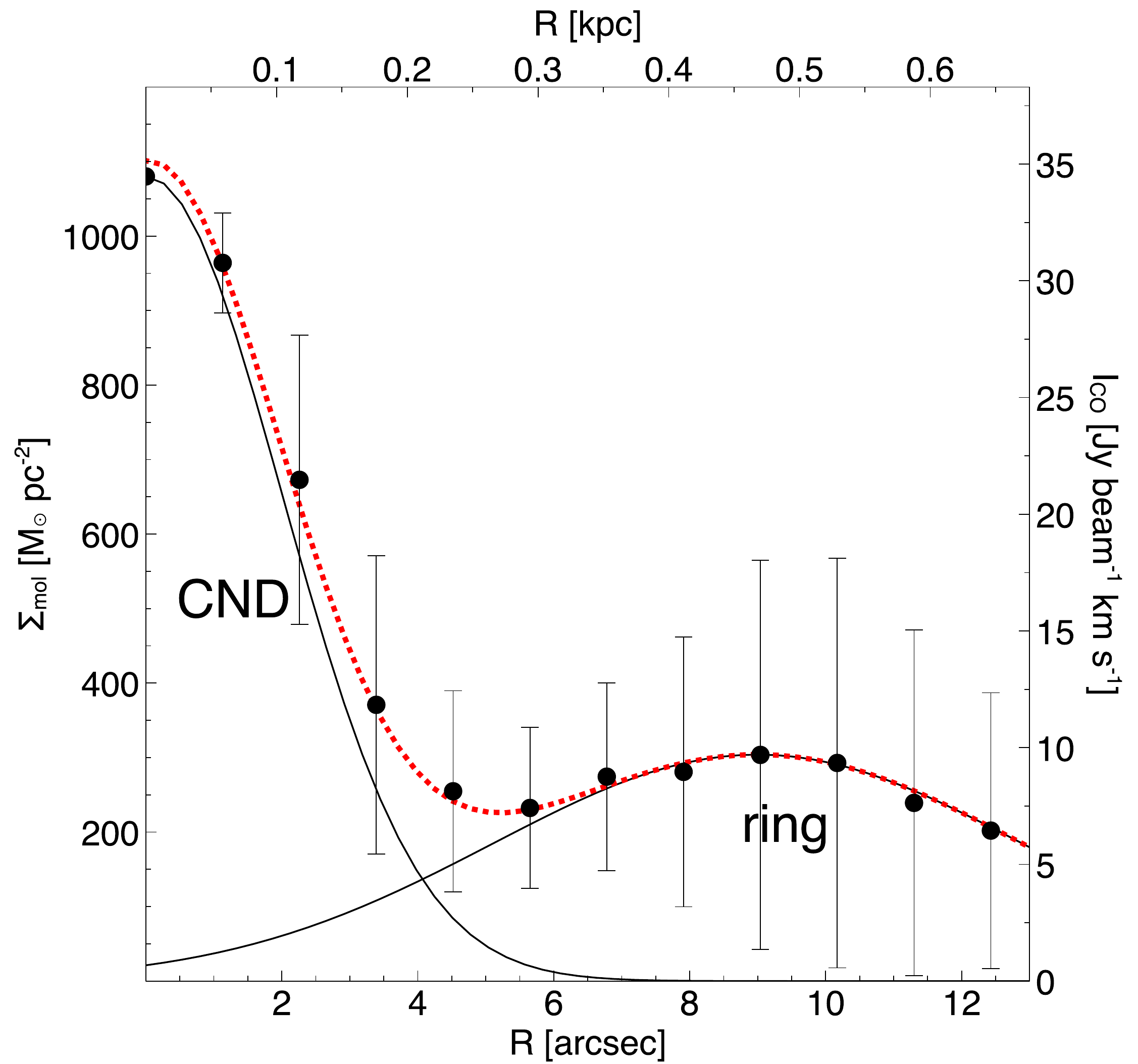}
\caption{Molecular gas surface density (or integrated flux density) averaged over azimuth as a function of radius, \(\Sigma_\mathrm{mol}(R)\). The data are fit with two Gaussian functions representing the circumnuclear disk (CND) and the 500-pc ring.}
\label{fig11}
\end{figure}

The CO (1-0) emission is prominent in the central \(r\lesssim800\) pc (figures \ref{fig10} and \ref{fig11}). Emission peaks in the CND (\(I_\mathrm{CO}=34\) Jy beam\(^{-1}\) km s\(^{-1}\) in the high-resolution image) and in a 500-pc pseudo-ring that is part of a two-arm spiral. Molecular gas is also present between the center and the ring, where infrared observations reveal an embedded nuclear stellar bar at a position angle of \(\simeq335\arcdeg\) \citep{Kot96,TG05}. Note that the nuclear bar is well aligned with the large-scale bar. This is not a usual configuration as the bars in double-barred galaxies show random relative position angles, while different angular velocities of the two bars are produced in simulations \citep{HSA07}. In the dynamical evolution, the nuclear bar is expected to change its pattern speed (e.g., accelerate its rotation during an alignment event) and possibly oscillate in thickness and ellipticity as the two bars constantly change the position angle with respect to each other \citep{DS07,MS10}. Interestingly, there is no clear spatial correlation between the nuclear bar and the molecular gas distribution traced by CO: molecular gas is not following the bar with prominent ridges (as it is on the kiloparsec scale), but instead a nuclear spiral pattern indicated in figure \ref{fig10}. The spiral arm can be seen west and south-west of the nucleus at a radius of \(r\sim200\) pc (3.8\arcsec); it is spatially correlated with a dust lane visible as extinction on the HST image.

The mass of molecular gas in the central region can be calculated from \(M_\mathrm{mol}=1.05\times10^4(X_\mathrm{CO}/X_\mathrm{MW})d^2S_\mathrm{CO}\Delta v~[M_\sun]\), where \(X_\mathrm{CO}\) is the CO-to-H\(_2\) conversion factor, \(d\) distance to the galaxy in Mpc, \(S_\mathrm{CO}\) measured flux density in Jy, and \(\Delta v\) velocity width (e.g., \citealt{Bol13b}). Adopting a low conversion factor of \(X=0.8\times10^{20}~\mathrm{cm}^{-2}~(\mathrm{K~km~s}^{-1})^{-1}\), estimated by \citet{Sal14} for the central 1-kpc region of this galaxy using a radiative transfer analysis (kinetic temperature \(T_\mathrm{k}=35\) K and gas density \(n_{\mathrm{H}_2}=10^{3.5}\) cm\(^{-3}\)), \(d=10.8\) Mpc, and measured \(S_\mathrm{CO}\Delta v=281\) Jy km s\(^{-1}\), we find a molecular gas mass (including He) of \(M_{\mathrm{mol}}(r<250~\mathrm{pc})\simeq1.4\times10^8~M_\odot\) in the CND region. With a standard Galactic conversion factor of \(2\times10^{20}~\mathrm{cm}^{-2}\mathrm{K}^{-1}(\mathrm{km~s}^{-1})^{-1}\), the mass is \(3.4\times10^8~M_\odot\). This is about 23\% of the total molecular gas mass in the central \(r<780\) pc detected with the 12-m array. The total CO (1-0) flux detected over the \(150\arcsec\times150\arcsec\) mosaic image is 1521 Jy km s\(^{-1}\) (corrected for primary beam attenuation) corresponding to \(1.9\times10^9~M_\odot\). This is comparable to the total atomic gas mass (\(M_\mathrm{Hi}=1.7\times10^9~M_\odot\)) over the central \(\sim4\arcmin\) estimated by \cite{Kor93} from their large-field H\textsc{i} maps. All flux values are lower limits because short-spacing correction has not been applied. The masses are summarised in table \ref{tab5}.

The surface density of the H\(_2\) gas when viewed face-on can be expressed as

\begin{equation}\label{SD}
\left(\frac{\Sigma_\mathrm{H_2}}{M_\sun~\mathrm{pc}^{-2}}\right)=3.3\times10^2\cos{i}\left(\frac{\Omega_\mathrm{A}}{\mathrm{arcsec}^2}\right)^{-1}\left(\frac{I_\mathrm{CO}}{\mathrm{Jy~beam^{-1}~km~s}^{-1}}\right)\left[\frac{X_\mathrm{CO}}{2\times10^{20}~\mathrm{cm^{-2}~(K~km~s^{-1})^{-1}}}\right],
\end{equation}
where \(\Omega_\mathrm{A}\) is the beam solid angle calculated as

\begin{equation}
\Omega_\mathrm{A}=\frac{\pi\theta_\mathrm{maj}\theta_\mathrm{min}}{4\ln{2}}=3.15~\mathrm{arcsec}^2
\end{equation}
for the high-resolution image (\(\theta_\mathrm{maj}=2\farcs26\) and \(\theta_\mathrm{min}=1\farcs23\)). Note that \(\Sigma_\mathrm{mol}\) depends only on the mass-to-light conversion factor \(X_\mathrm{CO}\propto \Sigma_\mathrm{mol}/I_\mathrm{CO}\propto M_\mathrm{mol}/L_\mathrm{CO}\), where \(L_\mathrm{CO}\) is the CO (1-0) luminosity. The total molecular gas mass corrected for He and heavy elements is \(\Sigma_\mathrm{mol}=1.41\Sigma_\mathrm{H_2}\) for the fractional abundance of hydrogen nuclei of 71\%.

The measured flux density at the galactic center (34.5 Jy beam\(^{-1}\) km s\(^{-1}\)) yields a peak surface density of \(\Sigma_\mathrm{mol}^\mathrm{max}=1.1\times10^3~M_\odot~\mathrm{pc}^{-2}\) if the low conversion factor \(X_\mathrm{CO}=0.8\times10^{20}~\mathrm{cm}^{-2}~(\mathrm{K~km~s}^{-1})^{-1}\) is applied \citep{Sal14}. In the central 1-kpc region, we derived azimuthally averaged surface density \(\Sigma_\mathrm{mol}\) as a function of radius by using the MIRIAD task ELLINT to find average flux densities in rings and then applied equation \ref{SD}. The result is shown in figure \ref{fig11}; the high surface density of \(\Sigma_\mathrm{mol}=(10^2-10^3)~M_\odot~\mathrm{pc}^{-2}\) is typical for starburst galaxies \citep{Ken98} and implies a column density of \(N_\mathrm{H}=(10^{22}-10^{23})\) cm\(^{-2}\) that explains the obscured nature of the central region. The parameters used to determine the ellipses were \(PA=324\arcdeg\), \(i=58\arcdeg\) (derived in section 4.1), and radius increments of \(\Delta R=1.13\arcsec\) (equal to \(\theta_\mathrm{maj}/2\)). The resulting curve can be fit with two Gaussian functions of the form \(a\exp[-(R-b)^2/(2c)^2]\) that correspond to the nuclear concentration (CND) and the ring. The offset parameter becomes \(b_\mathrm{CND}=0\arcsec\) for the CND (corresponding to the galactic center) and \(b_\mathrm{ring}=9\arcsec\) for the ring. This radius corresponds to 469 pc, which we adopt as the radius of the ring. The standard deviations are \(c_\mathrm{CND}=2\arcsec\) (104 pc) and \(c_\mathrm{ring}=3\farcs9\) (203 pc).

\begin{table}
\begin{center}
\caption{Lower limits\tablenotemark{a} of CO (1-0) fluxes and molecular gas masses.}\label{tab5}
\begin{tabular}{lrrrr}
\tableline\tableline
Region & Flux [Jy km s\(^{-1}\)] & Mass [\(M_\sun\)] & \(X_\mathrm{CO}\) [\(\mathrm{cm}^{-2}~(\mathrm{K~km~s}^{-1})^{-1}\)] & Fraction \\
\(r_\mathrm{CND}<250\) pc (\(4\farcs8\)) & 281 & \(1.4\times10^8\) & \(0.8\times10^{20}\) \\
& & \(3.4\times10^8\) & \(2.0\times10^{20}\) & 0.18 \\
\(r_\mathrm{ring}<780\) pc (\(15\arcsec\)) & 1224 & \(6.0\times10^8\) & \(0.8\times10^{20}\) \\
& & \(1.5\times10^9\) & \(2.0\times10^{20}\) & 0.80 \\
Total image & 1521 & \(1.9\times10^{9}\) & \(2.0\times10^{20}\) & 1.00 \\
\tableline
\end{tabular}
\tablenotetext{a}{The data are not corrected for missing short baselines. The fluxes are calculated by using the high-sensitivity data cube clipped at \(4~\sigma\) where \(1~\sigma=5.5\) mJy beam\(^{-1}\).}
\end{center}
\end{table}

\clearpage

\subsection{Kinematics}

\subsubsection{Global properties}

\begin{figure}
\epsscale{1}
\plotone{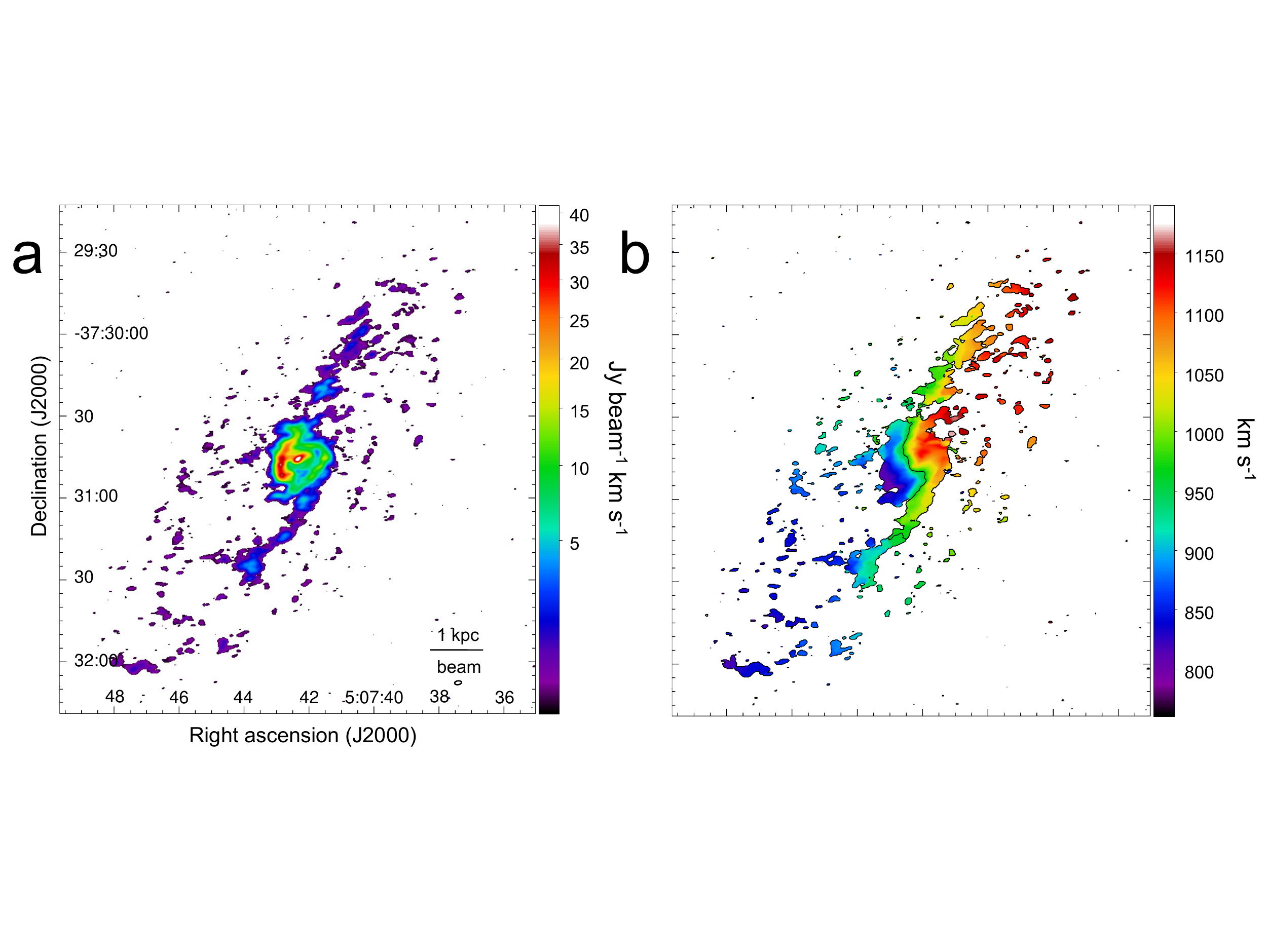}
\caption{CO (1-0) integrated intensity (a), as in figure \ref{fig7}, and intensity-weighted velocity field (b) (moment 1) displayed from \(v_\mathrm{LSR}=760\) to 1190 km s\(^{-1}\) on a linear scale. The central contour is plotted at \(v_\mathrm{LSR}=964\) km s\(^{-1}\).}
\label{fig12}
\end{figure}

\begin{figure}
\epsscale{1}
\plotone{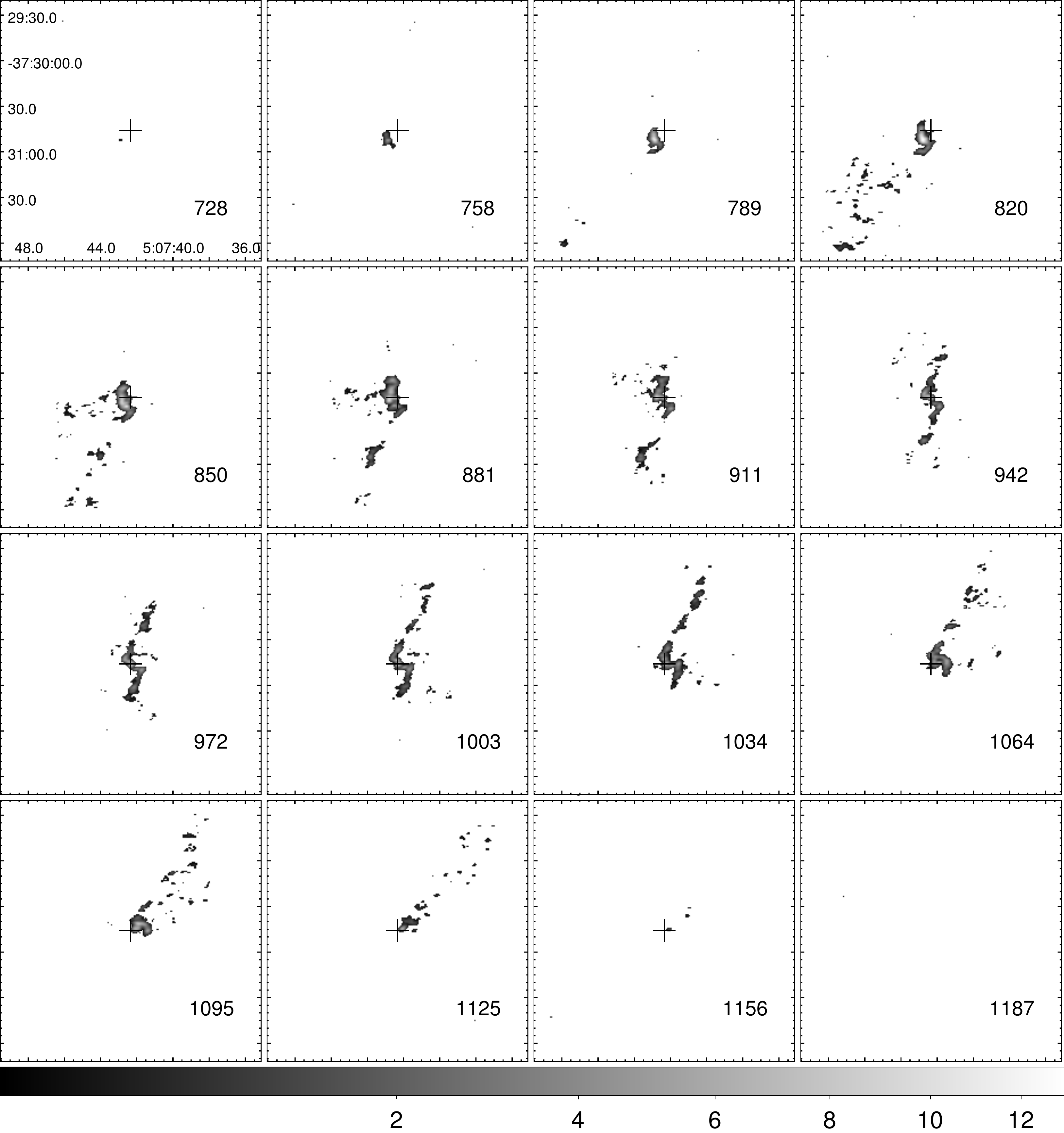}
\caption{Channel maps of the entire image - integrated intensity maps of three subsequent channels (separations of 30.6 km s\(^{-1}\)) starting from 728 km s\(^{-1}\); the start channel velocity is shown at the bottom right corner of each image. The intensity is displayed on a square root scale identical for all images (from 0.22 to 13 Jy beam\(^{-1}\) km s\(^{-1}\); grey scale). The cross in each image marks the galactic center.}
\label{fig13}
\end{figure}

\begin{figure}
\epsscale{1}
\plotone{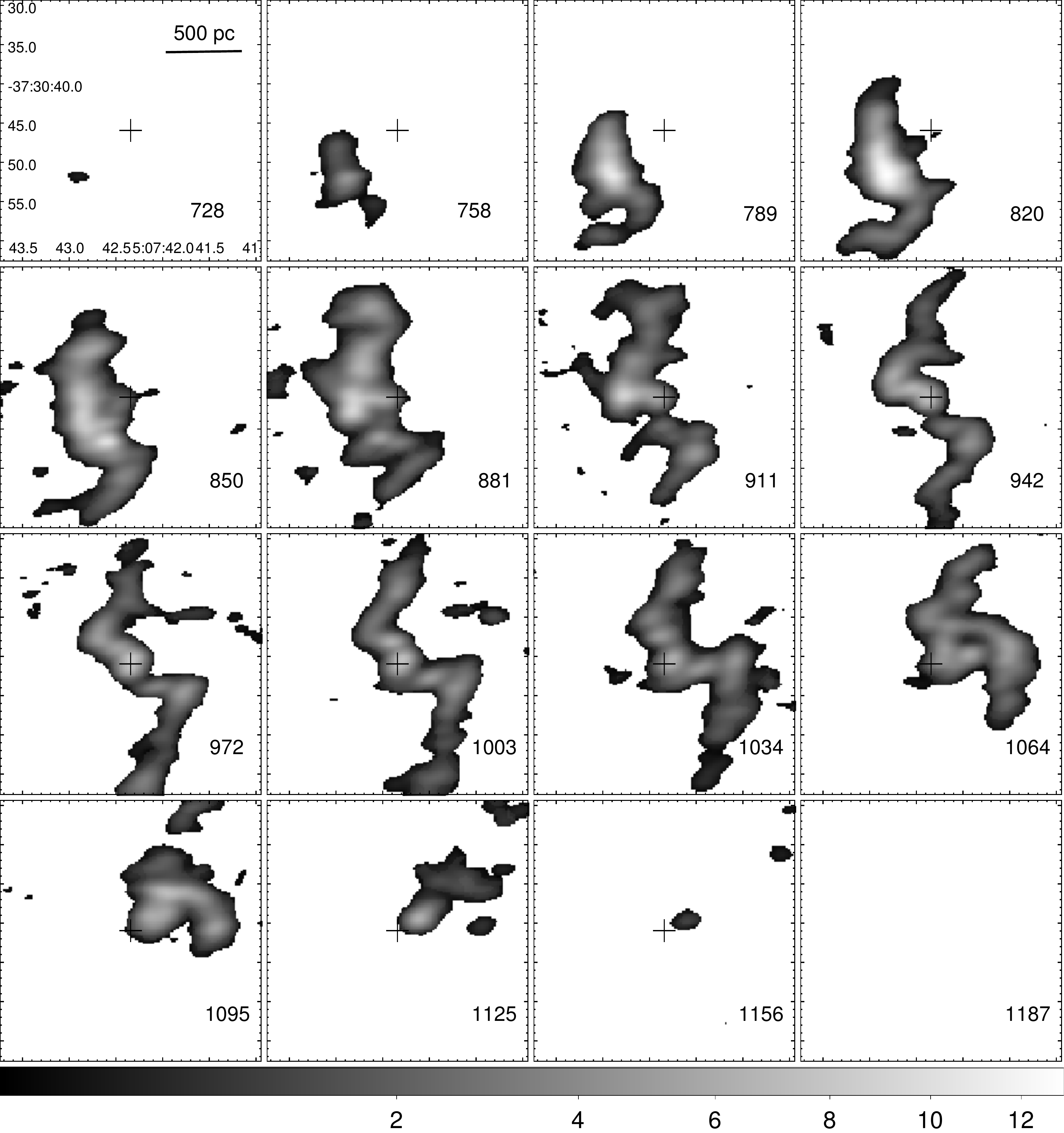}
\caption{Same as figure \ref{fig13} but showing the central region.}
\label{fig14}
\end{figure}

The high-sensitivity image presented in section 3.2 allows us to extract information on GMC velocities in the galactic disk and large-scale bar. An intensity-weighted velocity field image (moment 1), defined as \(M_1=\sum \mathcal{S}_i v_i /M_0\), is shown in figure \ref{fig12}. Intensity images integrated over narrow velocity bins (channel maps) of the entire mosaic and the central 2\arcmin region are shown in figures \ref{fig13} and \ref{fig14}, respectively. In addition to large-scale rotation, molecular gas in the bar region exhibits peculiar motion evident as a large velocity gradient over the entire ridges on the leading side of the bar. This was noticed in H\textsc{i} data by \citet{Kor96}, and we show that CO is following a similar trend. The velocity field in the offset ridges shows a velocity gradient (shear) of \(v_\mathrm{sh}\equiv\Delta v/\Delta l\sim0.2\) km s\(^{-1}\) pc\(^{-1}\), or \(\Delta v\sim100\) km s\(^{-1}\) across the \(\Delta l\sim500\) pc wide ridges (in azimuthal direction perpendicular to the galactocentric radius).

\subsubsection{Velocity field and dispersion in the central 1 kpc}

\begin{figure}
\epsscale{1}
\plotone{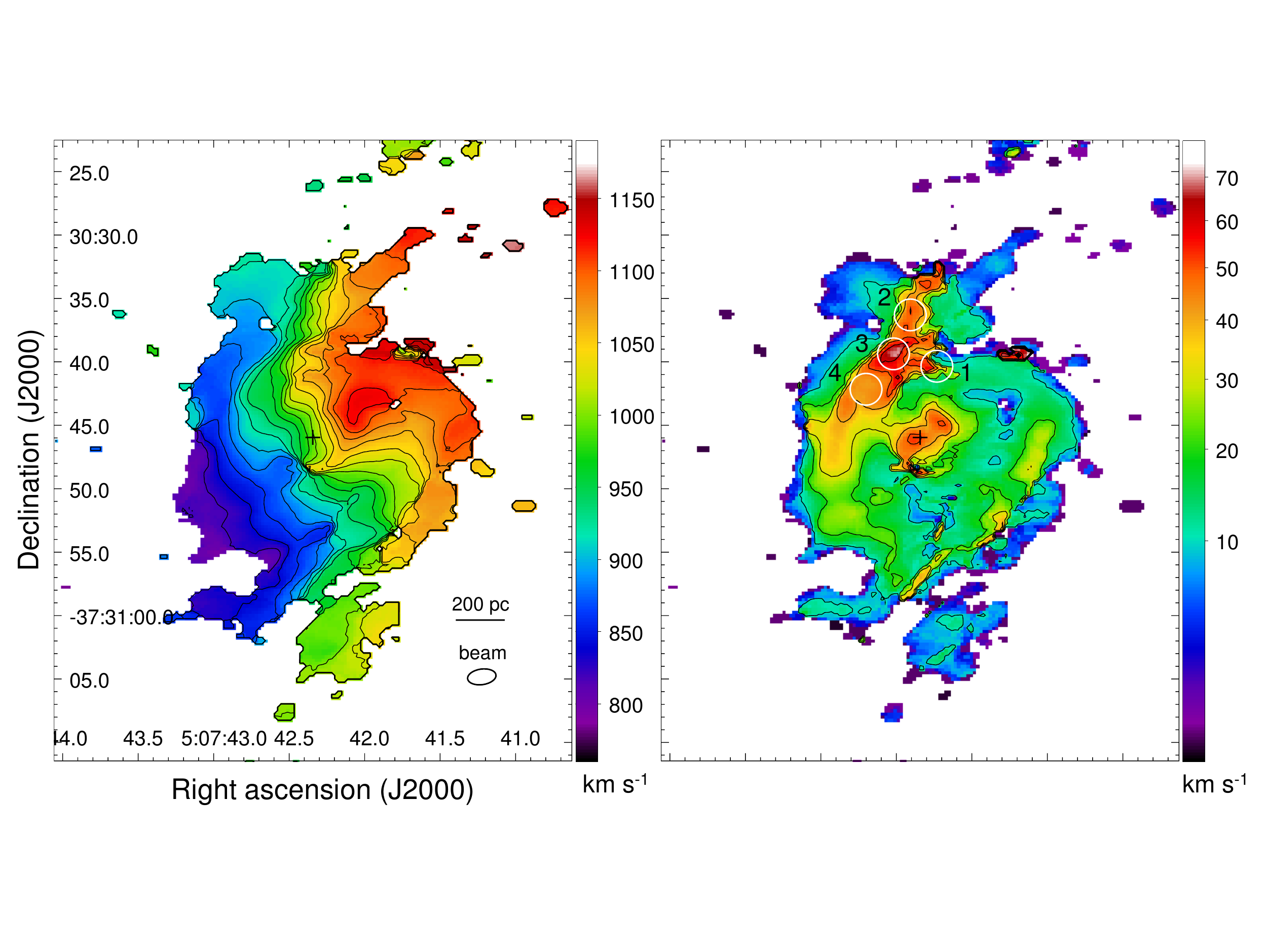}
\caption{\emph{Left.} Intensity-weighted velocity field of the central 1-kpc region with contours from \(v_\mathrm{LSR}=820\) to 1140 km s\(^{-1}\) in steps of 20 km s\(^{-1}\). \emph{Right.} Velocity dispersion image displayed on a square root scale with contours from \(\sigma_\mathrm{mol}=10\) to 60 km s\(^{-1}\). Pixels below \(5~\sigma\) in this high-resolution image, where \(1~\sigma=10\) mJy beam\(^{-1}\), were masked. The black cross marks the position of the 2.8-mm continuum peak. The circles 1-4 (radius \(1\farcs25\)) indicate the regions where \(\sigma_\mathrm{mol}\sim100\) km s\(^{-1}\) within the synthesized beam. The mean spectra within the circles are shown in figure \ref{fig16}.}
\label{fig15}
\end{figure}

The velocity field (moment 1) in the central region is shown in figure \ref{fig15} (left). It appears slightly distorted, with the largest deviation \(\sim3\arcsec\) south and south-west of the center. Note the overall ``S''-shape pattern, typical in galaxies with bars.

The velocity dispersion of molecular gas, defined as the second moment of intensity, \(M_2=\sqrt{\sum \mathcal{S}_i(v_i-M_1)^2/M_0}\) (where \(M_0\) and \(M_1\) are the zeroth and first moments defined above), is shown in figure \ref{fig15} (right). On average, \(M_2\) is high (above 40 km s\(^{-1}\)) in the nucleus and decreases to 10-20 km s\(^{-1}\) in the central disk. It rises again to 15-25 km s\(^{-1}\) in the 500-pc ring and then decreases once more beyond the ring in the galactic disk.

\begin{figure}
\epsscale{0.75}
\plotone{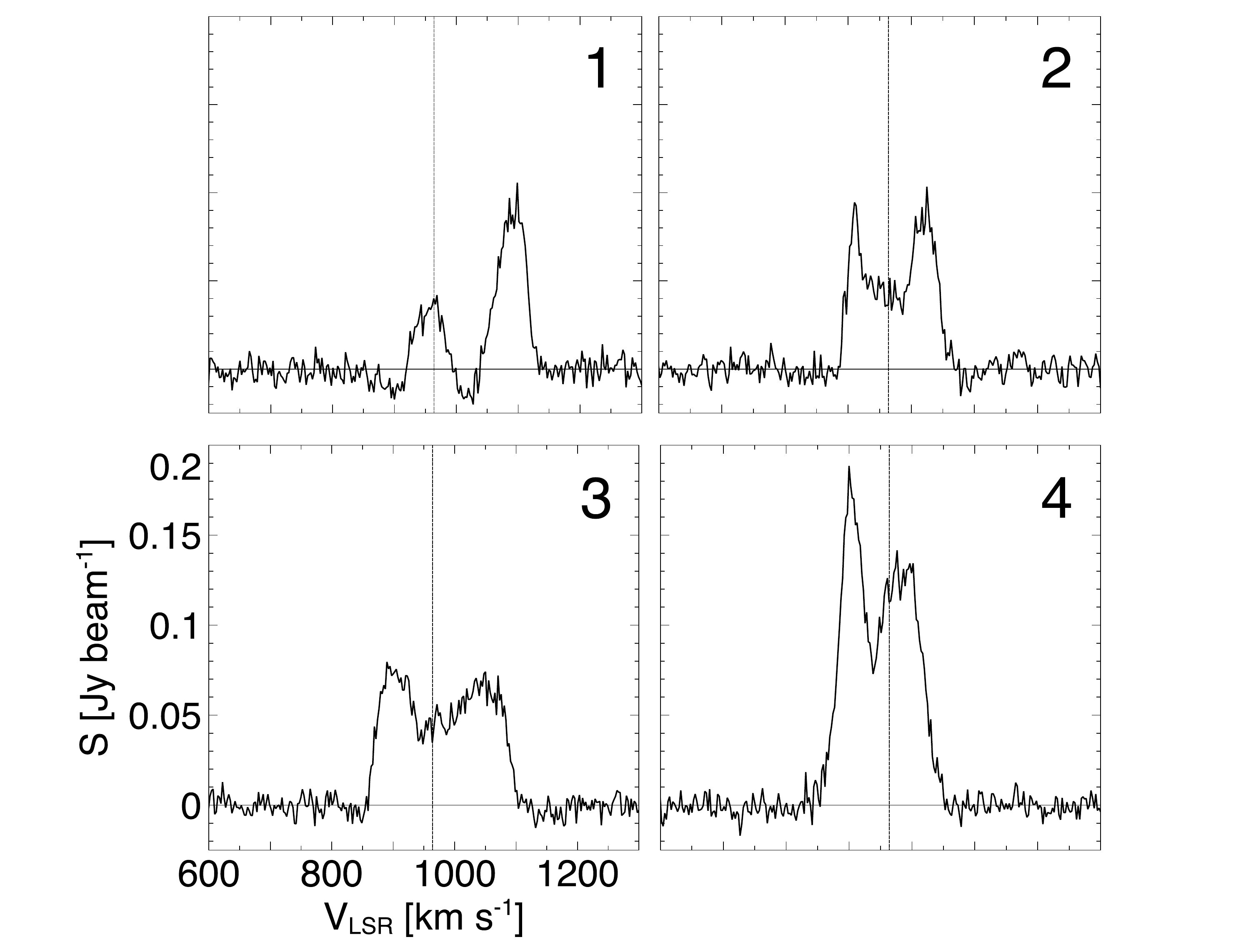}
\caption{CO (1-0) spectra from the positions 1-4 in figure \ref{fig15}. The separations between the peaks within the synthesized beam (\(\sim90\) pc) reach \(\sim100\) km s\(^{-1}\). The vertical line is plotted at \(v_\mathrm{LSR}=964\) km s\(^{-1}\).}
\label{fig16}
\end{figure}

Velocity dispersion can be an indicator of cloud (orbit) crowding. The dispersion is especially high (\(\sigma>40\) km s\(^{-1}\)) north and north-east of the nucleus. By inspecting the CO spectra in this region (figure \ref{fig16}), we find double peaks within the synthesized beam (\(\Delta l\simeq90\) pc). These features are not obvious in moment 1 images because the velocity is intensity-weighted. The moment 1 images do, however, show the S-shape and a steep velocity gradient north and south of the 500-pc ring. The profiles of the emission lines are somewhat affected by missing short-spacing baselines, but such effects cannot produce the observed line width because the negatives and sidelobes are weak. Note that this region is where the gas in the outer x1 orbits of the bar passes close to the 500-ring, where we expect to find an inner Lindblad resonance (ILR) and x2 orbits (section 4.2.2). The large velocity widths (separations between the peaks in the spectra, \(W\sim100\) km s\(^{-1}\)) indicate that the velocity gradient (shear) is high near the ILR, \(v_\mathrm{sh}\sim W/\Delta l\sim1\) km s\(^{-1}\) pc\(^{-1}\), a factor of several higher than in most of the ridge at larger radii. The gradient could disrupt the GMC structure, since the internal escape velocity of a cloud with radius \(R=10\) pc and \(\Sigma=200~M_\sun~\mathrm{pc}^{-2}\) is \(v_\mathrm{esc}=\sqrt{2\pi GR\Sigma}\approx7\) km s\(^{-1}\). The disrupted material may be in the form of low-density molecular gas unable to produce stars (e.g., \citealt{Dow96}). We will see in section 6 that most of star-forming activity in NGC 1808 appears to be inside the ILR, consistent with this suggestion. A similar trend is observed at the south end of the ring too, although the emission is weaker and \(W\approx80\) km s\(^{-1}\).

\clearpage

\section{Gas dynamics}

Detection of CO (1-0) in the central starburst region, kpc-scale bar, and spiral arms allows us to study gas dynamics in detail. In this paper, we focus primarily on the central 1-kpc region and derive basic geometric and kinematic parameters of molecular gas. On the large scale, we derive the bar pattern speed, and use it to locate the Lindblad resonances and corotation. Since the data presented here were taken with the 12-m array, i.e., without short-spacing baselines, extended structures such as arm/inter-arm regions and outflows are not fully sampled. We will discuss those features in detail in a future work that includes a short-spacing correction.

\subsection{Derivation of geometric and kinematic parameters in the central 1 kpc}

To determine the basic geometric and kinematic parameters of the central 1-kpc molecular zone, we used the program \(^{\mathrm{3D}}\)Barolo (3D-Based Analysis of Rotating Objects from Line Observations) that fits tilted-ring models to (3-dimensional) spectroscopic data \citep{dTF15}. The main parameters of the galaxy include the dynamical center at (\(x_0,y_0\)), systemic velocity (\(v_\mathrm{sys}\)), rotational velocity (i.e., rotation curve \(v_\mathrm{rot}\)), gas velocity dispersion (\(\sigma_\mathrm{mol}\)), galactic inclination (\(i\)), and position angle (\(PA\)). The program compares models with a data cube by fitting rings at various radii from the center. All parameters may vary from one ring to another, including inclination and position angle, thereby allowing a warp, i.e., a set of tilted rings with differential rotation (e.g., \citealt{RHW74}). The computation, however, does not handle non-circular motion (e.g., radial motion due to spiral arm and bar dynamics, outflows from starburst and AGN feedback). The aim to use the program in this work is to fit circular motion components so that we can produce a model and then subtract it from the data. The residuals would reveal non-circular features.

\subsubsection{Parameter computation}

The high-resolution data cube (\(\Delta v=2.5\) km s\(^{-1}\)) was used in this procedure. The calculation was carried out within 12 rings of equal width \(\Delta r=1\farcs13=\theta_\mathrm{maj}/2\) (where \(\theta_\mathrm{maj}\) is the FWHM of the synthesized beam along its major axis) from the center to \(12\farcs4\) (646 pc), i.e., to the radius that comfortably covers the central molecular ring; beyond this radius the signal-to-noise ratio is low and the influence of the galactic bar is too high to derive reliable results by assuming circular orbits. The gas scale height (disk thickness) was kept as a non-free parameter at \(h_z=150\) pc. In a disk with a Gaussian density profile perpendicular to the galactic plane, this parameter is related to the gas velocity dispersion and surface density as \(h_z=\sigma_{\mathrm{mol},z}^2(2\pi G\Sigma_\mathrm{mol})^{-1}\), where \(\sigma_{\mathrm{mol},z}\) is the velocity dispersion in the \(z\) direction. For \(\sigma_{\mathrm{mol},z}=20\) km s\(^{-1}\) and \(\Sigma_\mathrm{mol}=300~M_\sun~\mathrm{pc}^{-2}\), typical in the central region (e.g., figures \ref{fig11} and \ref{fig15}), the scale height is \(h_z\approx150\) pc. The parameter may vary by a factor of a few in the central region, though it makes negligible effect on the fitting result.

Although the program is able to guess all the parameters as functions of radius \(r\) simultaneously, we constrained the computations by providing the central position \((x_0,y_0)\) to be the location of the peak of 2.8-mm continuum emission (assuming that it coincides with the low-luminosity AGN candidate). This is a reasonable choice because the CO peak has been found offset from this position and more extended than the continuum source. Unlike continuum, CO is likely to exhibit a torus structure that can be offset [e.g., in the Seyfert galaxy NGC 1068; \cite{Kri11} and \cite{GB14}]. In addition, we have constrained the galactic inclination to the literature value of \(i=57\arcdeg\pm5\arcdeg\) (table \ref{tab1}), where the uncertainty is included to allow the program freedom to explore a range of \(10\arcdeg\).

\subsubsection{\textbf{Systemic velocity}}

\begin{figure}
\epsscale{0.75}
\plotone{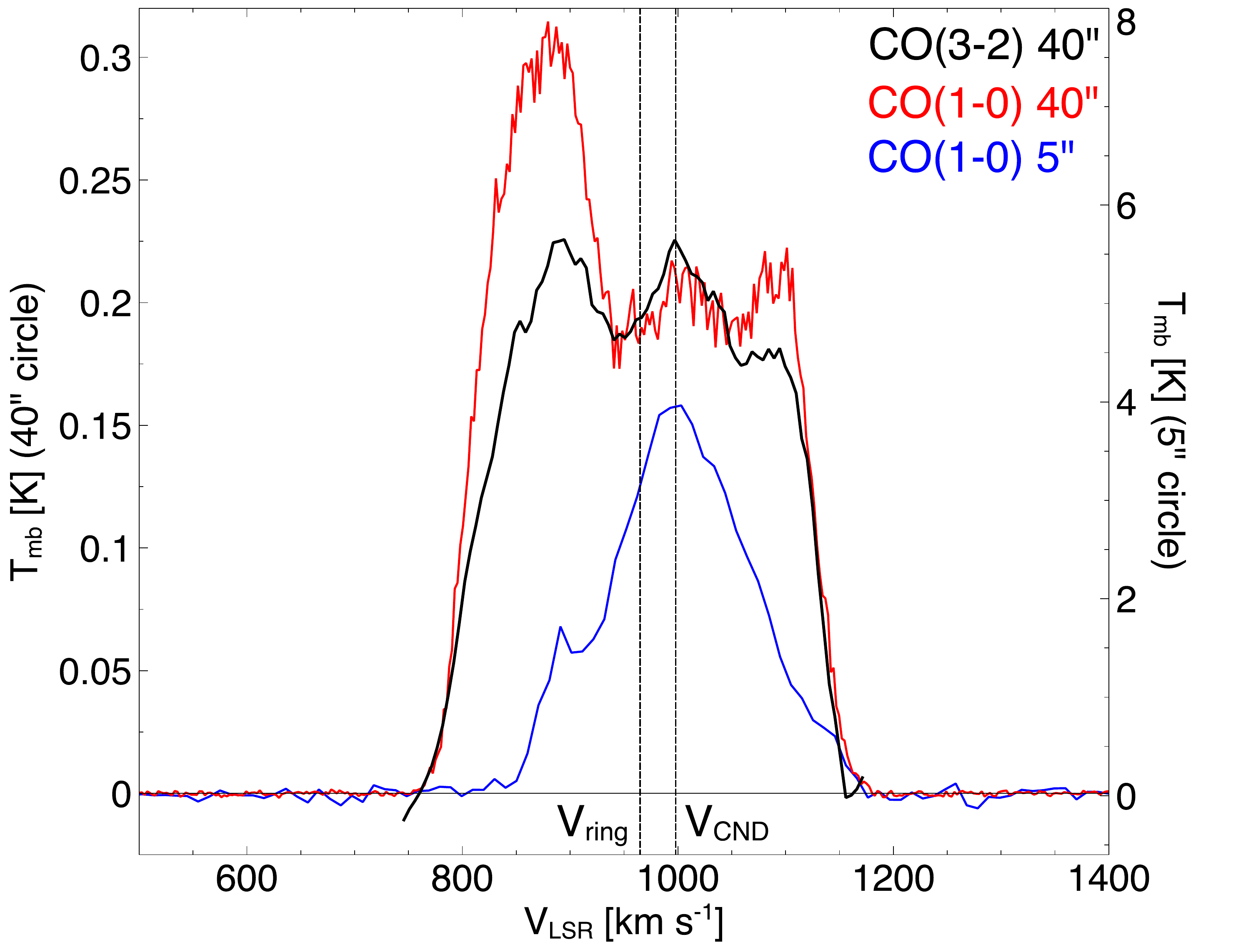}
\caption{Mean CO (3-2) spectrum acquired with ASTE (black), and CO (1-0) spectrum from the ALMA data presented in this work (red) plotted as the mean brightness temperature within the central \(r<20\arcsec\). The ALMA data were smoothed to the resolution of \(22\arcsec\) to match that of ASTE. Also shown is the ALMA CO (1-0) spectrum within \(r<2\farcs5\) (blue) at the original resolution (vertical axis on the right). Note the triple peak in the large-aperture spectra.}
\label{fig17}
\end{figure}

The program estimated the systemic velocity as a non-free parameter. This is calculated as the mid-point between the velocities that correspond to the \(20\%\) of the peaks of the CO (1-0) line profile. The profile within the central \(40\arcsec\times40\arcsec\) box (\(2.08\times2.08\) kpc\(^2\)) is shown in figure \ref{fig17} together with a spectrum of CO (3-2) acquired with the single dish telescope ASTE \citep{Sal14}. The derived systemic velocity is \(v_\mathrm{sys}=963.9\pm2.5\) km s\(^{-1}\) within the central region, in agreement with previously reported values measured with a large telescope beam that did not resolve the central 1-kpc region (965 km s\(^{-1}\); \citealt{Aal94}). A similar value of \(965\pm5\) km s\(^{-1}\) was derived when we applied the fitting procedure on the ASTE CO (3-2) data cube and when the center position was not fixed.

Note that the CO line profiles exhibit a triple peak in the central region (figure \ref{fig17}). The triple peak can be decomposed into the 500-pc ring and nucleus. It is clear from the CO (1-0) spectrum of the central \(r<2\farcs5\) in figure \ref{fig17} that the central peak corresponds to the nucleus. Applying the same procedure of systemic velocity computation, we fit 3 rings at radius increments that correspond to the beam width up to 235 pc (two times the major axis of the beam FWHM) and find a best fit at \(v_\mathrm{sys}=998.4\pm2.5\) km s\(^{-1}\), more than 30 km s\(^{-1}\) higher than \(v_\mathrm{sys}\) derived from global kinematics. This discrepancy, first noticed by \citet{BB68}, is also obvious in the position-velocity diagrams discussed below (figure \ref{fig24}).

A comparison of the CO (1-0) and CO (3-2) spectra in figure \ref{fig17} can be used to estimate the missing flux in the ALMA data. Although the ratio of the intensities of the two lines, defined as \(R_{31}\equiv I_{3-2}/I_{1-0}\), varies on small scale, \cite{Aal94} and \cite{Sal14} found that, on average, \(R_{31}\approx0.6\) in the central kiloparsec. Then, one should multiply the CO (3-2) spectrum in figure \ref{fig17} to obtain a CO (1-0) spectrum corrected for the short-spacing baselines. The resulting missing flux in the ALMA data could be as high as \(\sim40\%\) in the central 2 kpc region. Although the unrecovered flux is substantial, we note that this is likely much lower at the scale of the CND (\(2\arcsec\)), which is ten times smaller than the maximum recoverable scale of the imaging (21\arcsec). Also, the presence of the central peak in the single dish spectrum at a velocity shifted from the global systemic velocity by \(\sim30\) km s\(^{-1}\) suggests a kinematic offset.

Another method to calculate \(v_\mathrm{sys}\) is to use a dense gas tracer in the central galactic region. The HCN (4-3) line profile acquired with ASTE (see Appendix) was fitted with a Gaussian function that yielded a peak velocity of \(998.6\pm3.3\) km s\(^{-1}\), consistent with the nuclear component of the ALMA CO (1-0) spectrum and the central peak in the CO (3-2) spectrum. This velocity is also consistent with estimates from optical studies [e.g., nuclear heliocentric velocity of 1020 km s\(^{-1}\) in \cite{VV85} and \cite{Phi93}] Therefore, for the central galactic region (\(r<235\) pc), we suggest that the higher \(v_\mathrm{sys}\) is appropriate.

The origin of the kinematic offset is not clear. One possibility is that the bulge is composed of two components: a massive inner component that dominates the dynamics of gas in the CND, and a larger component that determines the gas motion in the central 2 kpc, and that these two do not have a single dynamical center, possibly a consequence of a tidal interaction in the past (also evident as a warped outer ring; e.g., \citealt{Kor93}).

\clearpage

\subsubsection{Position angle}

\begin{figure}
\epsscale{0.6}
\plotone{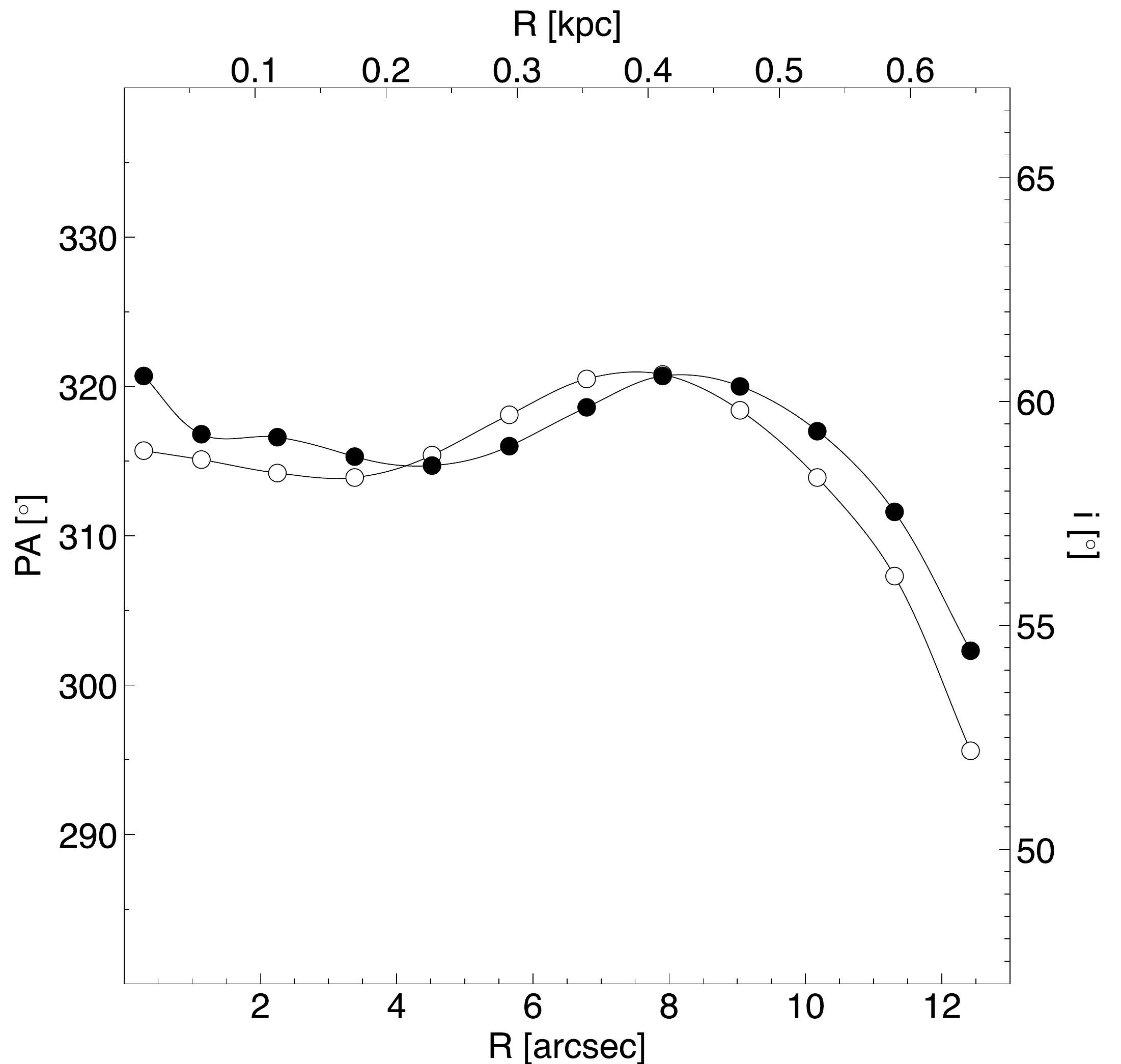}
\caption{Position angle (filled circles) and inclination (open circles) derived by using the systemic velocity \(v_\mathrm{sys}=998\) km s\(^{-1}\).}
\label{fig18}
\end{figure}

The position angle is determined by finding a line that maximizes the velocity gradient in the observed image. The computation constrained \(PA\) to the range of \(302\degr<PA<320\degr\) (when \(v_\mathrm{sys}=998\) km s\(^{-1}\)), as shown in figure \ref{fig18}. In the central \(r<400\) pc, the best fit is found for \(\langle PA\rangle\simeq316\degr\). By applying the same method to derive \(PA\) for \(v_\mathrm{sys}=964\) km s\(^{-1}\), an average value of \(324\arcdeg\) was found -- the central region exhibits a warp of \(\sim(5\degr\mathrm{-}10\degr)\) from the CND to the 500-pc ring. The position angle (and inclination) are consistent with the results based on H\textsc{i} data within \(r\lesssim120\arcsec\). The position angle remains \(PA\sim300\arcdeg\) until \(r\sim300\arcsec\), whereas inclination gradually decreases beyond \(r\gtrsim120\arcsec\) reaching \(i\sim30\arcdeg\) at \(r\sim300\arcsec\); this is manifested as the warped outer ring \citep{Kor93}. To compare this result with the distribution of stars in the galactic bulge, we fitted the 2MASS \(K_s\)-band image (figure \ref{fig7}b) with a two-dimensional Gaussian function within the central \(r<30\arcsec\). The resulting \(PA\) was \(323\fdg69\pm0\fdg48\), remarkably close to that derived from CO for \(v_\mathrm{sys}=964\) km s\(^{-1}\). We adopt the value of \(324\arcdeg\) for the major axis of the 500-pc ring and galactic bulge.

\clearpage

\subsubsection{Rotation curve and dynamical mass}

High-resolution CO line observations are useful for deriving rotation curves and mass distributions in the central regions of galaxies (e.g., \citealt{SR01,Sof13}). CO gas is often abundant and detectable with ALMA in the central regions where H\textsc{i} emission is weak and H\(\alpha\) affected by extinction. The data presented in this work have high resolution and sensitivity to probe the gas kinematics in the galactic bulge region with spatially continuous CO detection from the nucleus to the 500-pc ring. We first derive a rotation curve (\(v_\mathrm{rot}\)) by using \(^\mathrm{3D}\)Barolo and then analyze its composition.

\begin{figure}
\epsscale{1}
\plottwo{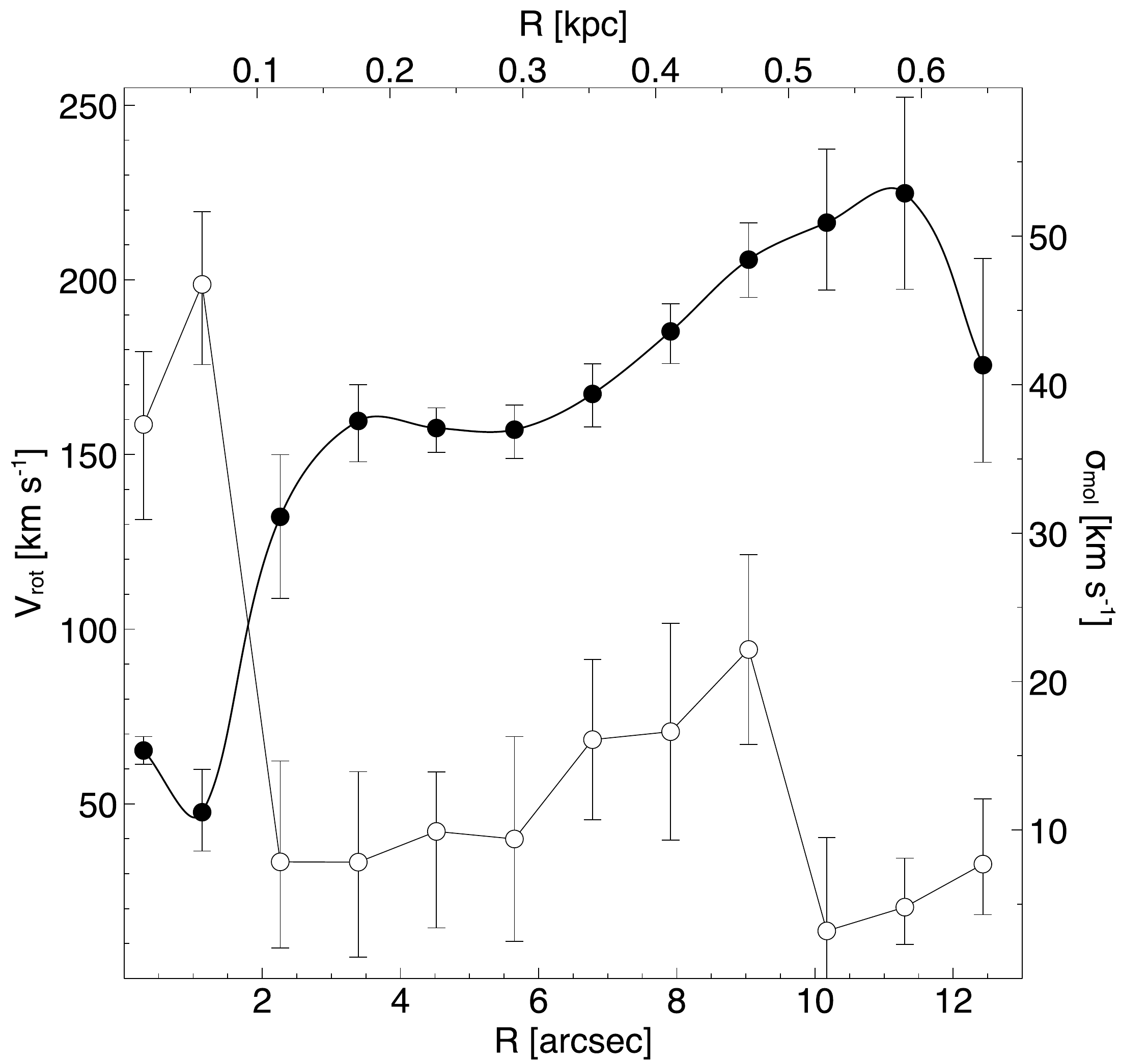}{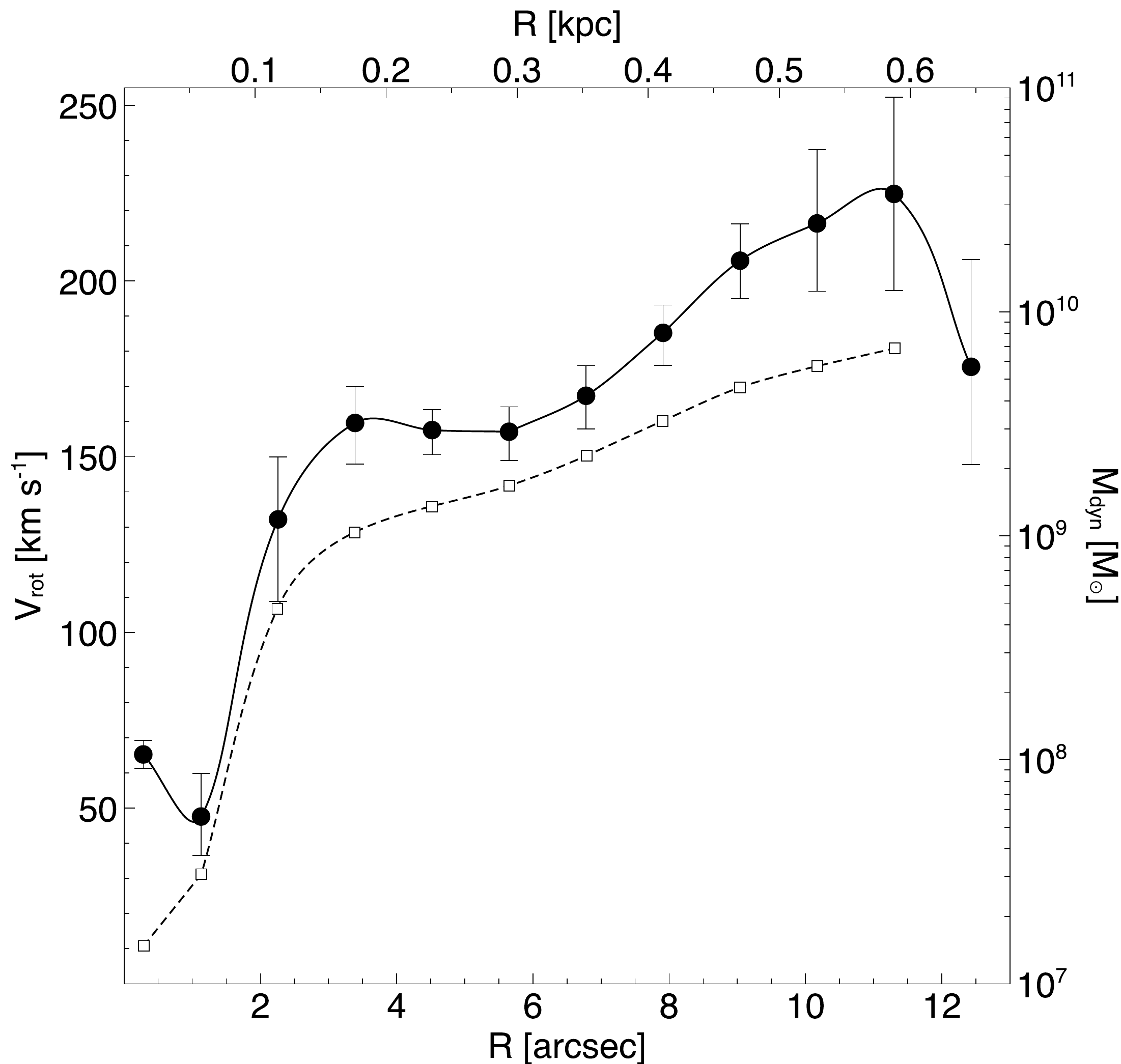}
\caption{\emph{Left.} Rotation curve (\(v_\mathrm{rot}\), full line) sampled in rings of width \(\Delta R=1\farcs13\) within \(R<676\) pc. It was determined by using the nuclear systemic velocity \(v_\mathrm{sys}=998\) km s\(^{-1}\) (filled circles) and corrected for the inclination and position angle (figure \ref{fig18}). Also plotted is the gas velocity dispersion (\(\sigma_\mathrm{mol}\), open circles). \emph{Right.} Rotation curve and dynamical mass distribution (\(M_\mathrm{dyn}\), dashed line).}
\label{fig19}
\end{figure}

The rotation curve derived by computation using \(v_\mathrm{sys}=998\) km s\(^{-1}\) is plotted in figure \ref{fig19}. It can be decomposed into three parts: the nucleus (\(r<1\arcsec\)), a steep rise at \(1\arcsec<r<3\arcsec\), and a more gentle rise beyond that radius. This change of slope is not a consequence of change in \(i\) or \(PA\) because these parameters are calculated separately and included in the derivation of \(v_\mathrm{rot}\). The last three measurement points are less reliable because the 500-pc ring exhibits a different systemic velocity of 964 km s\(^{-1}\) and because gas dynamics is dominated by streaming motion in the global bar. The outermost values, however, are in agreement with H\textsc{i} data from \citet{Kor93} and the result in \cite{Sof99} within uncertainty. The rotational velocity at the radius of \(r=705\) pc, derived by using the systemic velocity \(v_\mathrm{sys}=964\) km s\(^{-1}\), is \(v_\mathrm{rot}\simeq215\) km s\(^{-1}\). The results of modeling of gas kinematics inside the nuclear ring (derived by using \(v_\mathrm{sys}=998\) km s\(^{-1}\)) are summarized in table \ref{tab6}.

Assuming a simple spherically symmetric mass distribution, the derived rotational velocity yields a dynamical mass of \(M_\mathrm{dyn}(r<705~\mathrm{pc})=rv_\mathrm{rot}^2/G\simeq7.5\times10^9~M_\odot\). By comparison, the total velocity-integrated flux in the same region is \(S_\mathrm{CO}=991\) Jy km s\(^{-1}\) and the molecular gas mass derived by using the CO-to-H\(_2\) conversion factor \(X_\mathrm{CO}=0.8\times10^{20}~\mathrm{cm}^{-2}(\mathrm{K~km~s}^{-1})^{-1}\) becomes \(M_{\mathrm{H}_2}\simeq5\times10^8~M_\odot\) [see also \citet{Sal14} for a derivation using CO (3-2)]. The dynamical mass distribution is shown as a function of radius in figure \ref{fig19} (right).

\begin{figure}
\epsscale{1}
\plottwo{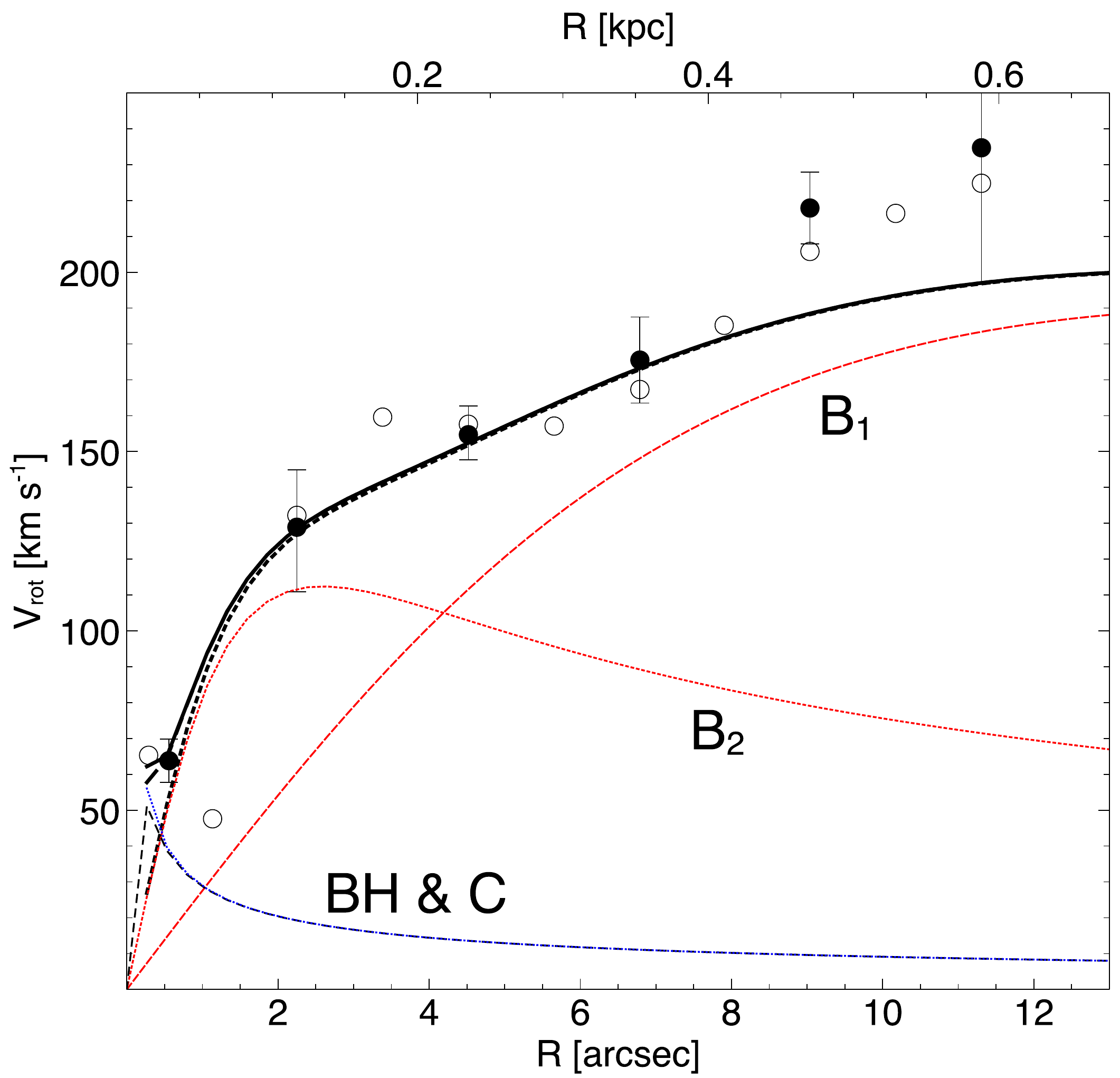}{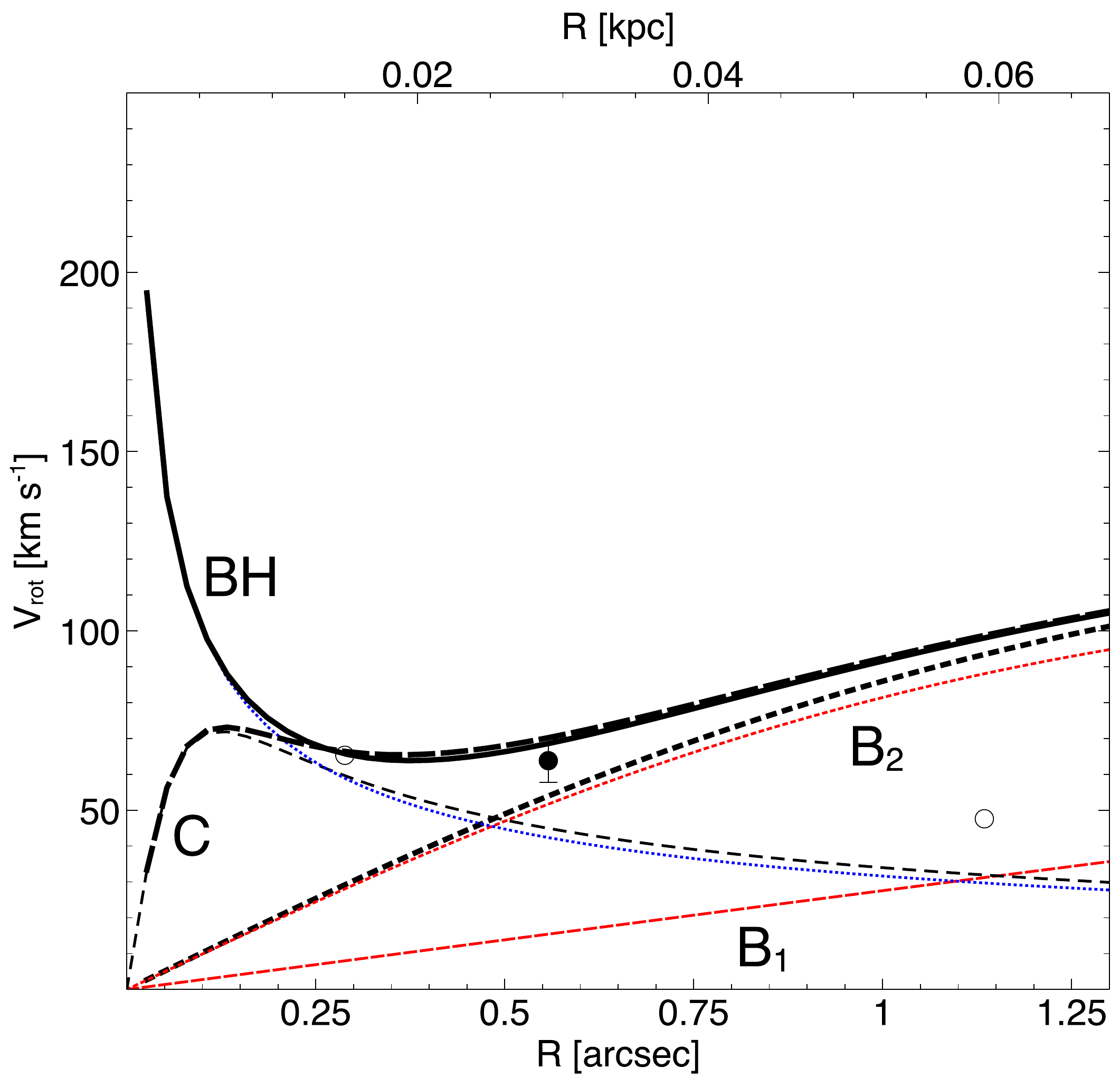}
\caption{\emph{Left.} Rotation curve within \(R<676\) pc fitted with three Plummer spheres and a black hole (Keplerian rotation). The curves represent: bulge (B\(_1\)) and major core (B\(_2\)) plotted with red dotted lines, nuclear cluster (C, dashed thin line), black hole (BH, blue dotted line), total curve with B\(_1\) and B\(_2\) only (dotted thick line), total curve including C (dashed thick line), and total curve including BH (full thick line). The open circles are the data from figure \ref{fig19}, while the filled circles are the data from the smoothed curve. \emph{Right.} Enlargement of the central 68 pc of the diagram in the left panel.}
\label{fig20}
\end{figure}

Next, we try to decompose the mass distribution into components that can explain the observed rotation curve. Since the rotation curve sampled at fine spacing of \(1\farcs13\) shows extreme changes in slope that require multiple mass components and are likely affected by non-circular motion (effective on a scale of \(\lesssim2\arcsec\); section 4.2 and figure \ref{fig24}), we ``smooth'' the curve by re-deriving the velocity at increments of \(2\farcs26\) (equal to \(\theta_\mathrm{maj}\)). The resulting curve can be fit by a crude model of two massive systems in the central 1 kpc. The larger one is the stellar bulge, while the smaller one denoted by major core (scale radius 95 pc) can be a simplified version of the nuclear bar and the molecular CND, which has a mass of \(\gtrsim1.4\times10^8~M_\odot\) in the central 300 pc and should be included in velocity curve calculations. The bulge and major core are represented by a simple Plummer sphere model \citep{Plu11} with the total gravitational potential given by

\begin{equation}
\Phi(r)=-\frac{G\mathcal{M}_1}{\sqrt{r^2+a_1^2}}-\frac{G\mathcal{M}_2}{\sqrt{r^2+a_2^2}}.
\end{equation}
Here, \(\mathcal{M}_1\) and \(\mathcal{M}_2\) are the total masses, and \(a_1\) and \(a_2\) the scale lengths of the bulge and major core, respectively. The mass distribution of a Plummer sphere with the total mass \(\mathcal{M}\) is

\begin{equation}
M(<r)=4\pi\int_0^r r'^2\rho(r')\,\mathrm{d}r'=\frac{\mathcal{M}r^3}{(r^2+a^2)^{3/2}},
\end{equation}
where \(\rho(r)=3\mathcal{M}(1+r^2/a^2)^{-5/2}(4\pi a^3)^{-1}\) is the volume density. Due to spherical symmetry, the rotational velocity is simply

\begin{equation}
v(r)=\sqrt{\frac{GM(<r)}{r}}=\sqrt{\frac{G\mathcal{M}}{(r^2+a^2)^{3/2}}}r.
\end{equation}
The total rotation curve due to both systems, expressed as \(v_\mathrm{rot}(r)=\sqrt{v_1^2(r)+v_2^2(r)}\), is plotted in figure \ref{fig20}. We also include a nuclear star cluster with mass \(\mathcal{M}_\mathrm{C}=1.4\times10^7~M_\odot\) and scale radius \(a_\mathrm{C}=4.5\) pc, indicated from high infrared luminosity of the core (e.g., \citealt{Oli95,TG96,GA08}). The selected scale radius corresponds to a core radius (where the surface density drops to the half of its maximum) of \(r_\mathrm{c}=\sqrt{\sqrt{2}-1}a\approx3\) pc, a typical value for nuclear clusters in nearby galaxies \citep{Wal05}. The derived mass of the bulge+major core (\(\mathcal{M}_\mathrm{B}=1.32\times10^{10}~M_\sun\)) is consistent within 5\% with the bulge mass derived by \cite{Sof99} by decomposing a kpc-scale rotation curve. In addition, a rotation curve that includes a tentative central supermassive black hole (mass \(\mathcal{M}_\mathrm{BH}=1.2\times10^7~M_\odot\), or about thrice as massive as the Galactic one; \citealt{Gil09}) is shown too; the black hole has a Keplerian rotation curve. It is not clear whether NGC 1808 has a supermassive black hole in its center, but figure \ref{fig20} shows that it would not be distinguished from the star cluster of the same mass at the present resolution of 100 pc. For a bulge mass of \(\mathcal{M}_\mathrm{B1}=1.25\times10^{10}~M_\sun\), the black-hole-to-bulge mass ratio is of the order \(\mathcal{M}_\mathrm{BH}/\mathcal{M}_\mathrm{B1}\sim1\times10^{-3}\), consistent with large samples in \cite{KH13}. Note that even if a black hole and a nuclear cluster coexist as a core\footnote{Supermassive black holes and nuclear clusters seem to coexist in many galaxies and the ratio of their masses can vary by three orders of magnitude. In the Galaxy, the mass of the black hole is only \(\sim15\%\) of the mass of the nuclear cluster, while in our neighbour M31, the black hole is \(\sim4\) times more massive. For a review, see \cite{KH13} and the references therein.} of mass \(\mathcal{M}_\mathrm{core}=\mathcal{M}_\mathrm{BH}+\mathcal{M}_\mathrm{C}\), the core-to-bulge mass ratio \(\mathcal{M}_\mathrm{core}/\mathcal{M}_\mathrm{B1}\) also scales in agreement with \cite{KH13} for composite nuclei. Higher-resolution data will resolve the nucleus and clarify whether the central rise of the rotation curve identified in figure \ref{fig19} is due to a black hole or a nuclear star cluster. In either case, the rise of the rotation curve within \(r<1\arcsec\) reveals an embedded molecular gas disk or torus with a radius less than 50 pc. An exponential disk and dark halo components are not included here because their contributions are small at \(r<500\) pc. The disk would raise the rotation curve at larger radii, but due to the presence of a galactic bar, detailed modeling of the bar potential would be required to probe rotation beyond the 500-pc ring. The parameters of the model are shown in table \ref{tab7}.

\begin{table}
\begin{center}
\caption{Geometric and kinematic parameters of the central 1-kpc molecular zone.}\label{tab6}
\begin{tabular}{lrrrrrr}
\tableline\tableline
Ring & Radius [\(\arcsec\)] & Radius [pc] & \(v_\mathrm{rot}\) [km s\(^{-1}\)] & \(\sigma_\mathrm{mol}\) [km s\(^{-1}\)] & \(i\) [\(\arcdeg\)] & \(PA\) [\(\arcdeg\)] \\
\tableline
1 & 0.28 & 15 & \(65.3_{-4.0}^{+4.0}\) & \(37.3_{-6.4}^{+4.9}\) & 58.9 & 320.7 \\
2 & 1.13 & 59 & \(47.6_{-11.1}^{+12.3}\) & \(46.7_{-5.4}^{+4.9}\) & 58.7 & 316.8 \\
3 & 2.26 & 117 & \(132.2_{-23.4}^{+17.8}\) & \(7.9_{-5.8}^{+6.8}\) & 58.4 & 316.6 \\
4 & 3.39 & 176 & \(159.6_{-11.7}^{+10.4}\) & \(7.8_{-6.4}^{+6.1}\) & 58.3 & 315.3 \\
5 & 4.52 & 235 & \(157.6_{-7.0}^{+5.8}\) & \(9.9_{-6.5}^{+4.0}\) & 58.8 & 314.7 \\
6 & 5.65 & 294 & \(157.1_{-8.2}^{+7.1}\) & \(9.4_{-6.9}^{+6.9}\) & 59.7 & 316.0 \\
7 & 6.78 & 353 & \(167.3_{-9.4}^{+8.6}\) & \(16.1_{-5.4}^{+5.4}\) & 60.5 & 318.6 \\
8 & 7.91 & 411 & \(185.2_{-9.2}^{+7.9}\) & \(16.6_{-7.3}^{+7.3}\) & 60.6 & 320.7 \\
9 & 9.04 & 470 & \(205.8_{-10.8}^{+10.5}\) & \(22.2_{-6.4}^{+6.4}\) & 59.8 & 320.0 \\
10 & 10.17 & 529 & \(216.4_{-19.3}^{+21.0}\) & \(3.2_{-6.3}^{+6.3}\) & 58.3 & 317.0 \\
11 & 11.30 & 588 & \(224.8_{-27.5}^{+27.5}\) & \(4.8_{-2.5}^{+3.3}\) & 56.1 & 311.6 \\
12 & 12.43 & 646 & \(175.6_{-27.8}^{+30.5}\) & \(7.7_{-3.4}^{+4.4}\) & 52.2 & 302.3 \\
\tableline
\end{tabular}
\end{center}
\end{table}

\begin{table}
\begin{center}
\caption{Dynamical parameters of the galactic model in the central 1 kpc.}\label{tab7}
\begin{tabular}{lrrr}
\tableline\tableline
Component & \(\mathcal{M}\) [\(M_\odot\)] & \(a\) [pc] & \(\Sigma_\mathrm{s}(0)~[M_\sun~\mathrm{pc}^{-2}]\) \\
\tableline
Bulge (B\(_1\)) &  \(1.25\times10^{10}\) & 575 & \(1.2\times10^4\) \\
Major core (B\(_2\)) & \(7.2\times10^8\) & 95 & \(2.0\times10^4\) \\
Nuclear bar (NB) & \(5.8\times10^8\) & 95 & \(2.0\times10^4\) \\
Nuclear cluster (C) & \(1.4\times10^7\) & 4.5 & \(2.20\times10^5\) \\
Central black hole (BH) & (\(1.2\times10^7\)) & -- & -- \\
Total & \(1.32\times10^{10}\) & & \(2.52\times10^5\) \\
\tableline
\end{tabular}
\end{center}
\end{table}

\clearpage

\subsubsection{Stellar surface density in the nucleus}

\begin{figure}
\epsscale{1}
\plotone{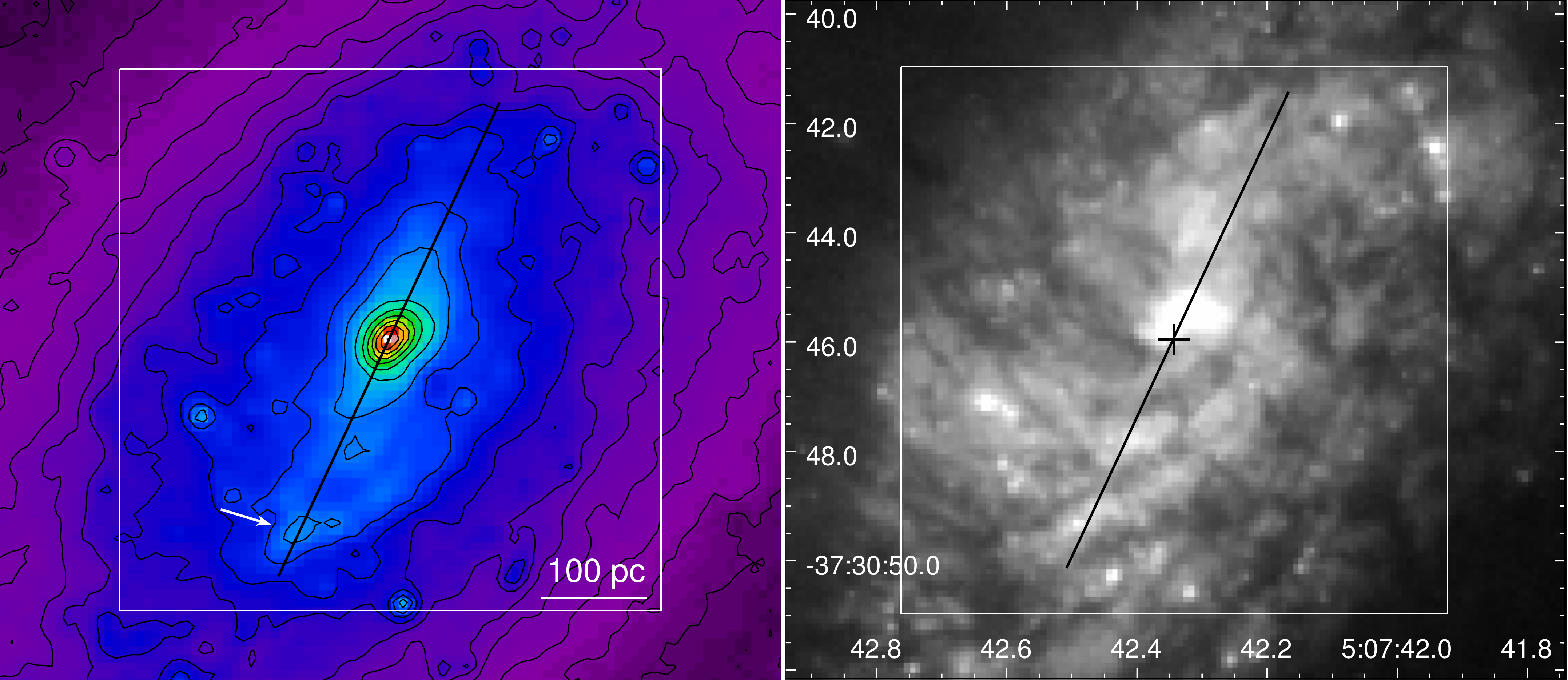}
\caption{\(K_s\)-band (left) and \(I\)-band (right) images of the central region. The white rectangle shows the central \(10\arcsec\times10\arcsec\). The black line is the projection slice at \(PA=335\arcdeg\) (orientation of the major axis of the nuclear bar) and length 500 pc. The white arrow marks the position of an arc-like feature (see figure \ref{fig22}).}
\label{fig21}
\end{figure}

\begin{figure}
\epsscale{1}
\plottwo{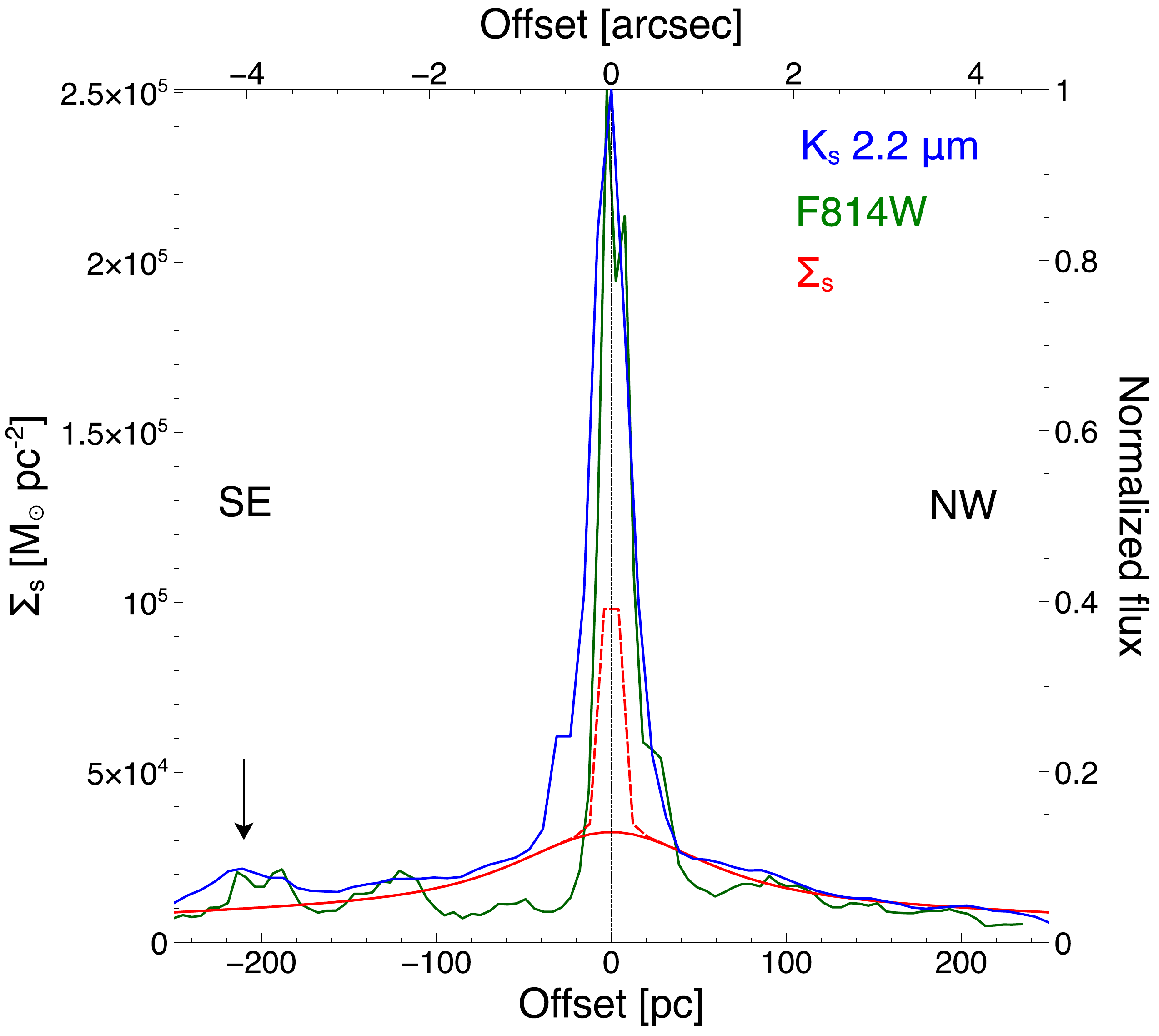}{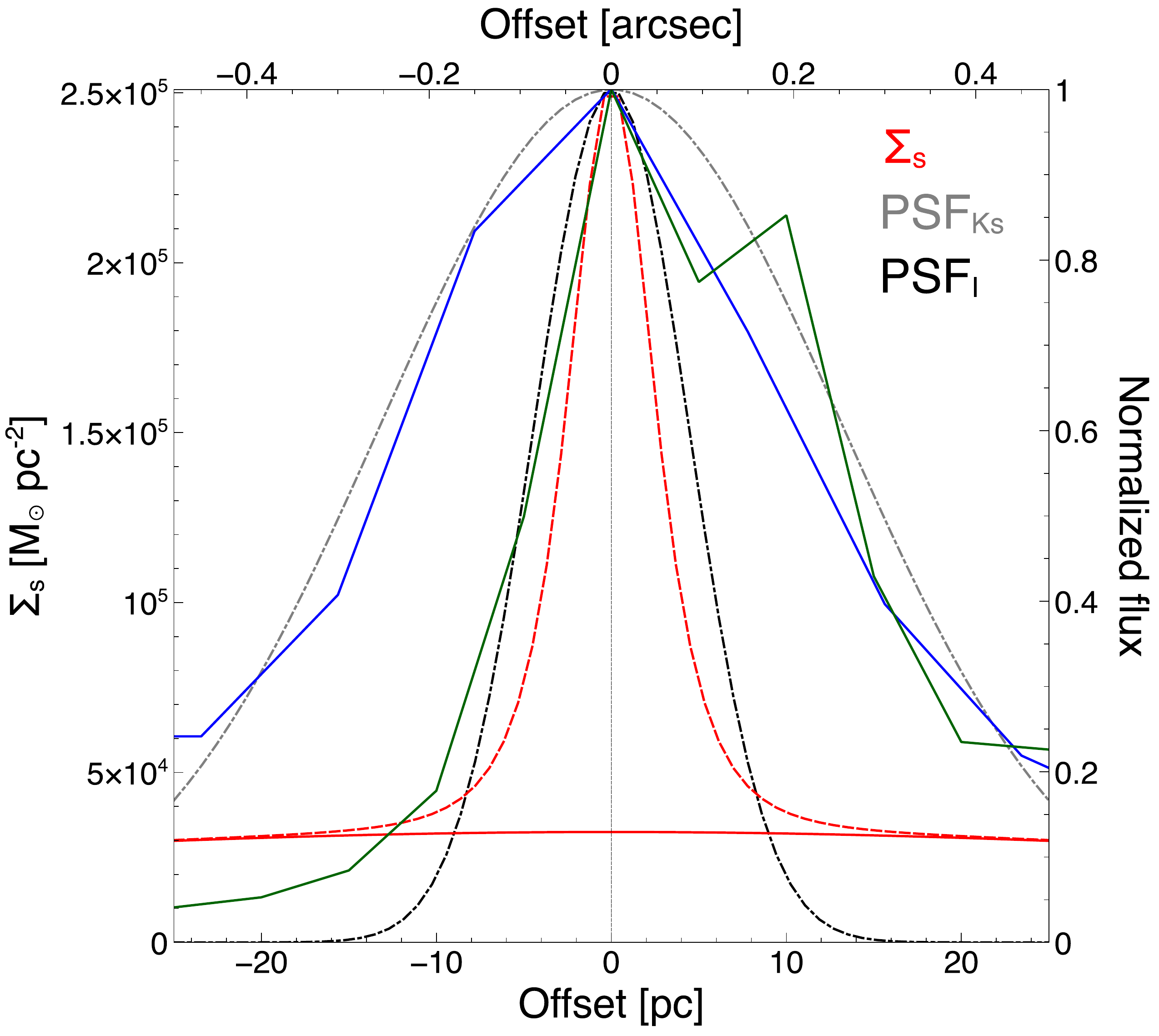}
\caption{\emph{Left.} Radial profiles of the normalized fluxes of the \(K_s\) (blue) and \(I\) (green) images plotted along a 500-pc slice (offset from the galactic center at \(PA=335\arcdeg\); figure \ref{fig21}). Also shown is the total surface density of old stars that comprises a bulge and nuclear bar (full red line: \(\Sigma_\mathrm{s}=\Sigma_\mathrm{B1}+\Sigma_\mathrm{NB}\)) calculated from gas dynamics (table \ref{tab7}). The dashed red line (cut at 1/3 its maximum for clarity) shows the total surface density that includes a nuclear cluster (\(\Sigma_\mathrm{s}=\Sigma_\mathrm{B1}+\Sigma_\mathrm{NB}+\Sigma_\mathrm{C}\)). The arrow marks the location of an arc feature (also marked in figure \ref{fig21}). \emph{Right.} Enlargement of the central 50 pc. The grey and black dot-dashed curves are the PSFs of the \(K_s\)-band and \(I\)-band images (FWHMs equal to \(0\farcs6\) and \(0\farcs2\), respectively).}
\label{fig22}
\end{figure}

In order to compare our mass estimate based on gas dynamics with the total mass budget of the old stars in the central region, we used \(I\)-band (F814W filter on \emph{Hubble Space Telescope}) and \(K_s\)-band [ISAAC on \emph{Very Large Telescope}; \cite{GA08}] images shown in figure \ref{fig21}.  High-resolution \(I\)-band images have been useful in estimating black hole masses in nearby galaxies \citep{Dav13,Oni15}, whereas the \(K_s\)-band image is a good choice because it is unaffected by dust extinction. The images were used to derive the radial profiles of surface brightness, \(I(R)=L/D^2\) (\(L\) is the luminosity and \(D\) is the physical size of the observed region), and estimate the total stellar surface density in the central \(r<250\) pc, assuming a uniform mass-to-light ratio \(M(r)/L(r)\). To do this, we normalized the peaks of the radial profiles to the peak of the surface density derived by using gas dynamics. For the Plummer sphere model discussed above, the stellar surface density is calculated as

\begin{equation}\label{sigma}
\Sigma_\mathrm{s}(R)=\int_{-\infty}^\infty\rho[r(z)]\,dz=2\int_{-\infty}^\infty\frac{3a^2\mathcal{M}\,dz}{4\pi(a^2+z^2+R^2)^{5/2}}=\frac{a^2\mathcal{M}}{\pi(a^2+R^2)^2},
\end{equation}
where \(z=\sqrt{r^2-R^2}\) is the axis along the line of sight, and \(R\) is the projected radius in the plane of the sky. The results are given in table \ref{tab7}, and the total surface density due to the bulge and nuclear bar, \(\Sigma_\mathrm{s}=\Sigma_\mathrm{B1}+\Sigma_\mathrm{NB}\), is plotted in figure \ref{fig22} along with the normalized flux profiles of the near-infrared images. The mass of the nuclear bar is calculated as \(\mathcal{M}_\mathrm{NB}=\mathcal{M}_\mathrm{B2}-\mathcal{M}_\mathrm{CND}\), where \(\mathcal{M}_\mathrm{CND}=1.4\times10^8~M_\sun\) (table \ref{tab5}). To deal with the central peak, we added a nuclear star cluster (mass \(\mathcal{M}_\mathrm{C}\) and surface density \(\Sigma_\mathrm{C}\) calculated using equation \ref{sigma}) to match the normalized flux peak (assuming at this point that the near-infrared emission is dominated by starlight rather than hot dust). The total surface density profile is then \(\Sigma_\mathrm{s}=\Sigma_\mathrm{B1}+\Sigma_\mathrm{NB}+\Sigma_\mathrm{C}\). The obtained \(\Sigma_\mathrm{s}\) is a good match for the observed surface brightness, except in a region that appears like an arc or a spiral arm at \(r\sim200\) pc marked in figure \ref{fig21}, suggesting that the mass of the core can be dominated by a compact stellar system. Furthermore, if the nucleus has undergone a starburst episode with a star formation rate of \(SFR\sim1~M_\sun~\mathrm{yr}^{-1}\) over the past \(t_\mathrm{SB}\sim10^7\) yr, as estimated by \cite{Kra94}, it is reasonable to expect a star cluster at the galactic center with a mass comparable to \(\mathcal{M}_\mathrm{C}\).

Alternatively, the central peak in the infrared images could be due to emission from dusty ISM heated to high temperatures by a low-luminosity AGN. In that case, the peak would be the product of a convolution of the AGN-dominated region with the point spread function (PSF) of the telescope. In the simplest case, the AGN is a point source (delta function), the PSF is a Gaussian, and the resulting convolution is equal to the PSF. Using the PSF of the \(K_s\)-band image, it is possible to reproduce the central peak in surface brightness shown in figure \ref{fig22} (right). However, the convolution of the PSF with the surface brightness of the compact cluster \(\Sigma_\mathrm{C}\) can also result in a function similar to the PSF.

\clearpage

\subsection{Non-circular motions in the galactic disk}

\begin{figure}
\epsscale{0.75}
\plotone{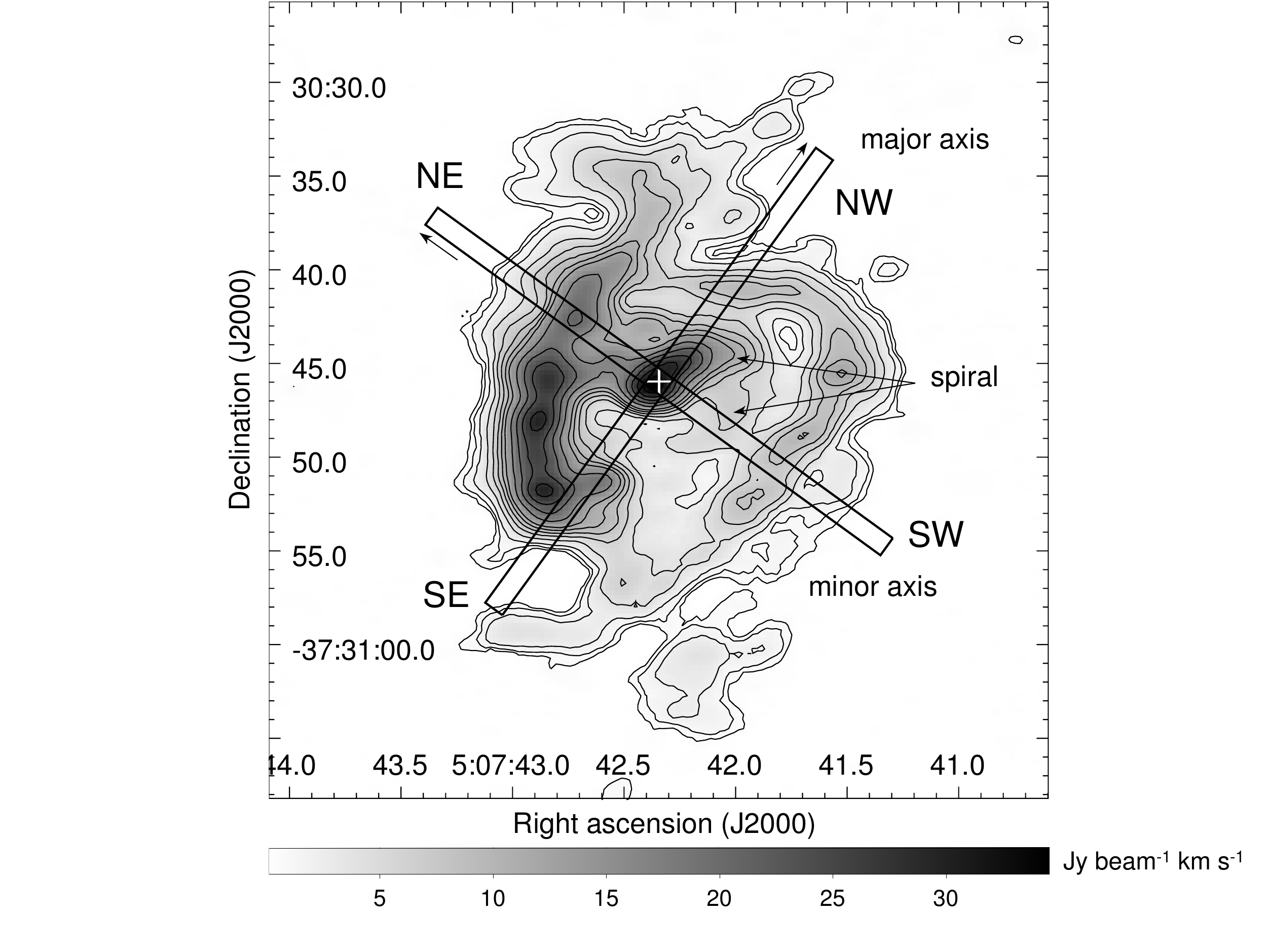}
\caption{PVD slices plotted on the intensity image. Major (\(PA=324\arcdeg\)) and minor (\(PA=54\arcdeg\)) axes are indicated. The width of each slice is 9 pixels (\(2\farcs25\)), corresponding to the major axis of the beam FWHM. The length is \(30\arcsec\) (1.6 kpc). Note the presence of a spiral arm between the CND and the 500-pc pseudo ring; the ring too extends as a two-arm pattern.}
\label{fig23}
\end{figure}

Gas kinematics can be further investigated with position-velocity diagrams (PVDs). PVDs along the major (\(X\)) and minor (\(Y\)) galactic axes can reveal azimuthal and radial streaming motions, respectively (e.g., \citealt{Aal99}). Using the parameters derived above, PVDs were plotted along the global kinematic major and minor axes at \(PA_\mathrm{maj}=324\arcdeg\) and \(PA_\mathrm{min}=54\arcdeg\), respectively (figure \ref{fig23}). The width of the slits is equivalent to the FWHM of the beam major axis.

\begin{figure}
\epsscale{1}
\plotone{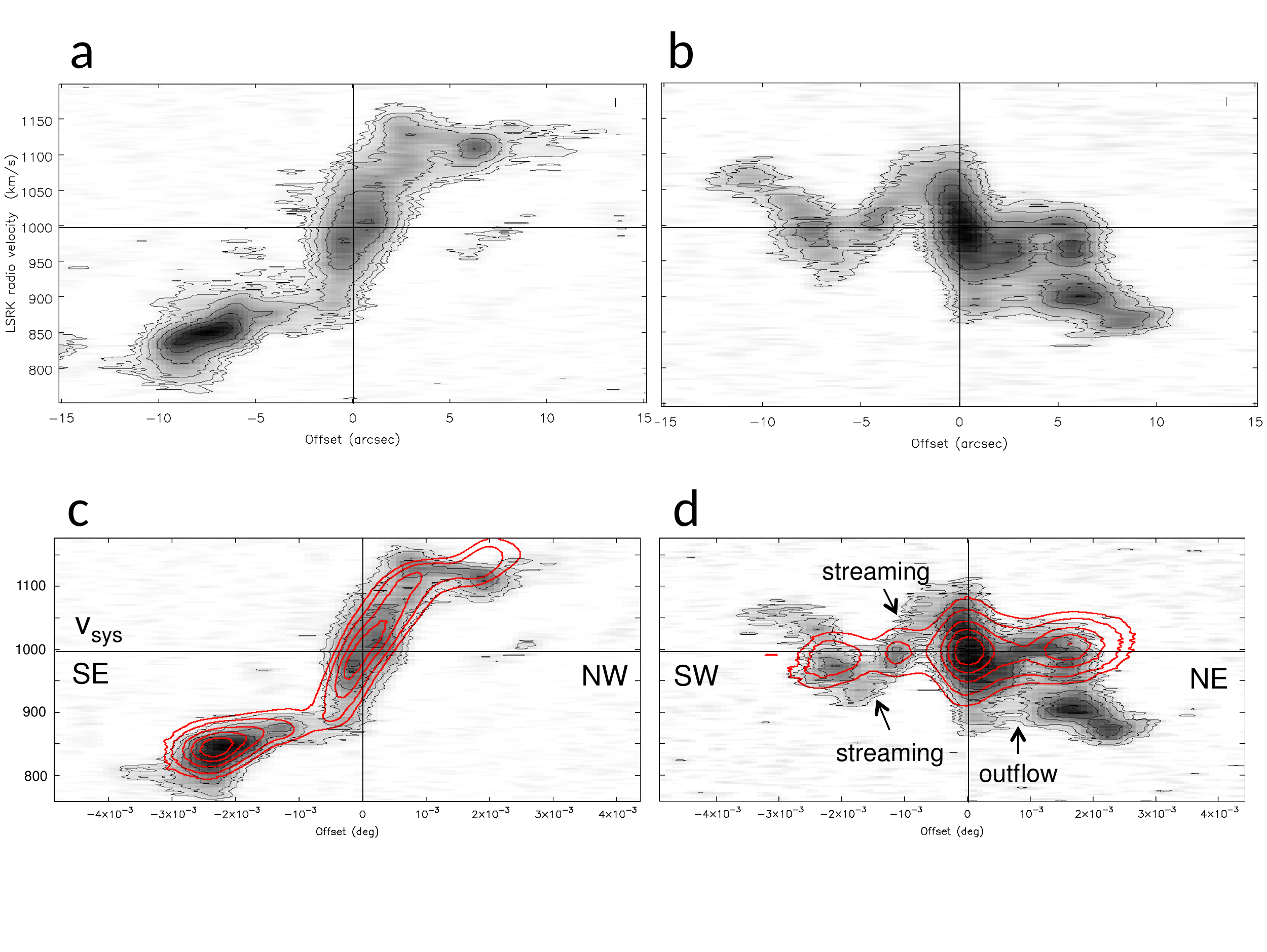}
\caption{(a) PVD along the adopted major axis (\(PA=324\arcdeg\)) at \(Y=0\) with the contours plotted at 0.05, 0.1, 0.2, 0.4, 0.6, 0.8 times the peak 0.36 Jy beam\(^{-1}\). (b) PVD along the adopted minor axis (\(PA=54\arcdeg\)) at \(X=0\) with the contours plotted at 0.1, 0.2, 0.4, 0.6, 0.8 times the peak 0.24 Jy beam\(^{-1}\). Panels c and d show examples of fitting with \(^\mathrm{3D}\)Barolo, where the red contours (normalized to 0.1, 0.2, 0.4, 0.6, 0.8 times the peak) are the best fits. The horizontal line is \(v_\mathrm{LSR}=998\) km s\(^{-1}\). Streaming motion and outflow are indicated.}
\label{fig24}
\end{figure}

\begin{figure}
\epsscale{1}
\plotone{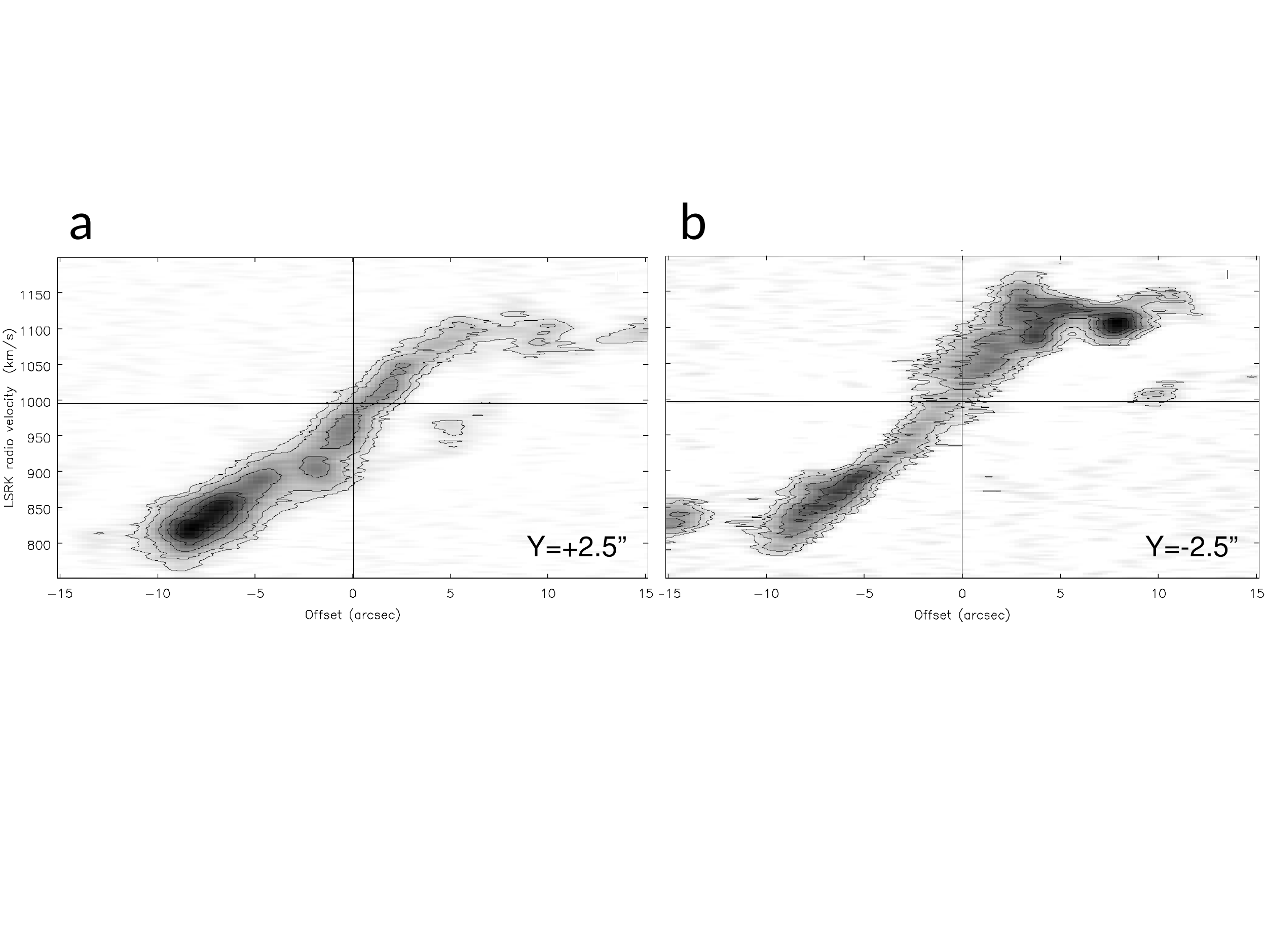}
\caption{PVD parallel to the adopted major axis (\(PA=324\arcdeg\)) at \(Y=2\farcs5\) (a) and \(Y=-2\farcs5\) (b). The contours are plotted at 0.1, 0.2, 0.4, 0.6, 0.8 times the peak 0.24 Jy beam\(^{-1}\). The horizontal line is \(v_\mathrm{LSR}=998\) km s\(^{-1}\).}
\label{fig25}
\end{figure}

The PVD along the major axis shown in figure \ref{fig24}a exhibits rotation that can be decomposed into three components (modeled in section 4.1.4): (1) rigid-body rotation within \(r<2\farcs5\), (2) nearly flat rotation curve within \(2\farcs5<r<6\arcsec\), followed by (3) additional rise in velocity. Inspection of figure \ref{fig25} shows that the deviation (1) is less prominent beyond \(|Y|=2\farcs5\) from the nucleus suggesting that it is confined to the central \(r\lesssim2\farcs5\), as expected from a massive core in the CND.

\subsubsection{Velocity gradient along the minor axis}

Figure \ref{fig24}b shows a PVD plotted along the minor axis (\(X=0\arcsec\)) that reveals striking non-circular motions in the central 1-kpc region: the ``zig-zag'' pattern as a signature of a velocity gradient along the minor axis. As shown in the next section, the velocity offsets at \(|Y|\gtrsim8\arcsec\) correspond to a component dominated by the global bar dynamics. Non-circular motions appear least important in the region \(4\arcsec< |Y|< 7\arcsec\), where gas kinematics is dominated by circular rotation of the disk (highest-intensity emission in figure \ref{fig24}b). Irregular motions intensify again at \(|Y|<4\arcsec\), in the vicinity of nucleus.

The panels (c) and (d) of figure \ref{fig24} show examples of fitting with \(^\mathrm{3D}\)Barolo: the red contour is the best fit obtained with a warp model superimposed on a the PVD diagram made from the data cube. We note that the program created a good fit for the major axis (panel c). The situation is, however, significantly different for the minor axis (panel d). The fit was successful only for the high-intensity emission that is near the systemic velocity (998 km s\(^{-1}\)), i.e., for the near-circular motion components. Since \(^\mathrm{3D}\)Barolo can handle warps in circular orbits by searching for symmetric features in the data cube, we regard the non-fitted emission in figure \ref{fig24}d as non-circular motion. Specifically, we recognise two patterns: (1) streaming motions in the SW with velocities \(|v_\mathrm{LSR}-v_\mathrm{sys}|\sim50\) km s\(^{-1}\) at \(Y\approx-2\farcs5\) and \(Y\approx-7\arcsec\) (marked with arrows in panel d), and (2) highly blueshifted component in the NE with \(|v_\mathrm{LSR}-v_\mathrm{sys}|\sim100\) km s\(^{-1}\) from \(Y=0\arcsec\) to \(Y\approx6\arcsec\). Most of the emitting gas on the NE side is near the systemic velocity (998 km s\(^{-1}\)) and there is signature of the zig-zag pattern, but the extreme velocities between \(v_\mathrm{LSR}=870\) and 950 km s\(^{-1}\) [corresponding to \(|v_\mathrm{LSR}-v_\mathrm{sys}|=(48-128)\) km s\(^{-1}\)] can be explained only as a large inflow in the disk or an outflow off the disk. We show below that the outflow scenario is more likely (section 5).

\subsubsection{Bar pattern speed and Lindblad resonances}

To investigate non-circular motions in the galactic disk, we created a model of a disk with pure circular rotation. The observed line-of-sight velocity \(v_\mathrm{obs}\) in cylindrical coordinates \((r,\phi,z)\) is given by

\begin{equation}\label{obs}
v_\mathrm{obs}=v_\mathrm{sys}+v_\phi\sin{i}\cos{\phi}+v_r\sin{i}\sin{\phi}+v_z\cos{i},
\end{equation}
where \(v_\phi\) is the rotational (azimuthal, \(\phi\) component) velocity, \(v_r\) radial velocity, \(v_z\) vertical velocity (perpendicular to the galactic plane negative toward the observer), and \(\phi\) the position angle measured with respect to the major galactic axis. The model velocity includes a rotation curve \(v_\mathrm{rot}\), and excludes radial and extraplanar motions (\(v_r=v_z=0\)):

\begin{equation}\label{mod}
v_\mathrm{mod}=v_\mathrm{sys}+v_\mathrm{rot}\sin{i}\cos{\phi}.
\end{equation}
Subtracting the model velocity field from the data, we get a residual image,

\begin{equation}\label{res}
v_\mathrm{res}=(v_\phi-v_\mathrm{rot})\sin{i}\cos{\phi}+v_r\sin{i}\sin{\phi}+v_z\cos{i},
\end{equation}
that carries information about non-circular and extraplanar motions. Assuming that most of the emission originates in the galactic plane (\(v_z=0\)), the residual image yields a relation between azimuthal and radial velocities. Note that negative values of \(v_\phi\) correspond to blueshifted velocities in images.

The parameters used to generate a model velocity field for the central region are \(PA=316\arcdeg\) (mean value; section 4.1.3), \(i=58\arcdeg\), \(v_\mathrm{sys}=998\) km s\(^{-1}\), and \(v_\mathrm{rot}(r)\) derived from numerical fitting described in subsection 4.1. In the case of a global model, we applied  \(PA=324\arcdeg\), \(i=57\arcdeg\), \(v_\mathrm{sys}=964\) km s\(^{-1}\), and a flat rotation curve of \(v=190\) km s\(^{-1}\) beyond \(13\farcs6\), consistent with the global rotation curve \citep{Sof99}. The position angle and inclination are kept constant in the global model because in the central 1 kpc they vary with radius less than about \(10\arcdeg\). The resulting residual velocity image is shown in figure \ref{fig26}. From the residual image, it is clear that the velocity of molecular gas in the offset ridges in the bar region deviates largely from the rest of the disk.

The position angle of the bar is \(PA_\mathrm{b}\approx335\arcdeg\), hence only \(\phi\approx11\arcdeg\) with respect to the galactic disk. Because of this, radial motion (\(v_r\sin{i}\sin{\phi}\)) can be ignored along the bar in the residual velocity image given by equation \ref{res}. This is a favourable situation for deriving the bar pattern speed \(\Omega_\mathrm{b}\) using CO kinematics because equation \ref{obs} simplifies to \(v_\mathrm{obs}\approx v_\mathrm{sys}+v_\phi\sin{i}\), where \(v_\phi\approx r\Omega_\mathrm{b}\) becomes the rotational velocity of the bar and is expected to linearly increase with \(r\) (e.g., \citealt{Sak00,Kun00,Kod02,KS06,Hir14}). This trend is clear from the PVD in figure \ref{fig27} where the offset ridges exhibit approximately linear behaviour. Measuring the velocities at emission peaks along the ridge within the bar (\(1~\mathrm{kpc}\leq r\leq3~\mathrm{kpc}\)) and fitting with a least-squares method, we derive a bar pattern speed of \(\Omega_\mathrm{b}=56\pm11\) km s\(^{-1}\) kpc\(^{-1}\).

Using the rotation curve derived in section 4.1.4, it is possible to calculate the angular velocity \(\Omega\) and investigate the dynamical resonances of stellar orbits in the galactic disk. The locations of Lindblad resonances are found from \(m(\Omega-\Omega_\mathrm{b})=\pm\kappa\), where \(\kappa=\sqrt{r\mathrm{d}\Omega^2/\mathrm{d}r+4\Omega^2}\) is the epicyclic frequency of stellar orbits calculated as \(\kappa\simeq\sqrt{2\Omega(\Omega+\Delta v/\Delta r)}\), and \(m\) is the rotational symmetry parameter (equal to 2 for a bar or two-arm pattern). Figure \ref{fig28} shows \(\Omega\), \(\Omega\pm\kappa/2\), \(\Omega-\kappa/4\), and \(\Omega_\mathrm{b}=56\) km s\(^{-1}\) kpc\(^{-1}\), where the rotation curve \(v_\mathrm{rot}\) was adopted from figure \ref{fig20} at \(r<600\) pc, and assumed constant, \(v_\mathrm{rot}=190\) km s\(^{-1}\), at \(r>1.5\) kpc \citep{Sof99}. If the disk rotation is flat at \(v_\mathrm{rot}=(160\mathrm{-}200)\) km s\(^{-1}\), corotation occurs at \(r_\mathrm{CR}=v_\mathrm{rot}/\Omega_\mathrm{b}\simeq(2.9-3.6)\) kpc, and \(r_\mathrm{CR}/a_\mathrm{b}\simeq(1.0-1.2)\), where \(a_\mathrm{b}\) is the length of the semi-major axis of the bar (section 3.2.1). This result is consistent with theoretical predictions for weak bars (e.g., \citealt{Con80,Ath92a,BT08}). A weak bar is ``allowed'' to exist only between the inner Lindblad resonance (ILR), where \(\Omega_\mathrm{b}=\Omega-\kappa/2\), and the corotation radius \(r_\mathrm{CR}\), where \(\Omega=\Omega_\mathrm{b}\) \citep{BT08}. From figure \ref{fig28}, the outer inner Lindblad resonance (oILR) appears at \((0.9-1.2)\) kpc. Additional ILRs are located near 450 pc where \(\Omega-\kappa/2\) has a local minimum and intersects with \(\Omega_\mathrm{b}\); they coincide with the 500-pc ring. The rotation curve includes the contribution from a tentative massive black hole at the galaxy center. As a consequence, \(\Omega-\kappa/2\) sharply rises within \(r<50\) pc [see, e.g., \cite{Com14} for NGC 1566]. There is also a resonance where \(\Omega-\kappa/4\) intersects with \(\Omega_\mathrm{b}\) near \(r\simeq2.2\) kpc; close to this radius in the SE ridge, we find a GMA and H\textsc{ii} region (briefly discussed in section 3.2.1; figure \ref{fig9}).

\subsubsection{Residual velocities in the central 1 kpc}

\begin{figure}
\epsscale{1}
\plotone{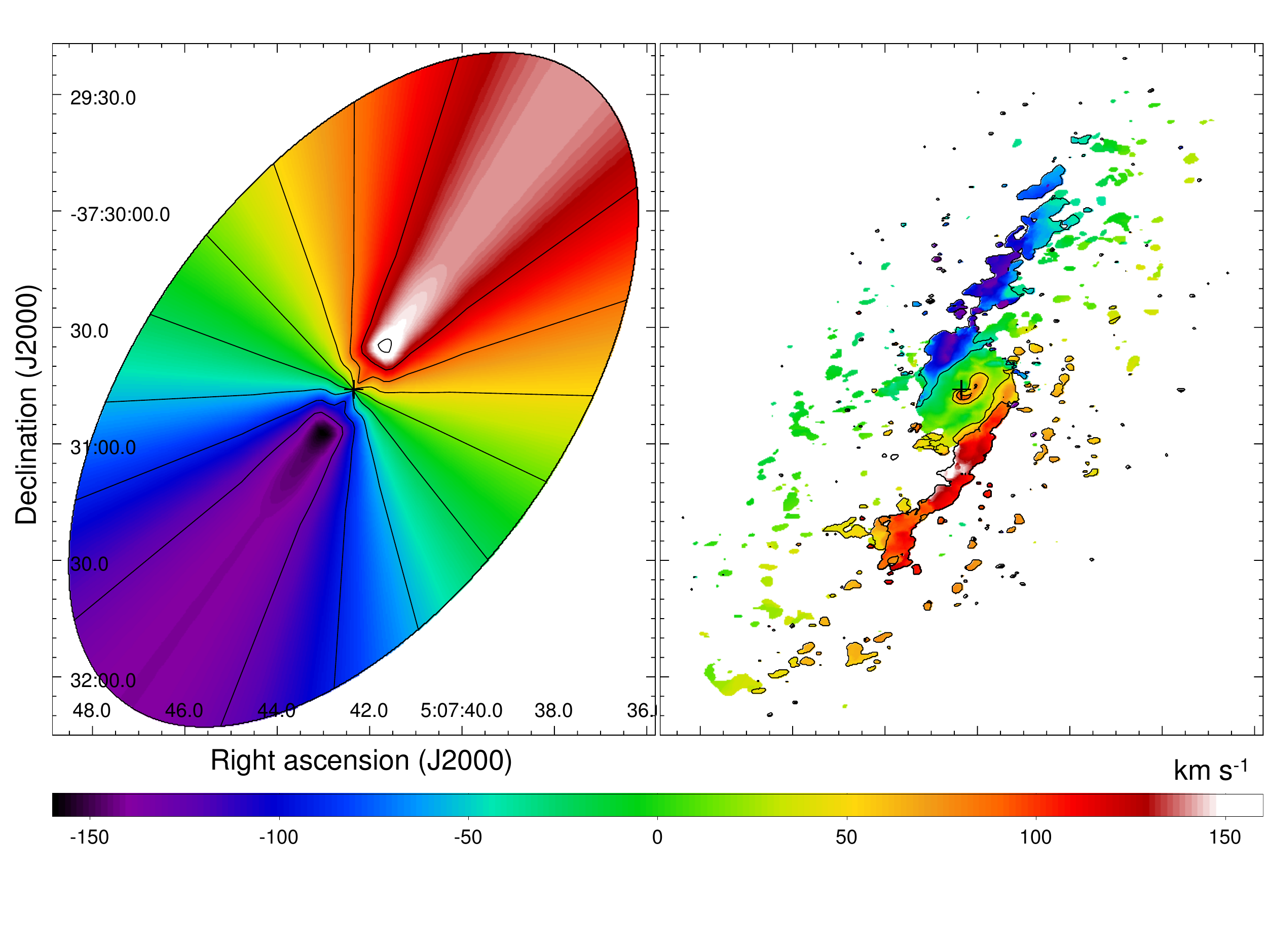}
\caption{\emph{Left.} Model velocity field (circular rotation) of the entire galaxy generated by using \(v_\mathrm{sys}=964\) km s\(^{-1}\) and \(PA=324\arcdeg\). \emph{Right.} Residual velocity image after subtracting the model from the observed data velocity. The contours are plotted at -80, -40, 40, 80 km s\(^{-1}\).}
\label{fig26}
\end{figure}

\begin{figure}
\epsscale{1}
\plotone{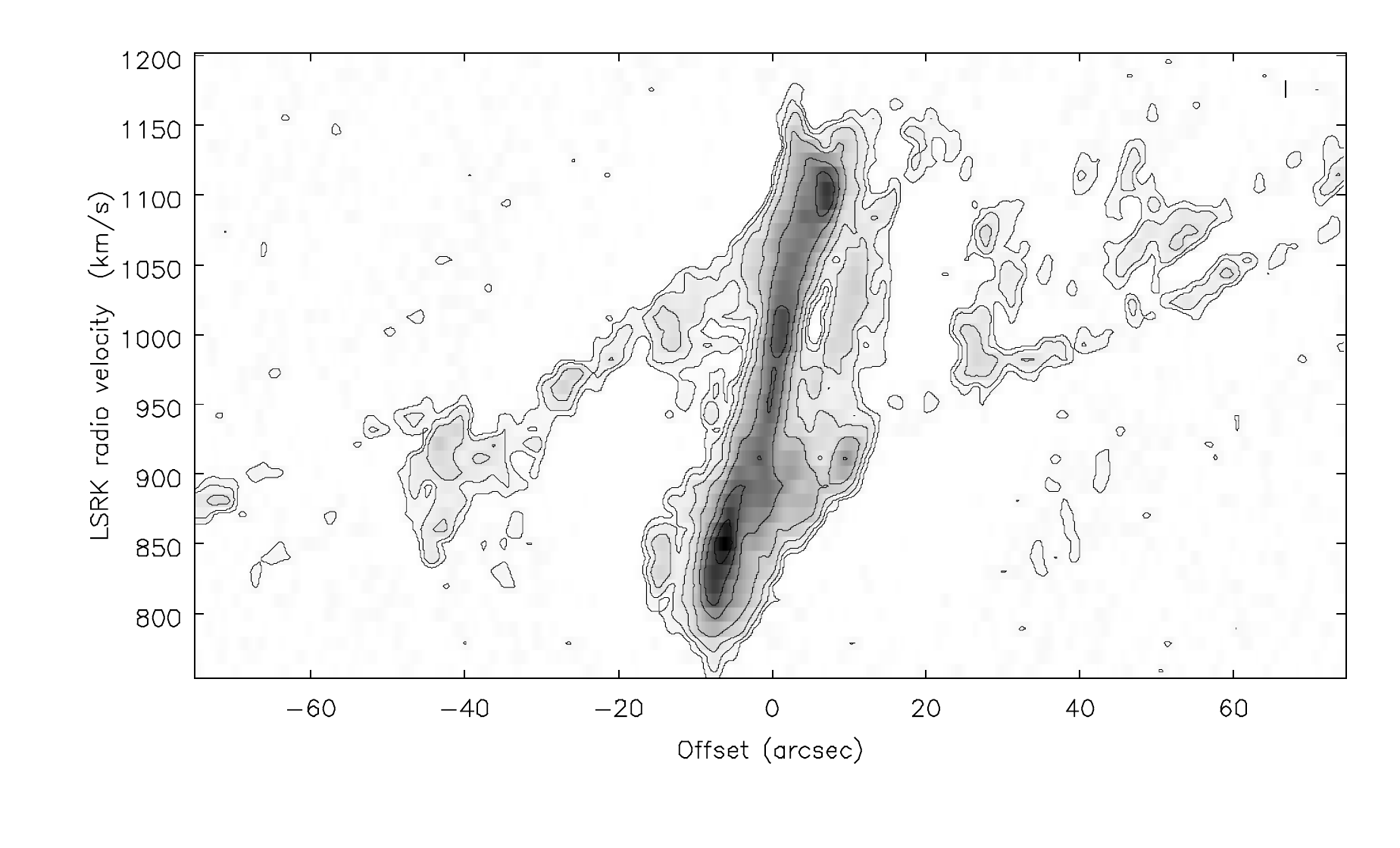}
\caption{Position-velocity diagram along the bar major axis \(PA_\mathrm{b}=335\arcdeg\) in the range \(-75\arcsec<X<75\arcsec\). The thickness of the slit is 99 pixels (\(25\arcsec\)) and the contours are plotted at 0.025, 0.05, 0.1, 0.2, 0.4, 0.6, and 0.8 times 127 mJy beam\(^{-1}\).}
\label{fig27}
\end{figure}

\begin{figure}
\epsscale{1}
\plotone{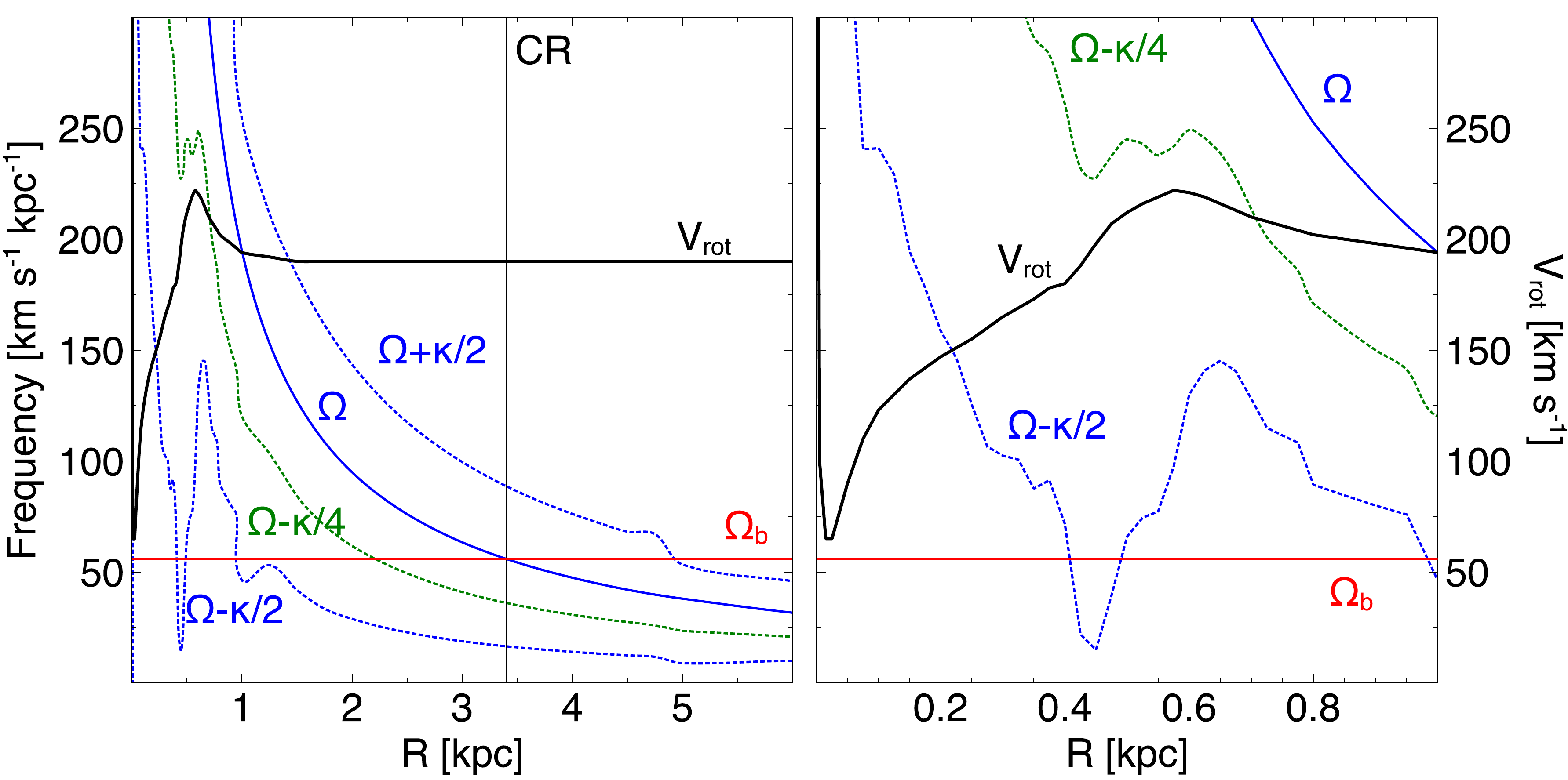}
\caption{Orbital resonances in the inner 6 kpc (left) and 1 kpc (right). The rotation curve \(v_\mathrm{rot}\) from figure \ref{fig20} is assumed constant at \(r>190\) km s\(^{-1}\). The contribution from a tentative supermassive black hole is included. The vertical line CR marks the corotation radius.}
\label{fig28}
\end{figure}

\begin{figure}
\epsscale{1}
\plotone{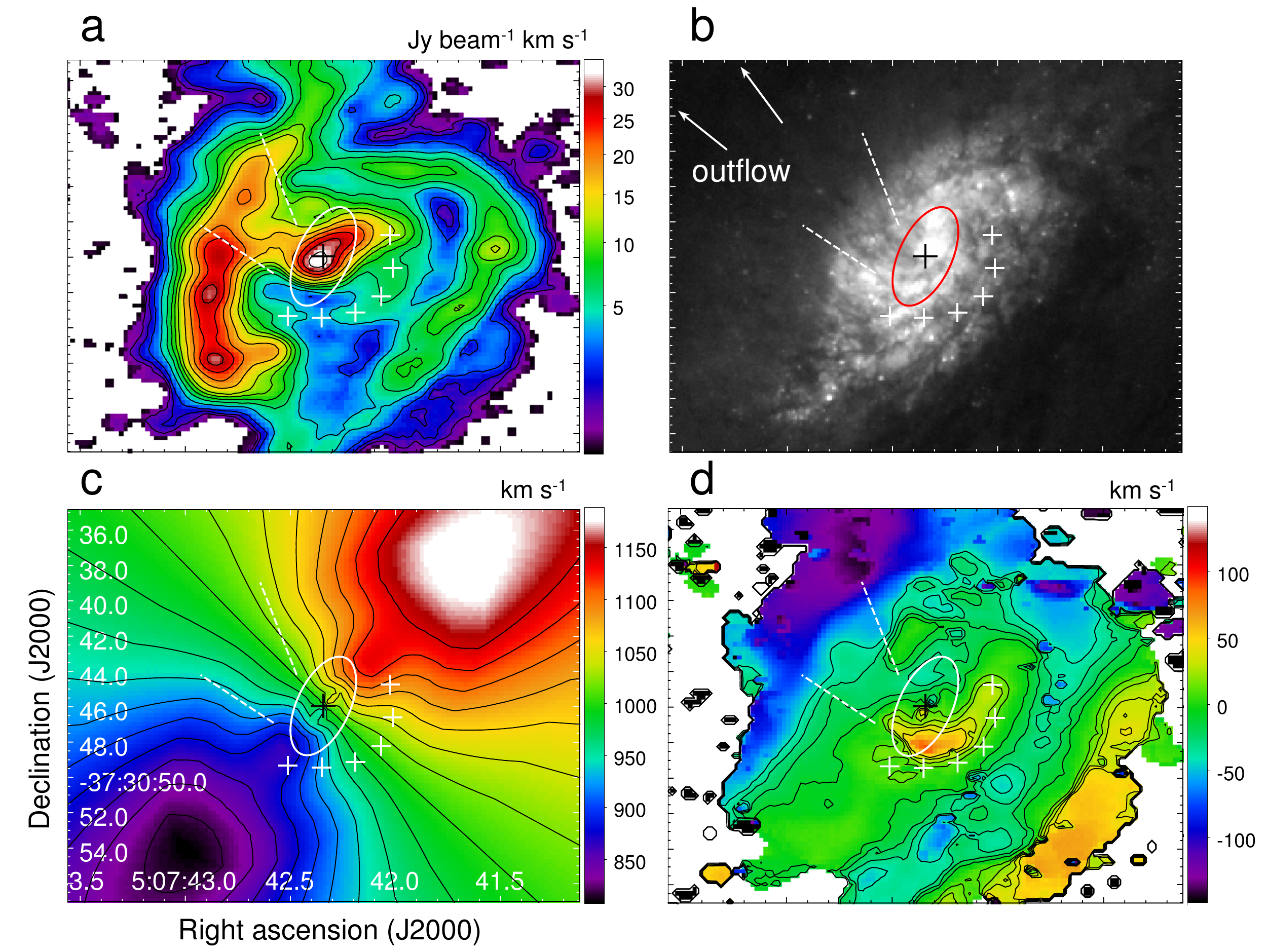}
\caption{a) CO (1-0) integrated intensity (same as figure \ref{fig10}, right); b) HST image (cf. figure \ref{fig1}, right); c) model velocity field with circular rotation generated by using \(v_\mathrm{sys}=998\) km s\(^{-1}\), \(PA=316\arcdeg\), and the rotation curve derived in section 4.1.4; d) residual velocity field calculated by subtracting the model from the data. The contours are plotted at -40, -30, -20, -10, 10, 20, 30, 40 km s\(^{-1}\). Note the non-circular motions as redshifted residual velocities up to \(\sim80\) km s\(^{-1}\) on the inner side of the nuclear spiral arm (indicated by white crosses to guide the eye) and the blueshifted emission north-east of the center tracing the nuclear outflow (white dashed lines marking the orientation of the extraplanar dust lanes). Also conspicious are the regions of streaming motion in the large-scale bar with residual velocities \(\sim(50\mathrm{-}100)\) km s\(^{-1}\) in the NE and SW. The galactic center and nuclear bar are marked by a black cross and ellipse, respectively.}
\label{fig29}
\end{figure}

In the nuclear region, the residual velocities of molecular gas, derived in the previous section and shown in figure \ref{fig29}, reveal the following patterns of non-circular motion (consistent with the PVD discussed in section 4.2.1). First, there is conspicuous streaming motion with magnitudes \(\sim50\) km s\(^{-1}\) on the inner edge of the nuclear spiral arm (\(2\arcsec\mathrm{-}3\arcsec\) south and south-west of the center). The magnitude of the streaming motion is comparable but larger than, e.g., in the grand-design spiral galaxy M51 by about a factor of two \citep{Mei13}. Second, north-east of the center, there is a blueshifted component spatially correlated with the extraplanar dust lane visible in the \(R\)-band (discussed in the next section). Finally, at radii of \(r>500\) pc in the NE and SW, the residual velocities abruptly reach \(\sim100\) km s\(^{-1}\); these features are related to the velocity gradient near the ILR, also apparent from the velocity field and dispersion images presented in section 3.3. A comparison with figure \ref{fig26} shows that the large residuals extend along the ridges in the bar region.

\clearpage

\section{Molecular gas outflow}

\subsection{Morphology and kinematics}

As shown in introduction, conspicuous dust lanes visible up to \(\sim3\) kpc above the galactic plane in optical images and the detection of Na\textsc{i} D in absorption and emission on opposite sides of the disk indicate the presence of a neutral gas outflow at velocities up to \(v_\mathrm{out}\sim380\) km s\(^{-1}\) \citep{Phi93}. Furthermore, \cite{McC13} found that the distribution of polycyclic aromatic hydrocarbons (PAHs) also extends into the halo to a similar distance, confirming the presence of a dusty wind. At the base of the outflow, recent integral-field observations revealed the presence of a cone of ionized gas \citep{SB10}. Since the central region of the galaxy is predominantly in molecular phase, one might expect that CO-emitting gas is entrained in the superwind as observed in galaxies like M82 and NGC 253 (e.g., \citealt{Sal13,Bol13a}). Is there any evidence of a CO outflow in NGC 1808?

\begin{figure}
\epsscale{1}
\plotone{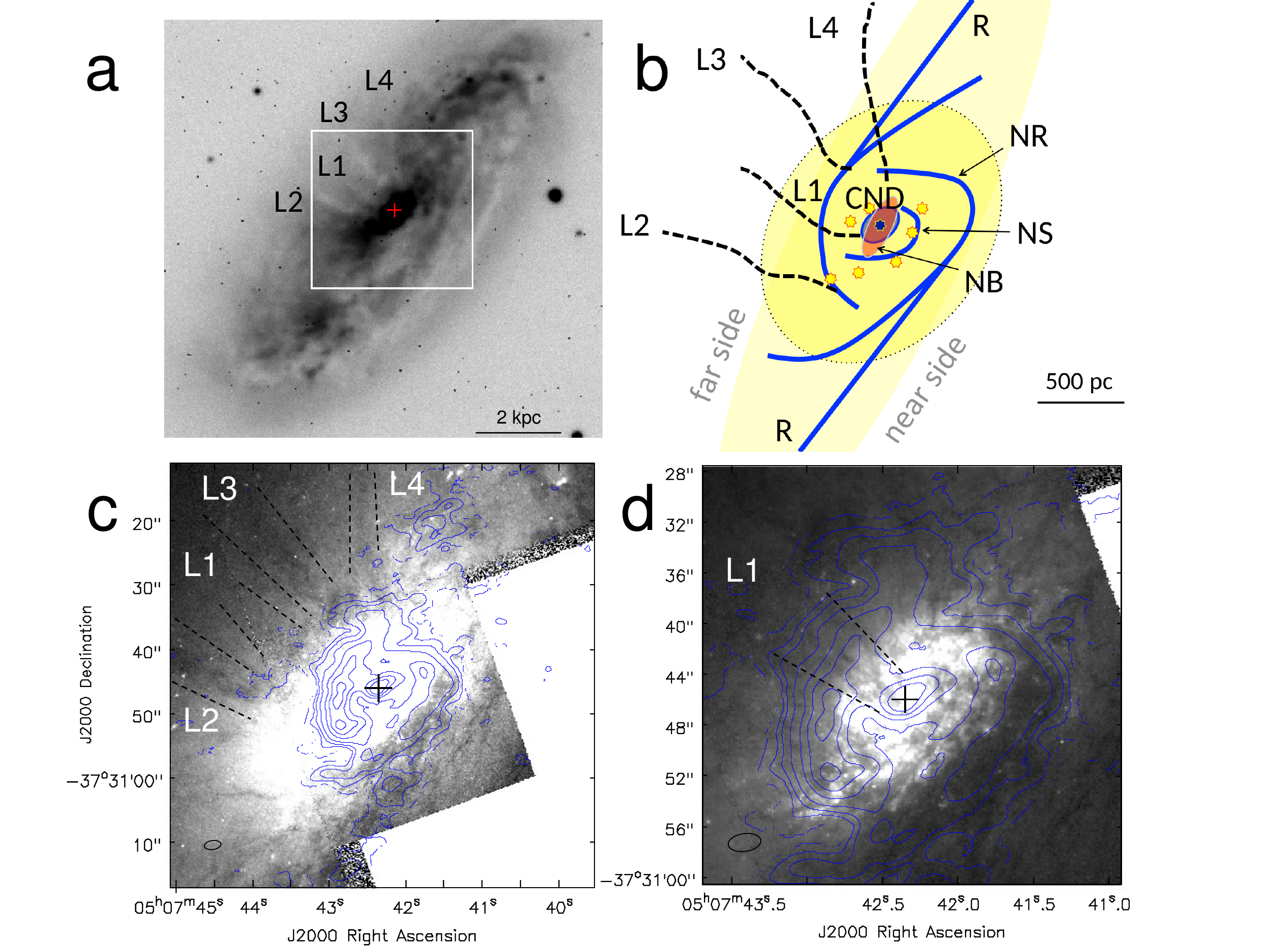}
\caption{Morphology of the outflow with respect to the galactic structure in NGC 1808. (a) Large-scale \(R\)-band image (same as figure \ref{fig1} with marked extraplanar dust lanes (L1-L4). (b) Illustration of the structure in the central 2 kpc. The main components are: bulge (dotted yellow filled ellipse), primary bar (elongated large-scale yellow filled ellipse), molecular gas ridges in the bar (R), star-forming regions and supernova remnants (stars), 500-pc ring (NR), nuclear spiral arm (NS), nuclear bar (red ellipse; NB), circumnuclear disk (blue ellipse; CND), and outflow of gas and dust (black dashed lines; L1-L4). (c) HST \(R\)-band image of the central region [rectangle field in (a); same scale as (b)]. The blue contours are CO (1-0) plotted at 0.01, 0.05, 0.1, 0.2, 0.4, 0.6, 0.8 times the peak of 41 Jy beam\(^{-1}\) km s\(^{-1}\). The optical image was enhanced to emphasize the dust lanes outside the central region. (d) Enlargement of the field in (c) to show the nuclear outflow L1. The cross marks the galactic center.}
\label{fig30}
\end{figure}

\begin{figure}
\epsscale{0.75}
\plotone{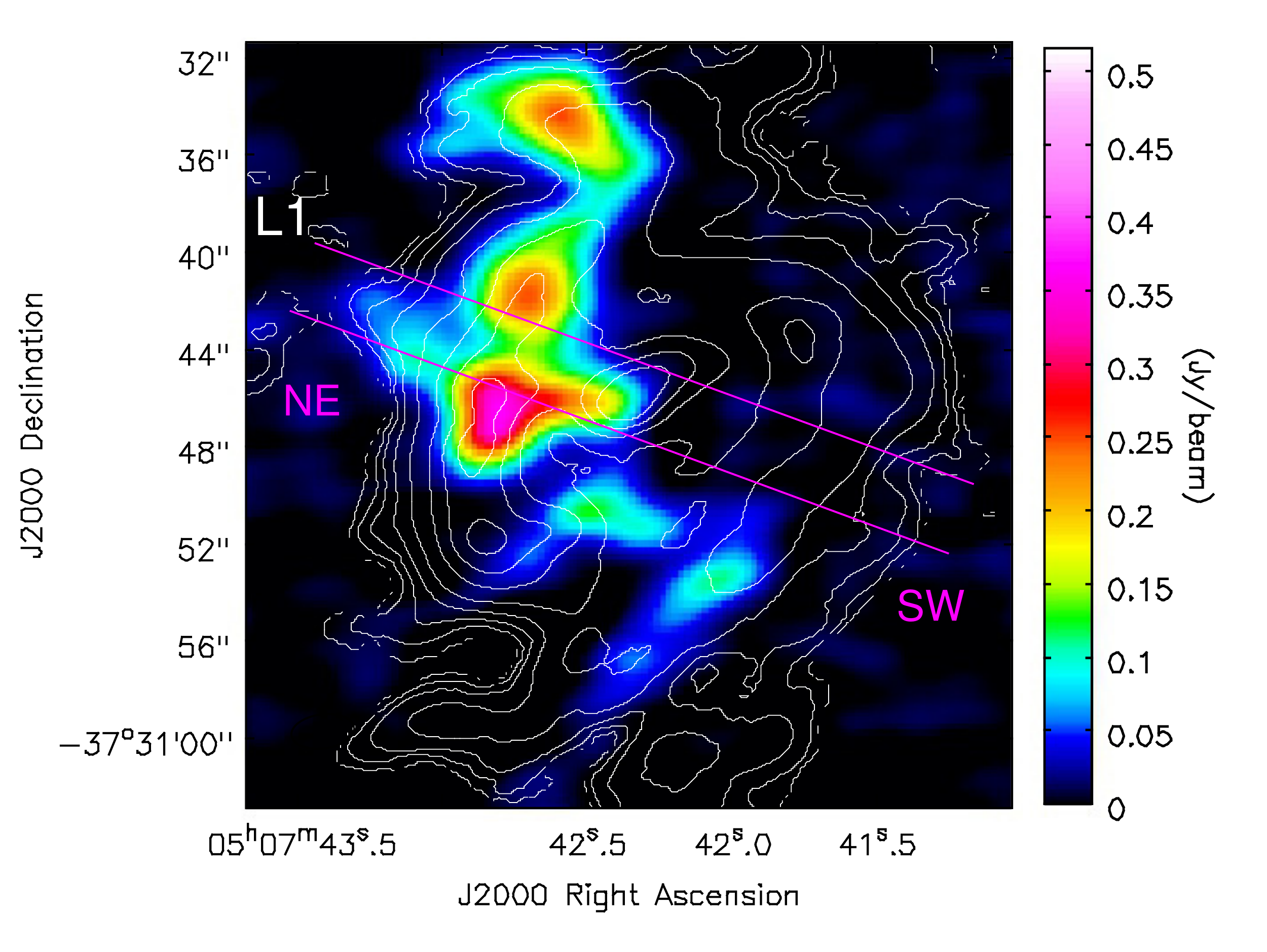}
\caption{Channel map of the high-sensitivity CO (1-0) data cube at 911 km s\(^{-1}\) (color) with contours of the integrated intensity plotted at 0.01, 0.05, 0.1, 0.2, 0.4, 0.6, 0.8 times 41 Jy beam\(^{-1}\) km s\(^{-1}\). The magenta lines mark the PVD slice width (3\farcs25) and orientation (70\arcdeg). The resulting PVD is shown in figure \ref{fig32}.}
\label{fig31}
\end{figure}

The optical image in figure \ref{fig29}b shows an extraplanar (polar) dust lane north-east of the nucleus, at a similar position angle as the kpc-scale dust lanes in figure \ref{fig30}a and the hot ionized cone revealed in the [N\textsc{ii}]/H\(\alpha\) ratio \citep{SB10}. The location of the outflow with respect to other galactic structures is marked in figure \ref{fig30}b. We also show a comparison of an HST image with the CO (1-0) intensity distribution (high-sensitivity image) in figure \ref{fig30} (panels c and d): there are four kpc-scale dust lanes denoted by L1-L4, all emerging from the central 1-kpc region. CO was not detected in the lanes L2-L4, except at their inner ends [base of the outflow; panel (c)]. The molecular gas is coincident, however, with an extraplanar dust lane L1 [panel (d)]. The L1 lane is blueshifted with respect to the systemic velocity (\(v_\mathrm{LSR}\approx900\) km s\(^{-1}\); see figure \ref{fig31} and the blue wing in the spectrum of the central \(5\arcsec\) region in figure \ref{fig17}) as would be expected from an outflow, though there is no unambiguous redshifted counterpart on the opposite SW side [the asymmetry is also apparent in the optical line in \cite{Phi93} and \cite{SB10}]. Figures \ref{fig30} and \ref{fig31} show that L1 is coincident with a spur of CO (1-0) emission that extends beyond the 500-pc ring to around \((\alpha,\delta)_\mathrm{J2000}=(5^\mathrm{h}7^\mathrm{m}43\fs2,-37\degr30\arcmin42\arcsec)\). Enhanced velocity dispersion (figure \ref{fig15}) and the velocity width as wide as 150 km s\(^{-1}\) along the minor axis north-east of the nucleus (figures \ref{fig24} and \ref{fig32} below) also indicate the presence of extraplanar gas. It is not clearly seen in the moment 1 image of figure \ref{fig15} (left) because the velocity field is intensity-weighted and the extraplanar emission is spatially coincident with the underlying circular-rotation in the disk that dominates the overall CO (1-0) emission. Assuming a nuclear outflow perpendicular to the galactic plane, the average deprojected outflow velocity becomes (equation \ref{obs})

\begin{equation}
v_z=\frac{|v_\mathrm{obs}-v_\mathrm{sys}|}{\cos{i}}\simeq180~\mathrm{km~s}^{-1},
\end{equation}
where \(i=57\arcdeg\), \(v_\mathrm{obs}=900\) km s\(^{-1}\), and \(v_\mathrm{sys}=998\) km s\(^{-1}\) (section 4.1). Note that if this gas were in the galactic plane, the observed velocity would yield an inflow of \(v_r=|v_\mathrm{obs}-v_\mathrm{sys}|/\sin{i}\simeq117\) km s\(^{-1}\) inside the \(r<500\) pc disk, which is comparable to the rotational velocity. The derived outflow velocity is consistent with typical values found in superwinds in galaxies like M82 and NGC 253 [100-200 km s\(^{-1}\); e.g., \cite{Nak87,WWS02,Sal13,Bol13a}]. Also, the detection of blueshifted emission NE from the center excludes the possibility of an extraplanar gas inflow. The opening angle of the outflow is uncertain, but the orientation of the dust lanes at the base of the outflow appears nearly perpendicular to the galactic major axis on optical images. The L4 lane reconnects with the galactic disk at a radius of \(r\sim3\) kpc, indicating the gas motion as a fountain - the ejected ISM material falling back onto the galaxy [see \cite{Phi93}]. The multitude of dust lanes is interesting because it suggests that extraplanar gas and dust may be confined to narrow regions as ``filaments'' in addition to the projected walls of a cylindric outflow as seen in, e.g., M82 \citep{Nak87,Sal13,Ler15}.

\begin{figure}
\epsscale{0.75}
\plotone{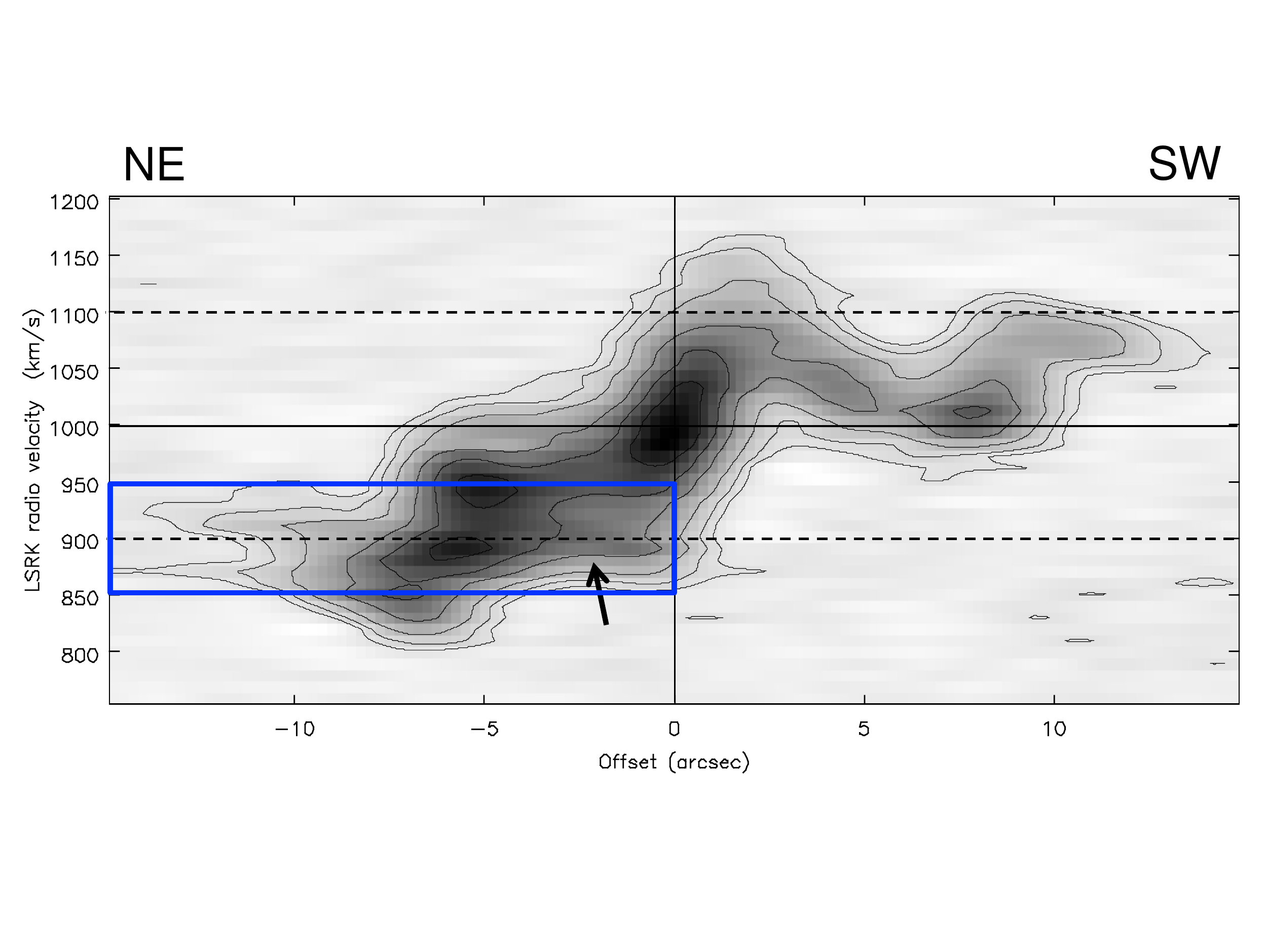}
\caption{PVD of the slice shown in figure \ref{fig31}. The outflow component, centered near 900 km s\(^{-1}\), is indicated with a black arrow. Note the double peak with separation 50 km s\(^{-1}\) near offset \(Y=-5\arcsec\) and line splitting at \(Y<-9\arcsec\). The contours are plotted at 0.05, 0.1, 0.2, 0.4, 0.6, 0.8 times the peak of 262 mJy beam\(^{-1}\).}
\label{fig32}
\end{figure}

\clearpage

\subsection{Mass and kinetic energy}

Infrared and radio observations show that the nucleus is ongoing a starburst episode with a star formation rate of \(SFR_\mathrm{core}\sim(0.5\mathrm{-}1)~M_\odot~\mathrm{yr}^{-1}\) \citep{Kra94,Kot96} (assuming that the energy output in the core is dominated by supernova explosions rather than AGN). As a consequence, the region is shocked by supernovae that generate non-thermal radiation and an ionization cone discussed above. Following \citet{CY90}, \citet{Kra94} estimated the supernova rate in the nucleus using the 5-GHz continuum as \(\mathcal{R}_\mathrm{SN}=0.024\) yr\(^{-1}\). The released energy during a starburst episode [\(t_\mathrm{SB}\gtrsim1\times10^7\) yr; \cite{Kra94} and \cite{Kot96}] is then\footnote{\cite{Kra94} estimated that the starburst has been going on for \(\gtrsim5\times10^7\) yr. We adopt a lower limit based on the mass of the nuclear cluster estimated in section 4.1.}

\begin{equation}
E_\mathrm{SN}\sim\mathcal{R}_\mathrm{SN}E_0 t_\mathrm{SB}\sim2\times10^{56}~\mathrm{erg},
\end{equation}
where \(E_0=10^{51}\) erg is the energy released by a single type-II supernova explosion. It is of importance to compare this energy with the kinetic energy of the molecular gas outflow. Below, we consider the outflow mass estimates based on two limiting cases.

\subsubsection{Optically thick emission}

Inspecting the channel maps (e.g., figure \ref{fig31}), we find that the outflow contribution to the CO (1-0) flux is no more than about 10\% in the central 1 kpc. If the mass of the molecular gas in the outflow is \(M_\mathrm{out}\sim0.1 M_\mathrm{mol}\sim6\times10^7~M_\sun\), where \(M_\mathrm{mol}\) is the total molecular gas mass in the central region (\(r<15\arcsec\); table \ref{tab5}), and \(v_z=180\) km s\(^{-1}\), the kinetic energy becomes \(E_\mathrm{out}\simeq M_\mathrm{out}v^2_z/2\sim2\times10^{55}\) erg. This is an upper limit because the CO (1-0) emission is assumed optically thick (the conversion factor \(X_\mathrm{CO}=0.8\times10^{20}~\mathrm{cm}^{-2}(\mathrm{K~km~s}^{-1})^{-1}\) was applied). Therefore, the derived energy released by supernova explosions is an order of magnitude higher than the upper limit of the molecular gas kinetic energy inferred from the measured CO (1-0) fluxes and velocities in the central region.

\subsubsection{Optically thin emission}

In the case of optically thin CO (1-0) emission, the total number of H\(_2\) molecules can be calculated as

\begin{equation}\label{col}
\mathcal{N}_{\mathrm{H}_2}=f\frac{3k}{4\pi^3\mu^2\nu} \frac{1}{J+1} \exp{\left[\frac{J(J+1)h\nu}{2kT_\mathrm{ex}}\right]} \left[1-\exp{\left(-\frac{(J+1)h\nu}{kT_\mathrm{ex}}\right)}\right]^{-1}I_\mathrm{CO}A.
\end{equation}
Here, \(f\equiv[\mathrm{H}_2]/[\mathrm{CO}]\) is the abundance ratio of H\(_2\) and CO molecules, \(h\) the Planck constant, \(k\) the Boltzmann constant, \(\mu\) the dipole moment, \(J\) the rotational level in a transition \(J+1\rightarrow J\), \(\nu\) the frequency of the line, \(T_\mathrm{ex}\) the excitation temperature, \(I_\mathrm{CO}\equiv\int T_\mathrm{R}dv\) the total integrated intensity (radiation temperature \(T_\mathrm{R}\) integrated over velocity) of the CO line, and \(A=\pi R^2\) the projected area (where \(R\) is the radius of the region). For a transition to the ground level, we have \(J=0\), \(\mu=0.11~\mathrm{Debye}\), and \(\nu=115.271\) GHz, and equation \ref{col} can be written as

\begin{equation}
\mathcal{N}_{\mathrm{H}_2}=2.31\times10^{14}f\left[1-\exp{\left(-\frac{5.53~\mathrm{K}}{T_\mathrm{ex}}\right)}\right]^{-1}\left(\frac{I_\mathrm{CO}}{\mathrm{K~km~s}^{-1}}\right)\left(\frac{A}{\mathrm{cm}^2}\right).
\end{equation}
We adopt an excitation temperature of \(T_\mathrm{ex}=31\) K \citep{Sal14}. The conversion between the brightness temperature and flux density is given by

\begin{equation}
\left(\frac{I_\mathrm{CO}}{\mathrm{K~km~s}^{-1}}\right)=92.5\left(\frac{\theta}{\mathrm{arcsec}}\right)^{-2}\left(\frac{\nu}{115~\mathrm{GHz}}\right)^{-2}\left(\frac{\int S_\mathrm{CO}dv}{\mathrm{Jy~km~s}^{-1}}\right),
\end{equation}
where \(\theta\) is the diameter of the region. For the adopted CO (1-0) flux in the outflow in the central region \(r<15\arcsec\), we get \(I_\mathrm{CO}=122.4\) Jy km s\(^{-1}=12.6\) K km s\(^{-1}\). Assuming an abundance ratio of \(f=10^4\), the total mass within \(r<R=15\arcsec\) becomes \(M_\mathrm{out}=1.41\times2m_\mathrm{H}\mathcal{N}_{\mathrm{H}_2}\sim8\times10^6~M_\sun\), where \(m_\mathrm{H}\) is the hydrogen atom mass and the factor 1.41 accounts for He and other elements. The kinetic energy of the nuclear outflow is then only \(E_\mathrm{out}\sim10^{54}\) erg.

\subsection{Dynamical evolution}

The molecular gas outflow traced by CO (spur in figure \ref{fig29}) is evident up to a deprojected height of \(z\sim1\) kpc. Assuming spherical symmetry, the lower limit of the escape velocity at this radius from the center is given by \(v_\mathrm{esc}=\sqrt{2|\Phi|}\), where \(\Phi=-GM/z\) is the gravitational potential (simplified as a point source). The total mass within this region is \(M(z<1~\mathrm{kpc})\sim1\times10^{10}~M_\sun\), hence \(v_\mathrm{esc}\sim300\) km s\(^{-1}\). The escape velocity is comparable but somewhat larger than the average outflow velocity, \(v_\mathrm{out}\lesssim v_\mathrm{esc}\), indicating that the molecular gas outflow may be trapped in the potential well of the galaxy. Note that \cite{Phi93} found a maximum outflow velocity of \(\sim380\) km s\(^{-1}\); the fastest gas clouds may be able to escape. For the adopted distance of \(z=1\) kpc and velocity \(v_\mathrm{out}=180\) km s\(^{-1}\), the dynamical age of the outflow is \(t_\mathrm{out}\sim z/v_\mathrm{out}\sim5\times10^6\) yr, consistent with the estimate by \cite{Phi93} based on Na\textsc{i} D spectra. This is shorter than the nuclear star-forming activity (\(t_\mathrm{SB}\gtrsim1\times10^7\) yr): the outflow is resupplied via supernova explosions during the starburst episodes.

If the masses and timescales are correct, the mass outflow rate from the central region of NGC 1808 is \(\dot{M}_\mathrm{out}\sim M_\mathrm{out}/t_\mathrm{out}\sim(1\mathrm{-}10)~M_\sun~\mathrm{yr}^{-1}\), comparable to the total star formation rate in the starburst 500-pc region [\(SFR\sim5~M_\sun~\mathrm{yr}^{-1}\) in all ``hot spots'' combined; \citep{Kra94}], yielding \(\dot{M}_\mathrm{out}/SFR\sim0.2\) for the optically thin case. Note that, in spite of violent star formation feedback, evidenced by numerous supernova remnants and polar dust lanes, the nucleus and the \(r<500\) pc disk (including the ring) are very abundant in molecular gas. In order to replenish the gas reservoir, a mass inflow rate comparable to \(\dot{M}_\mathrm{out}\) is required. This can be expected from the double bar and spiral arms. By comparison, the mass outflow rate in the nearby starburst galaxy NGC 253 is \(\dot{M}_\mathrm{out}\sim(3\mathrm{-}9)~M_\sun~\mathrm{yr}^{-1}\) and the ratio \(\dot{M}_\mathrm{out}/SFR\sim3\) \citep{Bol13a}. Similarly, the ratio in M82 is \(\dot{M}_\mathrm{out}/SFR\sim1\) \citep{WWS02,Sal13}: the quenching of star formation in NGC 1808 seems less efficient than in NGC 253 and M82.

It should be noted, however, that we cannot exclude the possibility that the outflow is too extended and diffuse to be sampled with the 12-m array without short-spacing correction, so the derivation of the outflow mass is largely uncertain. Combining 12-m array data with ACA (Atacama Compact Array) and total power data will clarify the spatial extent and energetics of the molecular gas outflow.

On the other hand, deep HST images of H\(\alpha\) (e.g., figure \ref{fig8}) show little sign of extraplanar ionized gas in polar direction as far as the kpc-scale dust lanes. Could the dust lanes further out in the halo be associated with atom-dominated gas, traced with H\textsc{i} and possibly [C\textsc{i}] or [C\textsc{ii}]? This question is related to the nature of superwinds in galaxies: what is their dominant ISM phase and what mechanisms govern phase transitions as the wind evolves?

\section{Star formation and disk stability in the central 1 kpc}

\begin{figure}
\epsscale{0.75}
\plotone{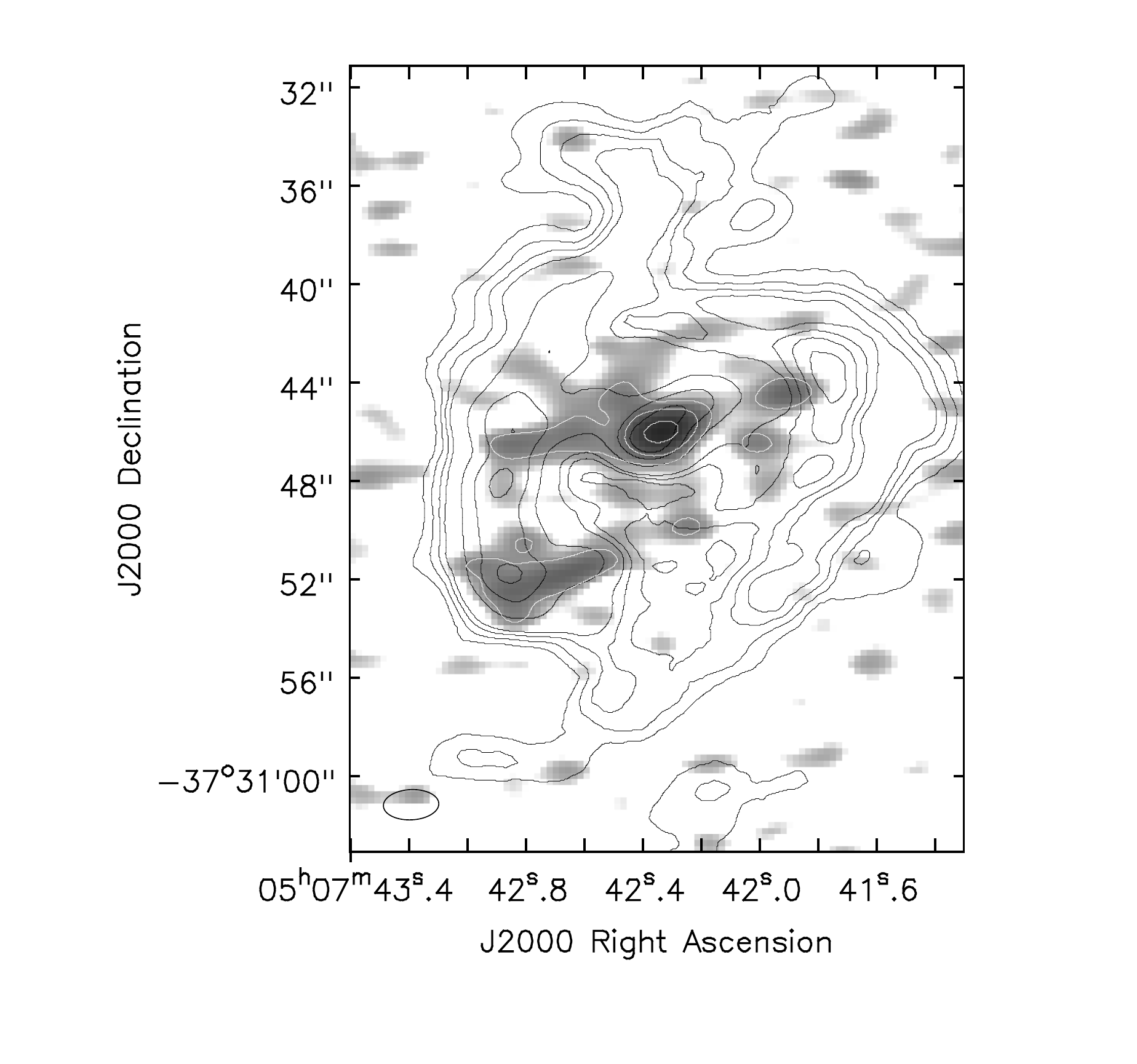}
\caption{CO (1-0) integrated intensity contours (0.05, 0.1, 0.15, 0.2, 0.4, 0.6, 0.8 times the peak of 34 Jy beam\(^{-1}\) km s\(^{-1}\)) superimposed on the 2.8-mm continuum image (grey scale and white contours at 0.1, 0.2, 0.4 times the peak of 2.6 mJy beam\(^{-1}\)).}
\label{fig33}
\end{figure}

In section 3.1 we presented 2.8-mm continuum data that reveal the presence of circumnuclear compact sources dominated by thermal emission. Note that the sources C1, C2, and C3 (figure \ref{fig4} and table \ref{tab4}) are located along the nuclear spiral arm on the inside of the 500-pc ring detected in CO (1-0) (figure \ref{fig33}). Sources C1, C3, and C4 are spatially correlated with the mid-infrared sources denoted by M3, M2, M6-M8 and interpreted as young star clusters and H\textsc{ii} regions embedded in dusty clouds \citep{GA08}. The comparison of the continuum and CO distributions (figure \ref{fig33}) also reveals that, apart from the SE region, there is no detected 2.8-mm continuum emission associated with the molecular 500-pc ring; star-forming activity and its feedback are more prominent in the inner \(r<400\) pc disk [see also, e.g., \cite{Sai90} for the distribution of supernova remnants that are also within the inner disk].

\begin{figure}
\epsscale{0.75}
\plotone{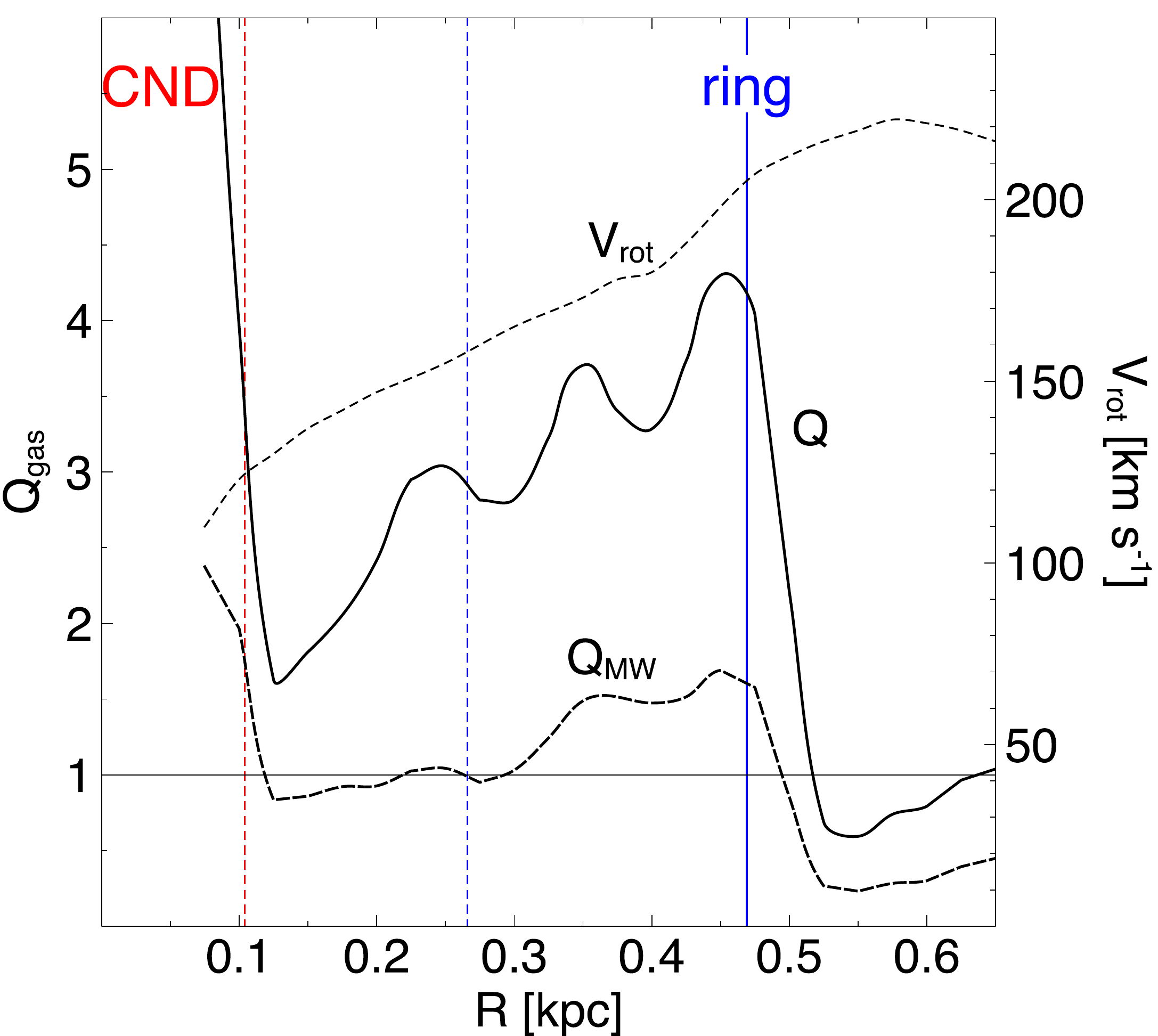}
\caption{Stability parameter \(Q_\mathrm{gas}\) for the gas disk in the central 0.6 kpc region derived by using two different CO-to-H\(_2\) conversion factors: the Galactic value \(X_\mathrm{MW}=2\times10^{20}~\mathrm{cm}^{-2}~(\mathrm{K~km~s}^{-1})^{-1}\) and \(X=X_\mathrm{MW}/2.25\). The vertical lines mark the 500-pc ring (blue) and the CND (red). The dashed vertical lines are the standard deviations of the two regions (section 3.2.2).}
\label{fig34}
\end{figure}

One of the long-standing problems in galactic dynamics is to understand the mechanism of transport of molecular gas from the nuclear ring near the inner Lindblad resonances, such as the 500-pc ring in NGC 1808, to the very center of the galaxy (\(<1\) pc). The main processes involved are nuclear bars and spiral arms. Both structures generate non-axisymmetric gravitational potential that exerts torque on the outer region (e.g., \citealt{Sch84,BC96,Com14}). Angular momentum is transferred outward, while the material decays to lower orbits. In NGC 1808, the presence of a nuclear bar and a spiral arm is indicative of gas inflow from the 500-pc ring. In order to clarify its relation with star formation unveiled with radio continuum and infrared observations, we investigated the gravitational stability of the disk region between the CND and the ring. The stability criterion can be expressed as a \(Q\) parameter for fluids \citep{Saf60,Too64},

\begin{equation}\label{q}
Q_\mathrm{gas}\equiv\frac{\alpha v_\mathrm{s}\kappa}{\pi G\Sigma_\mathrm{gas}},
\end{equation}
where \(\alpha\) is a constant (assumed equal to unity here), \(v_\mathrm{s}\) is the speed of sound related to the velocity dispersion of the ISM gas, \(\sigma_\mathrm{mol}\). The stability increases with velocity dispersion and epicyclic frequency \(\kappa\) (related to differential rotation), and decreases with gas surface density. When \(Q_\mathrm{gas}\lesssim1\), the gaseous disk is unstable and likely to increase star forming activity. The resulting \(Q_\mathrm{gas}\), calculated with \(v_\mathrm{s}=\sigma_\mathrm{mol}\), is shown in figure \ref{fig34}; the uncertainty of the derived values is high due to the variation in the CO-to-H\(_2\) conversion factor \(X_\mathrm{CO}\), velocity dispersion, surface density, and rotation curve. Nevertheless, the result shows that the disk stability is somewhat low (although \(Q_\mathrm{gas}\gtrsim1\)) between 100 and 300 pc where the rotation curve exhibits a gentle slope, qualitatively consistent with the observational results above. The sudden drop beyond \(r>500\) pc (beyond the ring) should be taken with reserve as the signal-to-noise ratio is low in that region.

This result is different from the one found, e.g., in the Seyfert galaxy NGC 7469, where the instability is highest in the resonant ring (\(Q_\mathrm{gas}\lesssim1\) and active star formation), while the gaseous disk is gradually more stable toward the galactic center \citep{Fat15}. The high \(Q_\mathrm{gas}\) in the central 500 pc of NGC 1808 is comparable to the results obtained for a sample of nearby spiral galaxies including NGC 6946 \citep{RF13,RF15}. Contrary to the inner \(r<400\) pc disk, star formation activity in NGC 1808 appears less prominent outside the resonant 500-pc ring due to a high velocity gradient imposed by bar dynamics (sections 3.3 and 4.2.2).

\clearpage

\section{Summary}

We have presented the first high-resolution ALMA observations of CO (1-0) and 2.8 mm continuum in the barred starburst galaxy NGC 1808. Below is a summary of the main findings.

\begin{enumerate}
\item{\emph{Continuum emission.} Radio continuum at 2.8 mm (synthesized over four bands at 101, 103, 113, and 115 GHz) was detected at the nucleus and circumnuclear regions within \(r<400\) pc. The spectral index (\(\alpha\)) image confirms that the nucleus is dominated by non-thermal (synchrotron) emission with \(\alpha\simeq-1\).}

\item{\emph{Molecular gas distribution.} High-resolution CO (1-0) images revealed four distinct components of molecular gas: (1) compact circumnuclear disk (CND) within \(r<200\) pc, (2) 500-pc ring, (3) gas-rich galactic bar with semi-major axis \(a_\mathrm{b}\simeq3\) kpc, and (4) global (kpc-scale) and nuclear (\(r<400\) pc) spiral structure. In the inner \(r<400\) pc disk, molecular gas is distributed in a spiral pattern.}

\item{\emph{Molecular gas kinematics.} Basic geometric and kinematic parameters (position angle, inclination, rotation curve, and velocity dispersion) were derived for the central 1 kpc region using the \(^\mathrm{3D}\)Barolo code. Two systemic velocities were found (998 km s\(^{-1}\) for the CND and 964 km s\(^{-1}\) for the 500-pc ring), indicating a kinematic offset. Streaming motion with magnitude \(\sim50\) km s\(^{-1}\) is revealed on the inner side of a nuclear spiral arm. On a large scale, molecular gas clouds exhibit conspicuous non-circular motions dominated by the primary bar, with a velocity gradient (shear) of \(v_\mathrm{sh}\simeq(0.2\mathrm{-}1)\) km s\(^{-1}\) pc\(^{-1}\) across the molecular gas ridges on the leading side of the bar.}

\item{\emph{Rotation curve decomposition and dynamical mass.} The rotation curve derived from the CO (1-0) emission was deconvolved into a stellar bulge (mass \(M_\mathrm{B1}\sim1.25\times10^{10}~M_\odot\)), a major core that comprises a nuclear bar and molecular medium (\(M_\mathrm{B2}\sim7.2\times10^8~M_\sun\)), and an unresolved minor core, possibly a nuclear star cluster or black hole with mass \(M_\mathrm{core}\sim1\times10^7~M_\odot\) surrounded by an embedded molecular gas disk with a radius \(r<50\) pc.}

\item{\emph{Bar dynamics and orbital resonances.} The bar pattern speed of the large-scale bar was derived as \(\Omega_\mathrm{b}=56\pm11\) km s\(^{-1}\) kpc\(^{-1}\) from the position-velocity diagram assuming a cloud-orbit model. For a flat rotation curve of 190 km s\(^{-1}\) at \(r>1.5\) kpc, the pattern speed yields a corotation radius of \(r_\mathrm{CR}\simeq3.4\) kpc and the ratio \(r_\mathrm{CR}/a_\mathrm{b}\simeq1.1\), consistent with theoretical predictions for weak bars. An inner Lindblad resonance is located at \(\sim500\) pc, coincident with the location of the molecular gas ring.}

\item{\emph{Molecular gas outflow.} We presented evidence of a molecular gas outflow from the nuclear starburst region (\(r<250\) pc). The outflow has a maximum velocity of \(v_\mathrm{out}\sim180\) km s\(^{-1}\) and a total kinetic energy two orders of magnitude smaller than the energy released by supernova explosions in the nucleus.}

\item{\emph{Gas disk stability and star formation.} In the region \(200~\mathrm{pc}<r<400~\mathrm{pc}\), the stability parameter of the gas disk, \(Q_\mathrm{gas}\), is lower than average, allowing enhanced star formation activity, consistent with observations of circumnuclear ``hot spots''.}
\end{enumerate}


\acknowledgments

The authors thank an anonymous referee for careful reading and many useful suggestions. We thank the graduate students Y. Yamada, S. Hisamatsu, and I. Tanaka for assistance during the observations with ASTE, and professor M. Seta for support in data reduction. D.S. is grateful to S. Imai for heartwarming discussions and to E. Galliano for providing the VLT \(K_s\) image.
This paper makes use of the following ALMA data:
ADS/JAO.ALMA\#2012.1.01004.S. ALMA is a partnership of ESO (representing
its member states), NSF (USA) and NINS (Japan), together with NRC
(Canada) and NSC and ASIAA (Taiwan), in cooperation with the Republic of
Chile. The Joint ALMA Observatory is operated by ESO, AUI/NRAO and NAOJ. Based on observations made with the NASA/ESA Hubble Space Telescope, and obtained from the Hubble Legacy Archive, which is a collaboration between the Space Telescope Science Institute (STScI/NASA), the Space Telescope European Coordinating Facility (ST-ECF/ESA) and the Canadian Astronomy Data Centre (CADC/NRC/CSA). This research has made use of the NASA/IPAC Extragalactic Database (NED) which is operated by the Jet Propulsion Laboratory, California Institute of Technology, under contract with the National Aeronautics and Space Administration.

\appendix

\section{HCN (4-3) observations with ASTE}

Single-dish observations with Atacama Submillimeter Telescope Experiment (ASTE)\footnote{The ASTE telescope is operated by National Astronomical Observatory of Japan (NAOJ).} were conducted in 2014 September and October with single pointing toward the galactic center (table \ref{tab8}). The frontend receiver CATS 345 was tuned to the HCN (\(J=4\rightarrow3\)) line at the rest frequency of 354.505473 GHz in the upper side band. The signal was down-converted to an intermediate frequency and resolved with an XF-type autocorrelation spectrometer with a total bandwidth of 512 MHz (443 km s\(^{-1}\)) and fine spectral resolution of 0.5 MHz. The bandwidth was wide enough to cover the velocity range of the H\textsc{i} and CO (3-2) lines \citep{Kor93,Sal14}, hence we assumed that the HCN (4-3) line does not exhibit a wider profile. Intensity calibration was performed by the standard chopper wheel method that yielded the antenna temperature \(T_\mathrm{A}^*\) corrected for atmospheric and ohmic losses \citep{UH76}. Each time the receiver was tuned, an intensity calibrator (Orion KL) was observed to measure a reference spectrum at the same frequency. An appropriate scaling factor was then applied to correct for intensity fluctuations and to convert \(T_\mathrm{A}^*\) to the main beam temperature, \(T_\mathrm{mb}\equiv T_\mathrm{A}^*/\eta_\mathrm{mb}\), where \(\eta_\mathrm{mb}\) is the main beam efficiency. The derived scaling factor of 1.6 yielded \(\eta_\mathrm{mb}\simeq0.6\), consistent with the telescope data provided online. Telescope pointing was checked after every observing run of about one hour and the relative offset was typically \(<3\arcsec\).

Data reduction was done by using the NEWSTAR tool developed by Nobeyama Radio Observatory of the National Astronomical Observatory of Japan. Raw data were converted to spectra, flagged, baseline-subtracted (first-order polynomial), and smoothed to a velocity resolution of 5.5 km s\(^{-1}\). The basic properties of the line are given in table \ref{tab8}. The uncertainty of the integrated intensity is estimated by \(\Delta I=\sqrt{\Delta I_\mathrm{rms}^2+\Delta I_\mathrm{b}^2}\), where \(\Delta I_\mathrm{rms}=\Delta T_\mathrm{rms}\sqrt{\Delta v_\mathrm{ch}v_\mathrm{int}}=0.10\) K km s\(^{-1}\) is due to the r.m.s. noise and \(\Delta I_\mathrm{b}=\Delta T_\mathrm{rms}\Delta v_\mathrm{int}\sqrt{\Delta v_\mathrm{ch}/\Delta v_\mathrm{b}}=0.18\) K km s\(^{-1}\) is the contribution from the baseline fitting. Here, \(\Delta v_\mathrm{int}=320\) km s\(^{-1}\) is the velocity width of the integrated spectrum, \(\Delta v_\mathrm{ch}=5.5\) km s\(^{-1}\) the channel width, and \(\Delta v_\mathrm{b}=80\) km s\(^{-1}\) the total range of the baseline fitting.

\begin{table}
\begin{center}
\caption{Observational summary.}\label{tab8}
\begin{tabular}{ll}
\tableline\tableline
ASTE 10 m & \\
\tableline
    Observation date & 2014 Sep. 16-17, Oct. 1-2  \\
    Intensity calibrator & Orion KL \\
    Pointing calibrator & R Dor \\
    Receiver & CATS 345 \\
    Spectrometer bandwidth & 512 MHz (445 km s\(^{-1}\)) \\
    Velocity resolution\tablenotemark{a} & 0.42 km s\(^{-1}\) \\
    Beam FWHM & \(22\arcsec\) (1.1 kpc) \\
    Integration time & 2.6 hr \\
    1 \(\sigma\) sensitivity in \(T_\mathrm{mb}\) & 2.5 mK \\
\tableline
\end{tabular}
\tablenotetext{a}{Smoothed to 5.5 km s\(^{-1}\).}
\end{center}
\end{table}

\subsection{HCN (4-3) emission in the galactic nucleus}

\begin{figure}
\epsscale{0.8}
\plotone{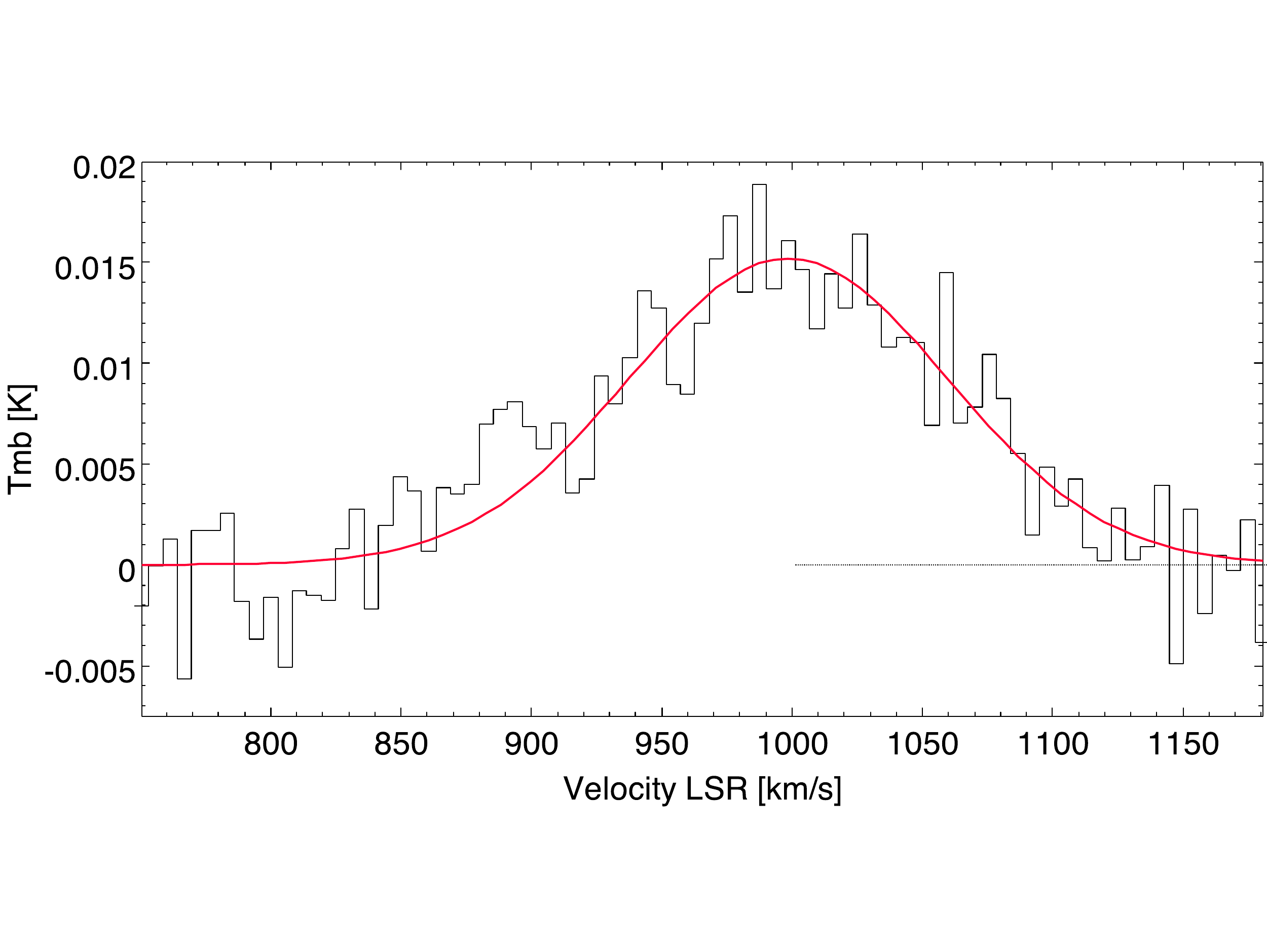}
\caption{HCN (4-3) spectrum toward the galactic center.}
\label{fig35}
\end{figure}

In figure \ref{fig35} we show the spectrum of HCN (4-3) toward the galactic center at \((\alpha,\delta)_\mathrm{J2000.0}=(5^\mathrm{h}7^\mathrm{m}42\fs06,-37\degr30\arcmin44\farcs42)\) measured using ASTE whose beam size was \(\mathrm{FHWM}=22\arcsec\) (1.1 kpc). The profile is notably different from the triple peak seen in CO (1-0) [as well as in CO (3-2) and \(^{13}\)CO (3-2); \cite{Sal14}] in the inner 1 kpc and much more similar to the CO profile within the central \(r<250\) pc (figure \ref{fig17}). The velocity of 998 km s\(^{-1}\) at the emission peak calculated with Gaussian fitting (table \ref{tab9}) also resembles the one derived from CO at the center. This result is not surprising if HCN (4-3) emission originates almost entirely in the CND and much less in the 500-pc ring. Given the expected effective critical density\footnote{The critical density is defined as the ratio of the Einstein coefficient for spontaneous decay between an upper and a lower energy level (u and l), \(A_\mathrm{ul}\), and the collisional cross section \(\sigma\). The effective critical density includes the effects of radiative trapping described by the photon escape probability \(\beta\), and can be defined as \(n_\mathrm{cr}\equiv\beta A_\mathrm{ul}/\sigma\).} of the order of \(n_\mathrm{cr}>10^4\) cm\(^{-3}\) for HCN (4-3) at any temperature between 10 and a few hundred K (e.g., \citealt{Shi15}), the line is known to trace extremely dense molecular medium commonly observed in galactic nuclei.

\begin{table}
\begin{center}
\caption{HCN (4-3) line parameters (Gaussian fit).}\label{tab9}
\begin{tabular}{lrr}
\tableline\tableline
Parameter & This work & \citet{Zha14} \\
\tableline
Peak \(T_\mathrm{mb}\) [mK] & \(15.2\pm2.8\)  & \\
Line FWHM [km s\(^{-1}\)] & \(144.2\pm8.7\) & \\
Peak velocity [km s\(^{-1}\)] & \(998.6\pm3.3\) & \\
Integrated intensity [K km s\(^{-1}\)] & \(2.3\pm0.2\) & \(2.5\pm0.2\) \\
\tableline
\end{tabular}
\end{center}
\end{table}

\clearpage


%



\begin{thebibliography}{}
\bibitem[Aalto et al.(1994)]{Aal94} Aalto, S., Booth, R. S., Black, J. H., Koribalski, B., \& Wielebinski, R. 1994, \aap, 286, 365
\bibitem[Aalto et al.(1999)]{Aal99} Aalto, S., H\"{u}ttemeister, S., Scoville, N. Z., \& Thaddeus, P. 1999, \apj, 522, 165
\bibitem[Athanassoula(1992a)]{Ath92a} Athanassoula, E. 1992, \mnras, 259, 328
\bibitem[Athanassoula(1992b)]{Ath92b} Athanassoula, E. 1992, \mnras, 259, 345
\bibitem[Awaki et al.(1996)]{Awa96} Awaki, H., Ueno, S., Koyama, K., Tsuru, T., \& Iwasawa, K. 1996, \pasj, 48, 409
\bibitem[Binney et al.(1991)]{Bin91} Binney, J., Gerhard, O. E., Stark, A. A., Bally, J., \& Uchida, K. I. 1991, \mnras, 252, 210
\bibitem[Binney \& Tremaine(2008)]{BT08} Binney, J., \& Tremaine, S. 2008, Galactic Dynamics, 2nd ed. (New Jersey: Princeton University Press), 528-539
\bibitem[Bolatto et al.(2013a)]{Bol13a} Bolatto, A., Warren, S. R., Leroy, A. K., et al. 2013, \nat, 499, 450
\bibitem[Bolatto et al.(2013b)]{Bol13b} Bolatto, A., Wolfire, M., and Leroy, A. K. 2013, \araa, 51, 207
\bibitem[Burbidge \& Burbidge(1968)]{BB68} Burbidge, E. M. \& Burbidge, G. R. 1968, \apj, 151, 99
\bibitem[Buta \& Combes(1996)]{BC96} Buta, R., \& Combes, F. 1996, Fund. Cosm. Phys., 17, 95
\bibitem[Chevalier \& Clegg(1985)]{CC85} Chevalier, R. A. \& Clegg, A. W. 1985, \nat, 317, 44
\bibitem[Collison et al.(1994)]{Col94} Collison, P. M., Saikia, D. J., Pedlar, A., Axon, D. J., and Unger, S. W. 1994, \mnras, 268, 203
\bibitem[Combes et al.(2013)]{Com13} Combes, F., Garc\'{i}a-Burillo, S., Casasola, V., et al. 2013, \aap, 558, A124
\bibitem[Combes et al.(2014)]{Com14} Combes, F., Garc\'{i}a-Burillo, S., Casasola, V., et al. 2014, \aap, 565, A97
\bibitem[Condon(1987)]{Con87} Condon, J. J. 1987, \apjs, 65, 485
\bibitem[Condon \& Yin(1990)]{CY90} Condon, J. J., \& Yin, Q. F. 1990, \apj, 357, 97
\bibitem[Condon(1992)]{Con92} Condon, J. J. 1992, \araa, 30, 575
\bibitem[Contopoulos(1980)]{Con80} Contopoulos, G. 1980, \aap, 81, 198
\bibitem[Dahlem et al.(1990)]{Dah90} Dahlem, M., Aalto, S., Klein, U., et al. 1990, \aap, 240, 237
\bibitem[Davis et al.(2013)]{Dav13} Davis, T. A., Bureau, M., Cappellari, M., Sarzi, M., \& Blitz, L. 2013, \nat, 494, 328
\bibitem[Debattista \& Shen(2007)]{DS07} Debattista, V. P. \& Shen, J. 2007, \apjl, 654, L127
\bibitem[de Vaucouleurs et al.(1991)]{deV91} de Vaucouleurs, G. de Vaucouleurs, A., Corwin, JR., H. G., et al. 1991, Third Reference Catalogue of Bright Galaxies, ver. 3.9
\bibitem[Di Teodoro \& Fraternali(2015)]{dTF15} Di Teodoro, E. M. \& Fraternali, F. 2015, \mnras, 451, 3021
\bibitem[Downes et al.(1996)]{Dow96} Downes, D., Reynaud, D., Solomon, P. M., \& Radford, S. J. E. 1996, \apj, 461, 186
\bibitem[Fathi et al.(2015)]{Fat15} Fathi, K., Izumi, T., Romeo, A. B., et al. 2015, \apjl, 806, L34
\bibitem[Forbes et al.(1992)]{For92} Forbes, D. A., Boisson, C., \& Ward, M. J. 1992, \mnras, 259, 293
\bibitem[Fujimoto et al.(2014)]{Fuj14} Fujimoto, Y., Tasker, E. J., Wakayama, M., \& Habe, A. 2014, \mnras, 439, 936
\bibitem[Galliano et al.(2005)]{Gal05} Galliano, E., Alloin, D., Pantin, E., Lagage, P. O., and Marco, O. 2005, \aap, 438, 803
\bibitem[Galliano \& Alloin(2008)]{GA08} Galliano, E. \& Alloin, D. 2008, \aap, 487, 519
\bibitem[Garc\'{i}a-Burillo et al.(2005)]{GB05} Garc\'{i}a-Burillo, S., Combes, F., Schinnerer, E., Boone, F., \& Hunt, L. K. 2005, \aap, 441, 1011
\bibitem[Garc\'{i}a-Burillo et al.(2014)]{GB14} Garc\'{i}a-Burillo, S., Combes, F., Usero, A., et al. 2014, \aap, 567, A125
\bibitem[Garc\'{i}a-Lorenzo et al.(1997)]{GL97} Garc\'{i}a-Lorenzo, B., Mediavilla, E., Arribas, S., \& del Burgo, C. 1997, \apj, 483, L99
\bibitem[Gillessen et al.(2009)]{Gil09} Gillessen, S., Eisenhauer, F., Trippe, S., et al. 2009, \apj, 692, 1075
\bibitem[Heike \& Awaki(2007)]{HA07} Heike, K. \& Awaki, H. 2007, \pasj, 59, 531
\bibitem[Heller et al.(2007)]{HSA07} Heller, C. H., Shlosman, I., \& Athanassoula, E. 2007, \apjl, 657, L65
\bibitem[Hirota et al.(2014)]{Hir14} Hirota, A., Kuno, N., Baba, J., et al. 2014, \pasj, 66, 46
\bibitem[Jarrett et al.(2003)]{Jar03} Jarrett, T. H., Chester, T., Cutri, R., Schneider, S. E., \& Huchra, J. P. 2003, \aj, 125, 525
\bibitem[Jim\'{e}nez-Bail\'{o}n et al.(2005)]{Jim05} Jim\'{e}nez-Bail\'{o}n, E., Santos-Lle\'{o}, M., Dahlem, M., et al. 2005, \aap, 442, 861
\bibitem[Kennicutt(1998)]{Ken98} Kennicutt, R. C., Jr. 1998, \araa, 36, 189
\bibitem[Koda et al.(2002)]{Kod02} Koda, J., Sofue, Y., Kohno, K., et al. 2002, \apj, 573, 105
\bibitem[Koda \& Sofue(2006)]{KS06} Koda, J. \& Sofue, Y. 2006, \pasj, 58, 299
\bibitem[Koribalski et al.(1993)]{Kor93} Koribalski, B., Dahlem, M., Mebold, U., \& Brinks, E. 1993, \aap, 268, 14
\bibitem[Koribalski et al.(1996)]{Kor96} Koribalski, B., Dettmar, R.-J., Mebold, U., \& Wielebinski, R. 1996, \aap, 315, 71
\bibitem[Kormendy \& Ho(2013)]{KH13} Kormendy, J. \& Ho, L. C. 2013, \araa, 51, 511
\bibitem[Kotilainen et al.(1996)]{Kot96} Kotilainen, J. K., Forbes, D. A., Moorwood, A. F. M., van der Werf, P. P., \& Ward, M.J. 1996, \aap, 313, 771
\bibitem[Krabbe et al.(1994)]{Kra94} Krabbe, A., Sternberg, A., \& Genzel, R. 1994, \apj, 425, 72
\bibitem[Krips et al.(2011)]{Kri11} Krips, M., Mart\'{i}n, S., Eckart, A., et al. 2011, \apj, 736, 37
\bibitem[Kuno et al.(2000)]{Kun00} Kuno, N., Nakai, N., Sorai, K., Vila-Vilar\'{o}, B., \& Handa, T. 2000, \pasj, 52, 775
\bibitem[Lauberts \& Valentijn(1989)]{LV89} Lauberts, A. \& Valentijn, E. A., The Surface Photometry Catalogue of the ESO-Uppsala Galaxies, ESO, Garching
\bibitem[Leroy et al.(2015)]{Ler15} Leroy, A., Walter, F., Martini, P. et al. 2015, \apj, 814, 83
\bibitem[Lindblad \& Lindblad(1994)]{LL94} Lindblad, P. O., \& Lindblad, P. A. B. 1994, ASPC, 66, 29L
\bibitem[Maciejewski \& Small(2010)]{MS10} Maciejewski, W. \& Small, E. E. 2010, \apj, 719, 622
\bibitem[McCormick et al.(2013)]{McC13} McCormick, A., Veilleux, S., \& Rupke, D. S. N. 2013, \apj, 774, 126
\bibitem[McMullin et al.(2007)]{McM07} McMullin, J. P., Waters, B., Schiebel, D., Young, W., \& Golap, K. 2007, Astronomical Data Analysis Software and Systems XVI (ASP Conf. Ser. 376), ed. R. A. Shaw, F. Hill, \& D. J. Bell (San Francisco, CA: ASP), 127
\bibitem[Meidt et al.(2013)]{Mei13} Meidt, S. E., Schinnerer, E., Garc\'{i}a-Burillo, S., et al. 2013, \apj, 779, 45
\bibitem[Meurer et al.(2002)]{Meu02} Meurer, G., Ferguson, H., et al. 2002, SINGG collaboration
\bibitem[Morgan(1958)]{Mor58} Morgan, W. W. 1958, \pasp, 70, 415
\bibitem[Murray et al.(2005)]{MQT05} Murray, N., Quataert, E., \& Thompson, T. A. 2005, \apj, 618, 569
\bibitem[Murray et al.(2011)]{MMT11} Murray, N., M\'{e}nard, B., \& Thompson, T. A. 2011, \apj, 735, 66
\bibitem[Nakai et al.(1987)]{Nak87} Nakai, N., Hayashi, M., Handa, T., Sofue, Y., \& Hasegawa, T. 1987, \pasj, 39, 685
\bibitem[Nath \& Silk(2009)]{NS09} Nath, B. B. \& Silk, J. 2009, \mnras, 396, L90
\bibitem[Oliva et al.(1995)]{Oli95} Oliva, E., Origlia, L., Kotilainen, J. K., \& Moorwood, A. F. M. 1995, \aap, 301, 55
\bibitem[Onishi et al.(2015)]{Oni15} Onishi, K., Iguchi, S., Sheth, K. \& Kohno, K. 2015, \apj, 806, 39
\bibitem[P\'{e}rez-Ram\'{i}rez et al.(2000)]{PR00} P\'{e}rez-Ram\'{i}rez, D., Knapen, J. H., Peletier, R. F., et al. \mnras, 317, 234
\bibitem[Phillips(1993)]{Phi93} Phillips, A. C. 1993, \aj, 105, 486
\bibitem[Plummer(1911)]{Plu11} Plummer, H. C. 1911, \mnras, 71, 460
\bibitem[Reif et al.(1982)]{Rei82} Reif, K., Mebold, U., Goss, W. M., van Woerden, \& H., Siegman, B., 1982, \aaps, 50, 451
\bibitem[Rogstad et al.(1974)]{RHW74} Rogstad, D. H., Lockhart, I. A., \& Wright, M. C. H. 1974, \apj, 193, 309
\bibitem[Romeo \& Falstad(2013)]{RF13} Romeo, A. B. \& Falstad, N. 2013, \mnras, 433, 1389
\bibitem[Romeo \& Fathi(2015)]{RF15} Romeo, A. B. \& Fathi, K. 2015, \mnras, 451, 3107
\bibitem[Safronov(1960)]{Saf60} Safronov, V. S. 1960, AnAp, 23, 979
\bibitem[Saikia et al.(1990)]{Sai90} Saikia, D. J., Unger, S. W., Pedlar, A., et al. 1990, \mnras, 245, 397
\bibitem[Sakamoto et al.(1999)]{Sak99} Sakamoto, K., Okumura, S. K., Ishizuki, S., \& Scoville, N. Z. 1999, \apjs, 124, 403
\bibitem[Sakamoto et al.(2000)]{Sak00} Sakamoto, K., Baker, A. J., \& Scoville, N. Z. 2000, \apj, 533, 149
\bibitem[Sakamoto et al.(2014)]{Sak14} Sakamoto, K., Aalto, S., Combes, F., Evans, A., \& Peck, A. 2014, \apj, 797, 90
\bibitem[Salak et al.(2013)]{Sal13} Salak, D., Nakai, N., Miyamoto, Y., Yamaguchi, A., \& Tsuru, T. G. 2013, \pasj, 66, 65
\bibitem[Salak et al.(2014)]{Sal14} Salak, D., Nakai, N., \& Kitamoto, S. 2014, \pasj, 66, 96
\bibitem[Shirley(2015)]{Shi15} Shirley, Y. L. 2015, \pasp, 127, 299
\bibitem[Schiano(1985)]{Sch85} Schiano, A. V. R. 1985, \apj, 299, 24
\bibitem[Schwarz(1984)]{Sch84} Schwarz, M. P. 1984, \mnras, 209, 93
\bibitem[S\'{e}rsic \& Pastoriza(1965)]{SP65} S\'{e}rsic, J. L. \& Pastoriza, M. 1965, \pasp, 77, 287
\bibitem[Sharp \& Bland-Hawthorn(2010)]{SB10} Sharp, R. G., \& Bland-Hawthorn, J. 2010, \apj, 711, 818
\bibitem[Sheth et al.(2000)]{She00} Sheth, K., Regan, M. W.., Vogel, S. N., \& Teuben, P. J. 2000, \apj, 532, 221
\bibitem[Sofue et al.(1999)]{Sof99} Sofue, Y., Tutui, Y., Homna, M., et al. 1999, \apj, 523, 136
\bibitem[Sofue \& Rubin(2001)]{SR01} Sofue, Y., \& Rubin, V. 2001, \araa, 39, 137
\bibitem[Sofue(2013)]{Sof13} Sofue, Y. 2013, \pasj, 65, 118
\bibitem[Tacconi-Garman et al.(1996)]{TG96} Tacconi-Garman, L. E., Sternberg, A., \& Eckart, A. 1996, \aj, 112, 918
\bibitem[Tacconi-Garman et al.(2005)]{TG05} Tacconi-Garman, L. E., Sturm, E., Lehnert, M., Lutz, D., Davies, R. I., and Moorwood, A. F. M. 2005, \aap, 432, 91
\bibitem[Terashima et al.(2002)]{Ter02} Terashima, Y., Iyomoto, N., Ho, L. C., \& Ptak, A. F. 2002, \apjs, 139, 1
\bibitem[Toomre(1964)]{Too64} Toomre, A. 1964, \apj, 139, 1217
\bibitem[Tully(1988)]{Tul88} Tully, R. B. 1988, Nearby Galaxy Catalog
\bibitem[Ulich \& Haas(1976)]{UH76} Ulich, B. L., \& Haas, R. W. 1976, \apjs, 30, 247
\bibitem[V\'{e}ron-Cetty \& V\'{e}ron(1985)]{VV85} V\'{e}ron-Cetty, M.-P. \& V\'{e}ron, P. 1985, \aap, 145, 425
\bibitem[Wada(1994)]{Wad94} Wada, K. 1994, \pasj, 46, 165
\bibitem[Walcher et al.(2005)]{Wal05} Walcher, C. J., van der Marel, R. P., McLaughlin, D., et al. 2005, \apj, 618, 237
\bibitem[Walter et al.(2002)]{WWS02} Walter, F., Weiss, A., \& Scoville, N. 2002, \apj, 580, L21
\bibitem[Veilleux et al.(2005)]{VCB05} Veilleux, S., Cecil, G., \& Bland-Hawthorn, J. 2005, \araa, 43, 769
\bibitem[Young \& Scoville(1991)]{YS91} Young, J. S. \& Scoville, N. Z. 1991, \araa, 29, 581
\bibitem[Zhang et al.(2014)]{Zha14} Zhang, Z.-Y., Gao, Y., Henkel, C., Zhao, Y., Wang, J., et al. 2014, \apj, 784, L31
\end{thebibliography}
\end{document}